\documentclass[12pt,letterpaper]{article}  

\usepackage[includeheadfoot,
            marginratio={1:1,2:3}, 
            width=440pt, 
            height=688pt,]{geometry}
\usepackage[utf8]{inputenc}
\usepackage[12pt]{moresize}
\usepackage{amsmath}
\usepackage{amsfonts}
\usepackage{amssymb}
\usepackage{stmaryrd}
\usepackage{graphicx}
\usepackage{booktabs}
\usepackage{multirow}
\usepackage{adjustbox}
\usepackage{bbold}
\usepackage{braket}
\usepackage[utf8]{inputenc}
\usepackage{tikz}
\usetikzlibrary{arrows}
\usetikzlibrary{decorations.markings}
\tikzset{
  big arrow/.style={
    decoration={markings,mark=at position 1 with {\arrow[scale=2,#1]{>}}},
    postaction={decorate},
    shorten >=0.4pt},
  big arrow/.default=black}
\usepackage{float}
\usepackage{empheq}
\usepackage{paralist}
\usepackage{cite}
\usepackage[hidelinks]{hyperref}

\newcommand{\nc}{\newcommand}
\nc{\lb}{\llbracket}
\nc{\rb}{\rrbracket}
\nc{\gl}{\llbracket}
\nc{\gr}{\rrbracket}
\nc{\del}{\partial}
\nc{\de}{\mathrm{d}}

\newcommand{\eq}[1]{\begin{equation}
                     \begin{split} #1 \end{split}
                     \end{equation}}

\newcommand{\ov}{\overline}

\allowdisplaybreaks[2]
\numberwithin{equation}{section}

\begin{document}

\vspace*{-1.5cm}
\begin{flushright}
  {\small
  MPP-2020-20\\
  }
\end{flushright}

\vspace{1.5cm}
\begin{center}
{\LARGE
dS Spaces and Brane Worlds   \\[0.2cm]
in Exotic String Theories\\[0.2cm]
} 
\vspace{0.4cm}

\end{center}

\vspace{0.35cm}
\begin{center}
  Ralph Blumenhagen$^{1}$, Max Brinkmann$^{1}$, Andriana
  Makridou$^{1,2}$, \\[0.1cm]
Lorenz Schlechter$^{1}$, Matthias Traube$^{1}$
\let\thefootnote\relax\footnote{\noindent \ssmall email: \texttt{ 
	\href{mailto:blumenha@mpp.mpg.de}{blumenha@mpp.mpg.de},
	\href{mailto:mbrinkm@mpp.mpg.de}{mbrinkm@mpp.mpg.de}, 
	\href{mailto:amakrido@mpp.mpg.de}{amakrido@mpp.mpg.de}, 
	\href{mailto:lschlech@mpp.mpg.de}{lschlech@mpp.mpg.de}, 
	\href{mailto:mtraube@mpp.mpg.de}{mtraube@mpp.mpg.de}} }
\\
\end{center}

\vspace{0.1cm}
\begin{center} 
\emph{$^{1}$ Max-Planck-Institut f\"ur Physik (Werner-Heisenberg-Institut), \\ 
   F\"ohringer Ring 6,  80805 M\"unchen, Germany } \\[0.1cm] 
\vspace{0.25cm} 
\emph{$^{2}$ Ludwig-Maximilians-Universit\"at M\"unchen, Fakult\"at f\"ur Physik,\\ 
               Theresienstr.~37, 80333 M\"unchen, Germany}\\

\vspace{0.2cm}

 \vspace{0.3cm} 
\end{center} 

\vspace{0.5cm}


\begin{abstract}\noindent
We investigate  string-phenomenological questions of Hull's exotic
superstring theories  with Euclidean strings/branes and 
multiple times.  
These  are known to be plagued by pathologies like the occurrence of  ghosts. 
On the other hand, these theories exhibit de Sitter solutions.
Our special focus lies on the question of the coexistence of such de Sitter solutions
and ghost-free brane worlds.
To this end, the world-sheet CFT description of Euclidean 
fundamental strings is generalized to include also the open string/D-brane sector.
Demanding that in the ``observable'' gauge
theory sector the gauge fields themselves are non-ghosts, 
a generalization of the dS swampland conjecture is found.
\end{abstract}

\setcounter{footnote}{0}
\clearpage

\tableofcontents


\section{Introduction}
\label{sec:intro}

Even though it is believed that there exists a huge landscape of
string compactifications, it turns out to  be notoriously difficult to realize  certain
four-dimensional properties. This led to the idea of the swampland 
\cite{Vafa:2005ui}
which contains those low-energy effective theories that cannot be 
UV completed to a consistent theory of quantum gravity.
The swampland program intends to extract a set of
relatively simple quantitative features 
that  low-energy effective field theories should satisfy to admit
such an embedding into a theory of quantum gravity
(see~\cite{Palti:2019pca} for a recent review). 

In the meantime several such swampland conjectures have been proposed
{\cite{ArkaniHamed:2006dz,Ooguri:2006in,Klaewer:2016kiy,Ooguri:2016pdq,Palti:2017elp,Obied:2018sgi,Andriot:2018wzk, 
Cecotti:2018ufg,Garg:2018reu,Ooguri:2018wrx,Gautason:2018gln,Klaewer:2018yxi,Heckman:2019bzm,Lust:2019zwm,Bedroya:2019snp,Kehagias:2019akr,Blumenhagen:2019vgj}},
triggering further new developments like  the  emergence proposal 
\cite{Heidenreich:2017sim,Grimm:2018ohb,Heidenreich:2018kpg} of infinite
distances in field space or the appearance of towers of light strings \cite{Lee:2018urn,Lee:2018spm,Lee:2019wij}.
One of these conjectures is the dS swampland
conjecture~\cite{Obied:2018sgi}
(see also \cite{Danielsson:2018ztv}), whose
classical version forbids dS minima altogether, even 
meta-stable ones with  a finite life-time. This conjecture has been
refined~\cite{Ooguri:2018wrx,Garg:2018reu} and subsequently related to a quantum argument, the 
trans-Planckian censorship conjecture~\cite{Bedroya:2019snp}. This quantum generalization
admits meta-stable dS minima as long as their life-time is
sufficiently small.

String theory as we know it is a background dependent formulation of
quantum gravity.  Therefore, most of the evidence for these
conjectures is derived from concrete examples of string
compactifications and their effective four-dimensional  field
theories. It should be mentioned that alternative arguments against de Sitter, based on the concept of quantum
breaking \cite{Dvali:2014gua,Dvali:2017eba,Dvali:2018jhn,Dvali:2018fqu}, have been formulated.
However, it could well be that the evidence
supporting the swampland conjectures is biased by looking just at a
certain, well understood,  subset of all consistent string theory backgrounds. 
Can one for instance imagine other backgrounds where de Sitter spaces
appear naturally?\footnote{On the same note, in \cite{Dasgupta:2019gcd,Dasgupta:2019vjn} the authors
  argued that time dependent backgrounds might circumvent the dS swampland conjecture.}

Indeed, such settings have been known since the early work of
C. Hull et al.~\cite{Hull:1998vg, Hull:1998ym,Hull:1998fh,Hull:1999mt} since 1998. 
By applying T-dualities along time-like directions, new string theories
were proposed. Their common features include that some of the
massless fields exhibit the wrong sign in their kinetic terms, and that extra
time-like directions can appear. These so-called exotic string theories were shown to form a
network related via T- and S-dualities. Roughly one half of these 10D 
exotic string theories still contain Lorentzian fundamental strings
while the other half features Euclidean string world-sheets. 
In many cases, the open string sector changes to include Euclidean D-branes\footnote{
Questions concerning the backreaction of spherical Euclidean D-branes have been studied 
lately in a series of papers \cite{Bobev:2018ugk,Bobev:2019bvq}.
}.
A comprehensive overview of the exotic theories can be found in Fig. \ref{fig:dualityweb1}.
Work towards a perturbative CFT
description of Euclidean fundamental strings was started
in~\cite{Dijkgraaf:2016lym}, where it was shown that
extra factors of $i=\sqrt{-1}$ appear and need to be taken into account.
In such a perturbative approach a number  of pathologies arise.
One is the aforementioned issue of ghost fields, while another is
the appearance of arbitrarily light states upon compactification of
time-like directions \cite{Dijkgraaf:2016lym}. It is believed that all these
pathologies are rooted in dealing with closed time-like curves, at
least in intermediate steps.
However, it has been argued \cite{Hull:1998vg} that in the UV complete theory 
these issues could resolve, and the IR pathologies are only a result of the 
perturbative approach.

Despite these open questions, we think
one should not immediately dismiss these models, since
their supergravity theories turn out to  admit dS solutions. 
Viewing the presence of ghost fields in the 10D supergravity actions 
as a feature rather than a bug, 
it is immediately clear that the standard tree-level dS no-go
theorem ~\cite{Hertzberg:2007wc}, which was extended in ~\cite{Obied:2018sgi}, does not apply.
Indeed, in~\cite{Hull:1998vg} it was already pointed out that one of these exotic string
theories does admit a $dS_5\times  {\cal H}_5$ solution, where
${\cal H}$ denotes the hyperbolic five-plane.
From this more general perspective,  one should at least be able to
learn something  about  which physical concepts need to be relaxed
to make dS possible. 

Moreover, in the past only a few, rather formal  aspects of
these exotic theories were investigated, so we think it is time
to also confront these exotic string theories with more phenomenological
questions.
For instance, one can ask whether such unconventional theories can
nevertheless contain a sector that phenomenologically does resemble
our low-energy world.  Not much is experimentally known about quantum
aspects of gravity, but gauge theories are tested experimentally to
very high precision. They are  free from physical ghosts and are unitary, 
or at least any deviation from these principles has escaped detection.

Therefore, in this paper we will start a string-phenomenological study 
of Hull's exotic string theories. Using complementary methods to the
ones  employed in \cite{Hull:1998fh,Hull:1999mt},
we will describe D-branes
for Euclidean exotic theories from a formal CFT point of view.
This provides the tools to analyze
whether such branes can support  bona fide gauge theories, though embedded
into a closed string background that has some of  the strange features
already mentioned.

This paper is organized as follows:
In a preliminary section \ref{sec_prelim} we will start by looking at the probably most studied 
string background, which is the famous type IIB $AdS_5\times S^5$ 
space supported by a self-dual five-form background, and  
observe  that  in  type IIB-like string theories with wrong kinetic terms  and/or time-like
directions this solution generalizes 
to\footnote{In our notation, signature $(q,p$) refers to $q$ space-like and $p$ time-like dimensions.}
$AdS_{5-m,m}\times dS_{5-n,n}$.
This little exercise provides some  motivation  to  contemplate cosmological and
phenomenological applications of exotic string theories, which are the
natural home for these dS solutions.
Next, we will review Hull's exotic string theories and 
the web of dualities that connects them to ordinary type II string theory. 

In section  \ref{sec_ghosts} we will discuss the appearance of closed
string ghosts and their consequences in more detail. We will also
encounter
other pathologies that have to do with the appearance of 
infinitely many ultra-light states once time-like directions are
compactified. Such compactifications seem inevitable, if we want 
to relate the multiple time exotic theories to our $3+1$
dimensional world.
While the usual approach of  gauging  extra world-sheet symmetries
(like for the $N=2$ heterotic string \cite{Ooguri:1991ie}) is not an option to get rid of the ghosts,
removing (part of) them  via an orbifold projection turns out to be feasible. 

Continuing the work of \cite{Dijkgraaf:2016lym}, in section \ref{sec_cft} we develop
CFT techniques for  the Euclidean 
exotic string theories and in particular provide  the
description of the open string sector.
Here subtle differences to the standard string theory 
with Lorentzian world-sheet signature appear, e.g. extra complex phases in the amplitudes.
This in turn allows us to constrain the D-brane spectrum of these Euclidean
theories by requiring real tensions. We provide a general formula 
which gives the spectrum of allowed D-branes in any signature.

Finally, in section \ref{sec_four}  we will discuss 
string-phenomenological aspects of these D-brane theories. 
We complement our results from the previous section with a 
different construction, employing a mapping motivated by
negative tension branes \cite{Dijkgraaf:2016lym} 
to derive their effective actions. This alternative
derivation verifies the  brane spectrum found using CFT methods. Both methods agree with 
the results obtained in \cite{Hull:1998fh}.
Among the branes we then search for brane-world theories that are 
phenomenologically viable, i.e.  free of massless ghosts and
featuring  a  $(3,1)$ subspace.
We will see that while there is such a brane in every exotic theory with 
Euclidean strings, the  O-planes necessary for tadpole
cancellation are precisely those from section~\ref{sec_ghosts}. 
Therefore, all massless ghosts of the 10D theory, including closed string ghosts, 
are projected out and the  loophole for dS solutions closes. 
Next we discuss brane worlds for Lorentzian exotic string theories.
They do  seem to admit a ghost-free massless 
brane sector. However here the problem of ultralight string modes
previously encountered for closed strings in section \ref{sec_ghosts} 
also applies to open string modes. 
This means, although the truly massless sector is ghost-free, 
there are infinitely many arbitrarily light states in the theory.

\section{Preliminaries}
\label{sec_prelim}

In this section we first recall that in theories with more time-like
directions the $AdS_5\times S^5$ solution of type IIB supergravity
generalizes to solutions containing de Sitter spaces (cf. \cite{Hull:1998fh}). The natural
habitat of these solutions are Hull's exotic string theories that we review
in the second part of this section.

\subsection{Fluxed $AdS \!\times\! dS$ solutions}
\label{sec_two}

The prototype solution of the type IIB superstring theory with flux is
$AdS_5\times S^5$ with self-dual five-form flux supported on
$AdS_5$ and $S^5$, respectively. Of course this theory has just a
single time-like coordinate which is part of the $AdS_5$ background.
The question that we would like to approach in this section is what happens if
more than one of the ten directions of type IIB were  time-like, i.e.
on a space with signature $(10-p,p)$.
For the five-form to still satisfy a self-duality  relation, one must
have $p$ odd.

The 10D effective (quasi-)action governing the dynamics of the metric and 
a form field $C_{n-1}$ reads
\eq{
\label{actioneffec}
  S\sim M_s^8 \int d^{10}x \sqrt{|G|}  \left( e^{-2\phi} R -{\kappa\over 2} |F_n|^2  \right)
}
where in the following we will  set the dilaton to a constant. This is justified for
the actual case of interest, namely the R-R four-form, for which in addition 
one has to impose the self-duality relation $F_5=\pm \star F_5$ by hand
and change the prefactor of $|F_n|^2$ to $\kappa/4$. Here we have left the
sign $\kappa=\pm 1$ of its kinetic term open, where  $\kappa=1$ is the
usual case.
The kinetic term of the n-form $F_n$ is defined as
\eq{
                  |F_n|^2= {1\over n!} \, G^{i_1 j_1} \ldots G^{i_n j_n}\,
                  F_{i_1 \ldots i_n}\,  F_{j_1 \ldots j_n}\,.
}
The resulting equation of motion for the metric reads
\eq{
\label{eomone}
            R_{ij} -{1\over 2} g_{ij}R  \;=\;{\kappa\over 2 (n-1)!} \,\Big( F_{i\,
              k_2\ldots k_n}\, F_{j}^{k_2\ldots k_n}
             -{1\over 2n}
            g_{ij} F_{k_1\ldots k_n} F^{k_1\ldots k_n} \Big)
}
and for $C_{n-1}$
\eq{
\label{eomtwo}
          \partial_i \left( \sqrt{|G|}\, F^{i\, k_2 \ldots
              k_n}\right)=0\,.
}
We can write the first relation \eqref{eomone}  as a matrix equation
${\bf R}=\kappa {\bf T}$.

Now, we want to consider these equations in a theory with more time-like
directions. Generalizing the $AdS_5\times S^5$ solutions, we make the ansatz
\eq{
                    AdS_{5-m,m}\times dS_{5-n,n}\,,\qquad {\rm with}\quad
                    m+n=p={\rm odd}\,.
}
The description of such multiple times $AdS$ and $dS$ spaces is reviewed in
appendix \ref{appendix_AdS}.
For a self-dual  five-form flux we can then solve \eqref{eomtwo}
simply by choosing $F_5$ to satisfy the Bianchi identity $dF_5=0$.
This is the case for 
\eq{
\label{fluxex}
          F_5=f E^1\wedge \ldots \wedge E^5 -f {\cal E}^1 \wedge \ldots\wedge {\cal
            E}^5
}
with constant $f$ and with  the 5-beins of $AdS_{5-m,m}$ and
$dS_{5-n,n}$ as reviewed in appendix \ref{appendix_AdS}.
Choosing the same curvature radius $\alpha$ for the $AdS$ and $dS$ factors, 
the Ricci scalar vanishes and the left hand side of \eqref{eomone} becomes
\eq{
       {\mathbf R}=\left(\begin{matrix}
                                    -{4\over \alpha^2} \eta^{(m,5-m)} &  0  \\
                                     0   & {4\over \alpha^2} \eta^{(n,5-n)} \end{matrix}\right)\,.
}
The right hand side then is
\eq{
       \kappa  {\mathbf T}=\kappa\, (-1)^n \left(\begin{matrix}
                                    -{f^2\over 4} \eta^{(m,5-m)} &  0  \\
                                     0   & {f^2\over 4}
                                     \eta^{(n,5-n)} \end{matrix}\right)\,.
}
Therefore, for $\alpha=4/f$ the equations of motion are satisfied if
we choose $\kappa=1$ for $n$ even and $\kappa=-1$ for $n$ odd.

Let us mention a few special cases: 
For  $m=1$, $n=0$ one gets the original $AdS_5\times S^5$ solution
and for $m=0$, $n=1$ one finds ${\cal H}_5\times dS_{4,1}$, where
${\cal H}_5$  denotes the hyperbolic 5-space. However, the price one
has to pay to get this simple solution is that the R-R five-form has
the wrong sign of the kinetic term.

We note that all these solutions  can also be understood by applying 
$(m\!-\!1,n)$ Wick-rotations to the respective coordinates of the original type IIB 
$AdS_5\times S^5$ background.
From this perspective,
in order to keep $F_5$ purely real or imaginary, one has to apply
either an $({\rm even},{\rm even})$ or an $({\rm odd},{\rm odd})$
number of Wick-rotations.
In the first case, $F_5$ remains real, giving the solutions with $n$
even and $\kappa=1$. 
In the second case however, $F_5$ becomes 
purely imaginary, so the sign of the kinetic term indeed changes
and one finds the  $n$ odd, $\kappa=-1$ solutions. 
If for the original type IIB the 5-form is chosen to be self-dual 
$\star F_5=F_5$, the Wick rotation changes this to
$\star F_5=\kappa F_5$. Thus, the sign of the kinetic term of $F_5$ 
and the one in the self-duality relation are related.

We have seen that in type IIB-like  supergravities with 
multiple time directions and possibly wrong signs of the kinetic term
for the 5-five form, $dS$ solutions do exist. Of course, our analysis 
was only applied to a subsector of the full initial type IIB
supergravity action so that one might wonder whether fully consistent
supergravity or  string theories exist that  exhibit precisely
those two features.

\subsection{Exotic superstring theories}
\label{sec_three}

Since the early work of Hull~\cite{Hull:1998vg, Hull:1998ym} it is
known that string theories 
with exotic signatures arise from the usual type II theories with  $(9,1)$ signature by
applying successive T-duality also along time-like directions. This leads
to an intricate web of dual theories in ten dimensions of more general
signature $(10-p,p)$, whose supergravity actions (quadratic in derivatives) are
similar to the type II actions but contain kinetic terms of opposite
sign. Despite these apparent ghosts, it was argued that each theory  
of this duality web represents a different limit of ordinary type II theories, 
and as a full non-perturbative theory should therefore be intrinsically well-behaved. 

However,  since these theories are reached via a circle
compactification of a time-like direction, they could also all be
severely pathological, as at an  intermediate step closed time-like curves are
encountered that are generally thought to be highly problematic. In the course
of  this paper we assume that this is not the case and that Hull's exotic theories
can make sense.

The de Sitter solutions from the previous subsection will find their
natural home in these exotic supergravity theories, meaning that they arise as solutions 
to the effective  theories at leading order in derivatives and at weak string
coupling.
Therefore, it is this limit that we are most interested in. 
The perturbative spectrum of the exotic closed string theories which arises 
via quantization of the corresponding fundamental string was recently 
worked out in~\cite{Dijkgraaf:2016lym}. As expected, the perturbative description
of exotic theories carries many pathologies, most prominently ghosts.
In this section we review the bouquet of exotic string
theories, for more details we refer to the original literature.

\subsubsection*{The zoo of type II$^{\alpha\beta}$ theories}

T-duality along a space-like direction exchanges type IIA and IIB string theory. 
Along a time-like direction, this cannot be the case. 
For instance, Dirichlet and Neumann boundary conditions of a D-brane are interchanged
in the direction that T-duality is applied to. Since regular type II
theories only contain Lorentzian D-branes, this means that the T-dual 
 theory can only have Euclidean branes.

Adopting the notation introduced in~\cite{Dijkgraaf:2016lym}, we label the
theories as IIA$^{\alpha\beta}_{(10-p,p)}$ and IIB$^{\alpha\beta}_{(10-p,p)}$ 
with two signs $\alpha,\beta\in\{+,-\}$  and the space-time signature $(10-p,p)$.
The first sign indicates whether the theory contains 
Lorentzian ($+$) or Euclidean ($-$) fundamental strings,
while the second indicates the same for D1/D2 branes. 
Here, we will call any even (odd) number of time-like directions Euclidean (Lorentzian).
If the signature is omitted we assume $(9,1)$. 
The usual string theories in this notation are IIA$^{++}$, IIB$^{++}$.
We will also use the notation IIA$^{\rm L}$ and IIB$^{\rm L}$
collectively for all theories with
Lorentzian fundamental strings and IIA$^{\rm E}$ and  IIB$^{\rm E}$
for the ones with Euclidean strings. 

Starting from the usual string theories, time-like T-duality as discussed above 
leads to Euclidean branes of one dimension less.
This means that (IIA$^{++} \leftrightarrow$ IIB$^{+-}$)
and (IIB$^{++} \leftrightarrow$ IIA$^{+-}$) 
are related by time-like T-duality, just as  
(IIA$^{++} \leftrightarrow$ IIB$^{++}$) 
and (IIB$^{+-} \leftrightarrow$ IIA$^{+-}$) 
are space-like T-duals.

Now taking the strong coupling limit of the type IIB theories,
S-duality acts by exchanging F1$\leftrightarrow$D1 and NS5$\leftrightarrow$D5, 
while D3 is self-dual.
Then while IIB$^{++}$ is self-dual, IIB$^{+-}$ has Euclidean D-branes 
that now get exchanged with the fundamental string and NS5-brane.
The resulting theory must be of type IIB$^{-+}$ with
Lorentzian D1 and D5 and Euclidean F1, NS5 and D3-branes.

One  can now complete the type II picture by considering the T-duals of this 
exotic IIB$^{\rm E}$ theory. However since the D-branes of type IIB$^{-+}$ are not 
homogeneously Lorentzian, one can see that type IIB$^{-+}$ compactified along 
a space-like circle must be dual to a theory compactified along a time-like 
direction! This theory has Euclidean D2-branes and is thus of 
type IIA$^{--}_{(8,2)}$.

All theories with Euclidean F1 have the property that T-duals are with respect
to different signature directions. Their respective D-brane spectrum is 
alternating between Euclidean and Lorentzian as was the case for 
type IIB$^{-+}$, and each T-dualization changes the signature. 
The list of T-dual theories with Euclidean fundamental strings is 
schematically given by
\eq{
\mathrm{IIA}^{-+}_{(10,0)} &\leftrightarrow \mathrm{IIB}^{-+}_{(9,1)} \leftrightarrow \mathrm{IIA}^{--}_{(8,2)}\leftrightarrow \mathrm{IIB}^{--}_{(7,3)} \leftrightarrow \mathrm{IIA}^{-+}_{(6,4)} \leftrightarrow ...\leftrightarrow \mathrm{IIB}^{-+}_{(1,9)} \leftrightarrow \mathrm{IIA}^{--}_{(0,10)} 
} 
where going to the right (left) means T-dualizing along a space-like
(time-like) direction. 

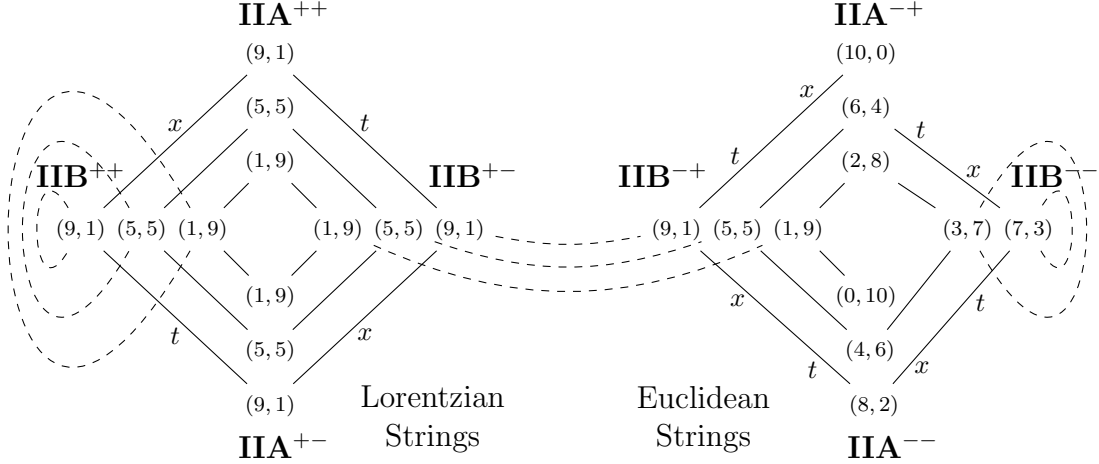
\begin{figure}[t]
	\begin{center}
		\begin{tikzpicture}[scale=1.8]
			\clip(-4.4,-2) rectangle (4,2);
			\node(IIA++91) at (-2.2,1.3) {\scriptsize $(9,1)$};
			\node(IIA++55) at (-2.2,0.9) {\scriptsize $(5,5)$};
			\node(IIA++19) at (-2.2,0.5) {\scriptsize $(1,9)$};
			\node(IIA+-91) at (-2.2,-1.3) {\scriptsize $(9,1)$};
			\node(IIA+-55) at (-2.2,-0.9) {\scriptsize $(5,5)$};
			\node(IIA+-19) at (-2.2,-0.5) {\scriptsize $(1,9)$};
			\node(IIB++91) at (-3.6,0) {\scriptsize $(9,1)$};
			\node(IIB++55) at (-3.15,0) {\scriptsize $(5,5)$};
			\node(IIB++19) at (-2.7,0) {\scriptsize $(1,9)$};
			\node(IIB+-91) at (-0.8,0) {\scriptsize $(9,1)$};
			\node(IIB+-55) at (-1.25,0) {\scriptsize $(5,5)$};
			\node(IIB+-19) at (-1.7,0) {\scriptsize $(1,9)$};
			\draw[] (IIA++91) to node[above,pos=0.5]{\footnotesize$x$} (IIB++91) 
				to node[below,pos=0.5]{\footnotesize$t$} (IIA+-91) 
				to node[below,pos=0.5]{\footnotesize$x$} (IIB+-91) 
				to node[above,pos=0.5]{\footnotesize$t$} (IIA++91);
			\draw[] (IIA++55) to (IIB++55) to (IIA+-55) to (IIB+-55) to (IIA++55);
			\draw[] (IIA++19) to (IIB++19) to (IIA+-19) to (IIB+-19) to (IIA++19);
			\node[align=center](LS) at (-1,-1.4) {Lorentzian\\Strings};
			\node[rectangle,rounded corners](IIA++) at (-2.1,1.6) {\bf IIA$^{++}$};
			\node[rectangle,rounded corners](IIA+-) at (-2.1,-1.6) {\bf IIA$^{+-}$};
			\node[rectangle,rounded corners](IIB+-) at (-.7,0.4) {\bf IIB$^{+-}$};
			\node[rectangle,rounded corners](IIB++) at (-3.6,0.4) {\bf IIB$^{++}$};
			
			\node(IIA-+100) at (2.2,1.3) {\scriptsize $(10,0)$};
			\node(IIA-+64) at (2.2,0.9) {\scriptsize $(6,4)$};
			\node(IIA-+28) at (2.2,0.5) {\scriptsize $(2,8)$};
			\node(IIA--82) at (2.26,-1.3) {\scriptsize $(8,2)$};
			\node(IIA--46) at (2.23,-0.9) {\scriptsize $(4,6)$};
			\node(IIA--010) at (2.2,-0.5) {\scriptsize $(0,10)$};
			\node(IIB-+91) at (0.8,0) {\scriptsize $(9,1)$};
			\node(IIB-+55) at (1.25,0) {\scriptsize $(5,5)$};
			\node(IIB-+19) at (1.7,0) {\scriptsize $(1,9)$};
			\node(IIB--73) at (3.4,0) {\scriptsize $(7,3)$};
			\node(IIB--37) at (2.95,0) {\scriptsize $(3,7)$};
			\draw[] (IIA-+100) to node[above,pos=.25]{\footnotesize$x$} 
					node[above,pos=0.75]{\footnotesize$t$} (IIB-+91) 
				to node[below,pos=.25]{\footnotesize$x$} 
					node[below,pos=0.75]{\footnotesize$t$} (IIA--82) 
				to node[below,pos=.25]{\footnotesize$x$} 
					node[below,pos=0.75]{\footnotesize$t$} (IIB--73) 
				to node[above,pos=.25]{\footnotesize$x$} 
					node[above,pos=0.75]{\footnotesize$t$} (IIA-+64) 
				to (IIB-+55) to (IIA--46) to (IIB--37) to (IIA-+28) to (IIB-+19) to (IIA--010);
			\node[align=center](ES) at (1,-1.4) {Euclidean\\Strings};
			\node[rectangle,rounded corners](IIA-+) at (2.3,1.6) {\bf IIA$^{-+}$};
			\node[rectangle,rounded corners](IIA--) at (2.4,-1.6) {\bf IIA$^{--}$};
			\node[rectangle,rounded corners](IIB-+) at (.7,0.4) {\bf IIB$^{-+}$};
			\node[rectangle,rounded corners](IIB--) at (3.6,0.4) {\bf IIB$^{--}$};	
			
			\draw[dashed] (IIB+-91) to[out=-11,in=191] (IIB-+91);
			\draw[dashed] (IIB+-55) to[out=-19,in=199] (IIB-+55);
			\draw[dashed] (IIB+-19) to[out=-24,in=204] (IIB-+19);
			\draw[dashed] (IIB++19) to[out=120,in=240,looseness=29] (IIB++19);
			\draw[dashed] (IIB++55) to[out=120,in=240,looseness=17] (IIB++55);
			\draw[dashed] (IIB++91) to[out=120,in=240,looseness=5] (IIB++91);
			\draw[dashed] (IIB--37) to[out=-60,in=60,looseness=17] (IIB--37);
			\draw[dashed] (IIB--73) to[out=-60,in=60,looseness=5] (IIB--73);
			
		\end{tikzpicture}
		\caption{T-dualities (solid lines) and S-dualities
                  (dashed lines) relating type II string theories. The
                  label $x$ ($t$) indicates dualities arising from
                  compactification on a spatial (time-like) circle. 
                  The left side consists of the type IIA$^{\rm L}$/IIB$^{\rm L}$
                  theories with Lorentzian fundamental strings, the theories with 
                  Euclidean fundamental strings (IIA$^{\rm E}$/IIB$^{\rm E}$)
                  are on the right.
			\emph{(Diagram adopted from~\cite{Dijkgraaf:2016lym})}}
		\label{fig:dualityweb1}
	\end{center}
\end{figure}

Now that we have found more signatures of type IIB$^{-+}$, we can S-dualize 
back to the theories with Lorentzian fundamental strings, where we find
that they correspond to type IIB$^{+-}$ with the same signatures. 
The full bouquet of dual theories and their relations is shown in Fig.~\ref{fig:dualityweb1}.

For completeness let us mention that the strong coupling limit  of the type IIA theories 
are two M-theory variants. In other words, the type IIA theories arise 
from M theories with Lorentzian or Euclidean M2-branes on various 
signatures, compactified on space- or time-like circles.
For more on the exotic M-theories see \cite{Dijkgraaf:2016lym,Hull:1998ym}.


\subsubsection*{Type II$^{\alpha\beta}$ supergravities}

The bosonic part of the low-energy SUGRA actions for the exotic theories have been worked out 
in~\cite{Hull:1998vg,Hull:1998ym}. Here we provide the compact presentation
given in~\cite{Dijkgraaf:2016lym}. As usual the 10D actions are given by a sum 
over  NS-NS, R-R and Chern-Simons  contributions
\eq{
S[{\rm IIA/B}^{\alpha\beta}]
= S_{\rm NS}^{\alpha\beta}
+ S_{\rm R}[{\rm A/B}]^{\alpha\beta}
+ S_{\rm CS}[{\rm A/B}]
}
where the NS-NS part is the same for type IIA and IIB and the CS part is 
independent of $\alpha,\beta$. The respective terms in the action are given by
\eq{
\label{eqn:sugraaction}
S_{\rm NS}^{\alpha\beta} 
&= \frac{1}{2 \kappa_{10}^2} \int \de^{10} x \sqrt{|\det G|} \ e^{-2 \Phi}\! \left[\mathcal{R} + 4(\nabla \Phi)^2 - \frac{\alpha}{2} |H_3|^2 \right] \,, 
\\[5pt] 
S_{\rm R}[{\rm A}]^{\alpha\beta} &= -\frac{1}{2 \kappa_{10}^2} \int \de^{10} x \sqrt{|\det G|} \left[\frac{\alpha \beta}{2} |F_2|^2 + \frac{\beta}{2} |\tilde{F}_4|^2 \right] \,, \\
S_{\rm R}[{\rm B}]^{\alpha\beta} &= -\frac{1}{2 \kappa_{10}^2} \int \de^{10} x \sqrt{|\det G|} \left[\frac{\alpha \beta}{2} |F_1|^2 + \frac{\beta}{2} |\tilde{F}_3|^2 + \frac{\alpha \beta}{4} |\tilde{F}_5|^2 \right] \,, 
\\[5pt]
S_{\rm CS}[{\rm A}] &= -\frac{1}{4 \kappa_{10}^2} \int B_2\wedge F_4 \wedge F_4 \,, \\
S_{\rm CS}[{\rm B}] &=  -\frac{1}{4 \kappa_{10}^2} \int B_2 \wedge F_3 \wedge F_5 \,,
}
where $H_3 = d B_2$, $F_p = d C_{p-1}$, and $\tilde{F}_p = F_p - H_3 \wedge C_{p-3}$. 
These actions are all independent of the respective  signature $(p,q)$, and as usual 
one has to additionally require (anti-)self-duality 
$\tilde F_5 = (\alpha\beta) \star \tilde F_5$ in the type IIB
variants.  Note that as in the previous subsection the sign of the
kinetic term of the 5-form flux is correlated with the sign in the
self-duality relation.
Moreover, we have set the Romans mass of type IIA variants  to zero for simplicity.

It is clear that in general these actions feature the appearance of
ghost states, i.e. states whose kinetic term has the wrong sign.
This sign is the result of two effects. First, there is the overall
sign of the kinetic terms in \eqref{eqn:sugraaction} and second
the combination of signs of the inverse metric factors in
\eq{
           {\cal L}_{\rm kin}\sim      -\kappa \, \sqrt{|G|}\, |F_n|^2= -{\kappa\over n!} \sqrt{|G|}\, G^{i_1 j_1} \ldots G^{i_n j_n}\,
                  F_{i_1 \ldots i_n}\,  F_{j_1 \ldots j_n}\,.
}
If $\kappa= + / \!- 1$, an odd/even number of time-like indices indicates a ghost. 
The presence of ghosts is of course strongly related to the existence
of dS solutions.
Recall that in section \ref{sec_prelim} we have seen that theories with ghosts can admit 
solutions to the SUGRA equations of motion that contain dS factors.
These ghosts could  arise  either due to wrong overall signs of the
kinetic terms or due to extra time-like directions. 
These are precisely the two issues that also appear for the 
 exotic superstring theories, making them  the natural framework for 
a string theory embedding of the dS solutions of section~\ref{sec_prelim}.

It is by now folklore that finding dS vacua in string theory is a highly nontrivial undertaking. 
As a matter of fact, the dS swampland conjecture \cite{Obied:2018sgi} 
explicitly forbids dS vacua in string theory as the scalar potential should always satisfy
\eq{
                 |\nabla V|> c\, V \,,
} 
where $c$ is an order one number.  In fairly general tree-level setups, there exist no-go theorems
\cite{Maldacena:2000mw,Hertzberg:2007wc} (and extensions in\cite{Obied:2018sgi}) that explicitly forbid dS vacua.
Moreover, in \cite{Junghans:2018gdb} dS was excluded in parametrically controlled regimes.
However, the derivation of the above arguments
implicitly assumes that all fields have the usual  kinetic terms.
The explicit dS solutions show that  violations of the dS
no-go theorems  can potentially arise from the presence of ghost fields.
For a concrete set-up,  an  effective 4D potential is generated  via dimensional
reduction of an exotic string theory on some internal space with
non-trivial fluxes turned on. Whether this effective potential indeed
admits
dS minima requires a more detailed investigation, but as long
as there are ghosts present we expect that the no-go theorems will
not hold.

\subsubsection*{Type II$^{\alpha\beta}$ brane spectrum} 
\label{branesector}

The brane spectrum of the exotic superstring theories has been worked out in \cite{Hull:1998fh,Hull:1999mt} 
by explicitly solving the equations of motion for the exotic supergravity theories. 
The results for branes coupling to RR-fields are summarized in tables \ref{table_HullIIA} 
and \ref{table_HullIIB} in the type~II$^{\alpha \beta}$ notation of \cite{Dijkgraaf:2016lym}. 
In addition to the branes listed here, in \cite{Hull:1998fh} an almost complete spectrum, 
including NS-NS branes, pp-waves etc., was derived. 
In the tables an entry $(p,q)$ stands for a $p+q$-dimensional brane with world volume of signature $(p,q)$. 
Table \ref{table_HullIIA} lists the Dp-branes for all type IIA$^{\alpha \beta}$ theories, 
while table \ref{table_HullIIB} shows the brane spectrum of the IIB$^{\alpha \beta}$ theories. 


\begin{table}[ht]
\centering
\begin{tabular}{l c c c c c}
  \toprule
  Theory   & D0 & D2 & D4 & D6 & D8   \\
    \midrule 
IIA$^{-+}_{(10,0)}$ & $(1,0)$		& - 					& $(5,0)$ 						& - 					& - \\ \addlinespace
IIA$^{++}_{(9,1)}$  & $(0,1)$   	& $(2,1)$  			& $(4,1)$				 		& $(6,1)$ 			& $(8,1)$ \\ \addlinespace
IIA$^{+-}_{(9,1)}$  & $(1,0)$   	& $(3,0)$ 			& $(5,0)$ 						& $(7,0)$     		& $(9,0)$ \\ \addlinespace
IIA$^{--}_{(8,2)}$  & $(0,1)$   	& $(3,0)$,$(1,2)$ 	& $(4,1)$	 					& $(7,0)$,$(5,2)$ 	& $(8,1)$ \\ \addlinespace    
IIA$^{-+}_{(6,4)}$  & $(1,0)$   	& $(2,1)$,$(0,3)$ 	& $(5,0)$,$(3,2)$,$(1,4)$    	& $(6,1)$,$(4,3)$ 	& $(5,4)$ \\ \addlinespace
IIA$^{++}_{(5,5)}$  & $(0,1)$   	& $(2,1)$,$(0,3)$ 	& $(4,1)$,$(2,3)$,$(0,5)$    	& $(4,3)$,$(2,5)$ 	& $(4,5)$ \\ \addlinespace
\bottomrule
\end{tabular}
\caption{Brane spectrum of IIA$^{\alpha\beta}$ theories.}
\label{table_HullIIA}
\end{table}


\begin{table}[ht]
\centering
\begin{tabular}{l c c c c c}
  \toprule
  Theory   & D(-1) & D1 & D3 & D5 & D7   \\
    \midrule
IIB$^{++}_{(9,1)}$  & -		& $(1,1)$			& $(3,1)$					& $(5,1)$					& $(7,1)$ \\ \addlinespace
IIB$^{+-}_{(9,1)}$  & $(0,0)$	& $(2,0)$			& $(4,0)$					& $(6,0)$					& $(8,0)$ \\ \addlinespace
IIB$^{-+}_{(9,1)}$  & $(0,0)$	& $(1,1)$			& $(4,0)$					& $(5,1)$					& $(8,0)$ \\ \addlinespace
IIB$^{--}_{(7,3)}$  & -		& $(2,0)$,$(0,2)$	& $(3,1)$,$(1,3)$			& $(6,0)$,$(4,2)$			& $(7,1)$,$(5,3)$ \\ \addlinespace    
IIB$^{++}_{(5,5)}$  & -		& $(1,1)$			& $(3,1)$,$(1,3)$			& $(5,1)$,$(3,3)$,$(1,5)$	& $(5,3)$,$(3,5)$ \\ \addlinespace
IIB$^{+-}_{(5,5)}$  & $(0,0)$ 	& $(2,0)$,$(0,2)$	& $(4,0)$,$(2,2)$,$(0,4)$	& $(4,2)$,$(2,4)$			& $(4,4)$ \\ \addlinespace
IIB$^{-+}_{(5,5)}$  & $(0,0)$	& $(1,1)$			& $(4,0)$,$(2,2)$,$(0,4)$	& $(5,1)$,$(3,3)$,$(1,5)$	& $(4,4)$ \\ \addlinespace
\bottomrule
\end{tabular}
\caption{Brane spectrum of IIB$^{\alpha\beta}$ theories.}
\label{table_HullIIB}
\end{table}


The ``mirror" theories\footnote{See \eqref{eqn:mirrors} in the next section for more details.} 
with reversed signatures are not included in the tables,
as their supergravity solutions differ only by an overall minus sign, flipping the world volume signature of the branes. 
Moreover, for type IIB theories allowing D$(-1)$-branes the dual space-filling D$9$-brane is expected to be present. 
Although not stated in the tables, in \cite{Hull:1998fh} it has been argued via T-duality 
that in IIA$^{-+}_{(10,0)}$ a D$8^{(9,0)}_{(10,0)}$ should also exist.

Based on dimensional reduction of super Yang-Mills (SYM) in 10 dimensions,  
it has been argued in \cite{Hull:1999mt} that in the world volume theories of the branes 
in IIA/B$^{-\pm}$ the kinetic term for the gauge field should appear
with the opposite sign.

\subsection{Map of type II action to exotic actions}
\label{sec_map}

As has already been observed in~\cite{Dijkgraaf:2016lym}, by studying
negative tension branes and the geometry around them, one can determine 
a map from the usual type IIA/B actions to all exotic actions of type 
IIA$^{\rm E}$/IIB$^{\rm E}$. Since we will employ this map during the
course of this paper, let us review it in more detail.

Such negative branes are  extended objects that differ
from the ordinary branes by carrying opposite tension and R-R
charge\footnote{Note that  negative branes $\widehat{\rm Dp}$ are
  distinguished  from anti-branes $\ov{\rm Dp}$. The latter
  carry opposite R-R charge but still positive tension.}.
Studying the black D-brane geometry of a stack of $N_i^+$
Dp-branes and $N_i^-$ negative $\widehat{\rm Dp}$-branes, one can easily see that
there exists an interface, at a finite distance from the stack, where
the curvature becomes singular and the harmonic function $H$ appearing in
the backreacted geometry vanishes. Within this region $H$ becomes
negative. Following~\cite{Dijkgraaf:2016lym}, one can perform the
analytic continuation of the initial background
${\de{s^2}={H}^{-\frac{1}{2}}\de{s^2_{p+1}}+H^{\frac{1}{2}}\de{s^2_{9-p}}}$
beyond the interface and get
\eq{
\label{mapping}
        \de{s^2}&=\omega^{-1}\bar{H}^{-\frac{1}{2}}\de{s^2_{p+1}}+\omega\bar{H}^{\frac{1}{2}}\de{s^2_{9-p}} \, \\
        e^{-2\phi}&=\omega^{p-3}g_s^{-2}\bar{H}^{\frac{p-3}{2}} \, \\
        F_{p+2}&=-g_s^{-1}\de{\bar{H}}^{-1}\wedge\Omega_{p+1}
}
\noindent
with $\Omega_{p+1}$ denoting the volume form along the branes $\Sigma_{p+1}$ and $g_s$ the (asymptotic) string 
coupling. Moreover, one has  $\bar{H}\equiv -H$, with the precise form of $H$ not being relevant for our purposes. 
What is important here is the factor $\omega=\pm i$, where the sign depends on 
the way one goes around the singularity. 
A Weyl transformation is sufficient to get rid of the imaginary factors in the metric :
\eq{
        \de{s^2}=-\bar{H}^{-\frac{1}{2}}\de{s^2_{p+1}}+\bar{H}^{\frac{1}{2}}\de{s^2_{9-p}} .
}
It is evident that the directions parallel to the brane pick up a
relative minus sign, hence the signature of the space-time dynamically
changes from $(9,1)$ to $(10-p,p)$. One could think of this as having two
different theories at the two sides of the singular interface: one is
a usual string theory and the other is an exotic string theory living
in the modified space-time. Within the latter theory, the signature of
the brane changes from $(p,1)$ to $(1,p)$. Probe branes, under the 
requirement that they are BPS, can be used to find the nature of the
theory inside the bubble, which depends solely on $p$: 
\eq{
 p\ {\rm even}: \ \textrm{IIA}^{-(-)^\frac{p}{2}}_{(10-p,p)}\,,\qquad\quad
  p\ {\rm odd}:\  \textrm{IIB}^{-(-)^\frac{p-1}{2}}_{(10-p,p)}\, .
}
One can also perform a slightly different Weyl transformation of
\eqref{mapping}, leading to a metric of opposite overall 
sign:
\eq{
        \de{s^2}=\bar{H}^{-\frac{1}{2}}\de{s^2_{p+1}}-\bar{H}^{\frac{1}{2}}\de{s^2_{9-p}} .
}
Adopting this convention, the directions parallel to the negative
brane keep their signature, while the transverse directions pick up a
minus sign. The space-time signature now becomes $(p,10-p)$. This might
at first seem like an inconsistency, as the choice of conventions for
the signature crossing leads to different theories. However, it is
understood that there is some kind of ``mirror"\footnote{The term
  ``mirror" here refers to mirroring the signature of the theories
  (exchanging time and space), \emph{not} to the usual mirror symmetry
  of Calabi-Yau manifolds.} symmetry between theories in ``mirror" 
space-times, which is
\eq{
	\label{eqn:mirrors}
	\rm{IIA}^{\alpha\beta}_{(10-p,p)}\leftrightarrow \rm{IIA}^{\alpha(-\beta)}_{(p,10-p)}\, ,
	\qquad \rm{IIB}^{\alpha\beta}_{(10-p,p)}\leftrightarrow \rm{IIB}^{\alpha\beta}_{(p,10-p)} \,.
}

Now, taking into account  the phases $\omega$  in the solution
\eqref{mapping},  one can define a general (off-shell) map between the 
standard and exotic type II actions. We start by defining the transformation of
the vielbein determinant $\det{e^a_\mu}=\sqrt{|\det{G_{(9,1)}}|}$ 
in order to avoid branch cut issues later.
The determinant receives one factor of $\omega^{-1/2}$ ($\omega^{+1/2}$) 
for each direction parallel (transverse) to the negative brane.
Then mapping the vielbein in an appropriate manner reproduces the 
metric after singularity crossing \eqref{mapping},
and finally the dilaton field is redefined to get back to the standard dilaton profile.
In this manner, the map can be determined as
\eq{
\label{map-offschell}
	\det e^a_\mu 	\,&\to\, 	\omega^{-p} \det e^a_\mu  \,,
	\\
	e^a_\mu=\left(e^a_\parallel,e^a_{\perp}\right) 		\,&\to\,	
		\omega^{\frac{1}{2}} \left(\omega^{-1} e^a_\parallel, \,e^a_{\perp}\right) \,,
	\\
	G_{(9,1)} 		\,&\to\,	\omega \, G_{(10-p,p)} \,,
	\\
	e^{-\phi} \,&\to\, \omega^{\frac{p-3}{2}} e^{-\phi}\,,\\
        (C_{p+1})|_{\Sigma_{p+1}} \,&\to\, -(C_{p+1})|_{\Sigma_{p+1}} \,.
}

Note that the vielbein does not map to the vielbein of the new theory directly. 
Indeed, the metric of signature $(10-p,p)$ can be written  as
\eq{ 
\label{metricea}
	G_{(10-p,p)} = \left(\omega^{-1} e^a_\parallel, \,e^a_{\perp}\right) \,\eta_{(9,1)} \,
	\left(\omega^{-1} e^a_\parallel, \,e^a_{\perp}\right)^\mathsf{T} = 
	\left( e^a_\parallel, \,e^a_{\perp}\right) \,\eta_{(10-p,p)}\, 
	\left(e^a_\parallel, \,e^a_{\perp}\right)^\mathsf{T} \,.
}
Taking into account that the original metric is
\eq{
\label{metriceb}
   	G_{(9,1)} =	\left( e^a_\parallel, \,e^a_{\perp}\right) \,\eta_{(9,1)}\, 
	\left(e^a_\parallel, \,e^a_{\perp}\right)^\mathsf{T} ,
}
we first realize that the second and the third line in \eqref{map-offschell} are
compatible, i.e. the transformation of the vielbein implies $G_{(9,1)}\to \omega G_{(10-p,p)}$. 
Note that in the two expressions \eqref{metricea} and
\eqref{metriceb}  for the metrics 
exactly the same vielbein $e^a_\mu$ appears so that also the measures are both given by
\eq{
 \sqrt{|G_{(9,1)}|} = \det e^a_\mu = \sqrt{|G_{(10-p,p)}|}\,.
}
Then we can avoid branch cuts in the measure factor\footnote{
Had we defined the map on the vielbein or the metric instead of $\det e^a_\mu$ as first principle, 
the measure would map to something like $\sqrt{-\det(\omega G_{(10-p,p)})}$, 
which in principle gives the same factor $\sqrt{\omega^{-2p}}$
but now appropriate branch cuts have to be chosen.
}, 
since the measure must map in
the same way as the vielbein determinant
\eq{
	\sqrt{|G_{(9,1)}|} 	\,&\to\,	\omega^{-p} \sqrt{|G_{(10-p,p)}|}\,.
}
These  maps can be shown to map the supergravity actions according to
\eq{\label{TheoryMappings}
      \textrm{IIB} \to
      \textrm{IIB}^{-(-)^\frac{p-1}{2}}_{(10-p,p)}\,,\qquad
      \textrm{IIA} \to \textrm{IIA}^{-(-)^\frac{p}{2}}_{(10-p,p)}\,.
}
The analysis so far concerns the closed string actions. It is one of
the objectives  of this paper to generalize this analysis to the open
string sector and the corresponding DBI actions on the D-brane
world-volumes. Before we move to these questions in section \ref{sec_four},
we need to work out a couple of general aspects of ghosts in the
exotic string theories.


\section{Ghosts in exotic string theories}
\label{sec_ghosts}

In this section we continue the discussion of ghosts in the exotic
string theories. Ghosts are states of negative norm in the Hilbert space, preventing
a probabilistic interpretation and, even when removed by hand from the set
of physical states, leading to a violation of unitarity. 
While they might be an important ingredient to find dS solutions, 
massless or light ghosts are phenomenologically excluded.

A standard procedure to get rid of such states is to
gauge more symmetries on the world-sheet, hence introducing new
$(b,c)$ ghost systems that change the critical central charge of the theory
and cancel the contributions of the problematic ghosts.
A famous example is the $N=2$ superstring with a critical
central charge of $c=6$ and a four-dimensional target-space of
signature $(2,2)$ or $(0,4)$~\cite{Ooguri:1991fp}. Due to the extra gauge symmetry,
more target-space directions can be gauged away.

However, since we do not want to change
the critical dimension, there will only be the usual gauge invariances
leading to a critical central charge of $26\,(15)$ for the bosonic (super)
string theory. This means there will be a  single distinctive time direction
and the corresponding $bc$  (and $\beta\gamma$) ghost system. Let us analyze the
appearance  of ghosts for the Lorentzian IIA$^{\rm L}$/IIB$^{\rm L}$
theories and the Euclidean IIA$^{\rm E}$/IIB$^{\rm E}$
theories in more detail. For simplicity, we present the string
world-sheet arguments only for the bosonic string, while they
analogously hold for the superstring theories.

\subsection{Ghosts for  the Lorentzian string}

The Lorentzian fundamental strings in the IIA$^{\rm L}$/IIB$^{\rm L}$
theories can be quantized in complete analogy to the 
usual IIA and IIB (super-) strings with signature $(9,1)$.  
This means that the mode algebra  for the bosonic fields $X^\mu$ reads
\eq{
 \label{LorentzModeAlgebra}
[\alpha_m^\mu,\alpha_n^\nu]= m\, \eta^{\mu\nu}_{(10-p,p)} \,\delta_{m,-n}\;,
}
where $\eta^{\mu\nu}_{(10-p,p)}$ denotes the flat metric of signature 
$(10-p,p)\in\{(9,1),(5,5),(1,9)\}$. In the following let us denote the
space-like directions and the single universal time direction with indices
$m,n,...$ and the additional new time-like directions by $a,b,...\,$.
Note that the universal  time and one space direction can be gauged away as usual.
Then for instance the  off-diagonal graviton states
\eq{
|V^{st}_G(0)\rangle= \epsilon_{\mu a} \alpha^m_{-1} \tilde\alpha^a_{-1}|0\rangle
}  
have negative norm (for $\langle 0|0\rangle=1$) and give physical ghosts that cannot be gauged away.
Note that the graviton modes 
$|V^{ss}_G(0)\rangle=\epsilon_{\mu\nu} \alpha^m_{-1} \tilde\alpha^n_{-1}|0\rangle$ and
$|V^{tt}_G(0)\rangle=\epsilon_{ab} \alpha^a_{-1} \tilde\alpha^b_{-1}|0\rangle$ have
positive norm.

 In the NS-NS sector of the superstring, one only has to
replace the $X^\mu$ by their fermionic superpartners $\psi^\mu$ and the
logic goes through analogously. In the R-R sector, there is the
distinction between the IIA/B$^{++}$ and the  IIA/B$^{+-}$ theories,
where
the latter carry a wrong overall sign for the kinetic terms of the
massless R-R fields. This can be taken care of in the world-sheet
theory by flipping by hand the overlap
between the R-R ground states
\eq{                     
            \langle 0|0\rangle_{\rm RR}^{+-}=-\langle 0|0\rangle_{\rm RR}^{++}\,.
}

\subsubsection*{Light cone gauge and Lorentz symmetry}
\label{appendix_LCG}

The (time-like) T-duality arguments suggest that  a change of the target-space signature
does not affect the critical dimension of the string theory.
Let us check this explicitly on the world-sheet.
This can be  readily seen for the bosonic string in light cone gauge by checking for
anomalies of the $SO(p,q)$ Lorentz symmetry.
The world-sheet metric is fixed as
$h_{\alpha \beta}=\eta_{\alpha \beta}$ and we introduce
light cone coordinates in space time $X^+=1/\sqrt{2}\left(X^0+X^{1}\right)$, 
$  X^-=1/\sqrt{2}\left(X^0-X^{1}\right)$, where
we singled out one time and spatial direction $X^0$, $X^{1}$. 
The target-space metric becomes $\eta_{+-}=\eta_{-+}=-1$ for the light cone, $\eta_{ab}=-\delta_{ab}$ for 
$a,b=1,\dots, p-1$ remaining time directions and 
$\eta_{mn}=\delta_{mn}$ for the $m,n=1,\dots, q-1$ spatial directions. 

We follow the standard procedure and look at the open string with (NN) boundary conditions. 
The remaining gauge freedom is fixed by setting $X^+\left(\sigma, \tau\right)=x^+ + p^+ \tau$.
Using the Virasoro constraint equation $\eta_{\mu\nu}(\dot{X}^\mu\pm X^{'\mu})(\dot{X}^\nu\pm X^{'\nu})=0$
to express the oscillator modes of $X^-$ in terms of the transverse modes yields
\eq{
	\alpha^-_n=\frac{1}{p^+}\left(\frac{1}{2} \sum_{k=-\infty}^{\infty}\, 
	:\eta_{ij}\, \alpha^i_{n-k}\alpha^j_{k}:\,-\, a\,\delta_{n,0}\right)
}
with $i,j$ running over the transverse directions and for simplicity setting $\alpha'=1 / 2$. The modes still 
satisfy a ``transverse" Virasoro algebra
\eq{\label{-Virasoro}
\left[p^+ \alpha^-_{m}, p^+ \alpha^-_{n}\right]=&(m-n)\,p^+\alpha^-_{m+n}+\\
&
      \left(\frac{D-2}{12}(m^3-m)+2am \right)\delta_{m+n}
}
and have commutation relations with the transverse oscillator modes
\eq{\label{-Virasoro2}
\left[ \alpha^i_{n},p^+ \alpha^-_{k}\right]= n\,\alpha^i_{n+k}\, .
}
The only relevant appearance of the space-time metric is in commutation relations
$\left[\alpha^\mu_m,\alpha^\nu_n\right]= k\, \eta^{\mu\nu}\delta_{m+n,0}$. We can use these 
commutation relations and follow the standard computation for the potentially anomalous 
commutator $\left[ J^{i-}, J^{j-}\right] $ of Lorentz generators $J^{\mu\nu}$. Doing so we find
\eq{
\left[ J^{i-}, J^{j-}\right] =0\quad \Leftrightarrow \quad D=26, \, a=1
}
but no additional constraints on the number of time respectively spatial dimensions.
Hence Lorentz symmetry $SO(p,q)$  is preserved for a total of $p+q=26$ space-time
dimensions.

\subsubsection*{Orbifolding ghosts}

Generally, having a theory that has too many degrees
of freedom one can proceed in two ways. Either one gauges extra
symmetries or one projects out the unwanted states. 
Since gauging symmetries completely removes the time-like directions, we want to take the second route.
Can one remove the massless  ghosts by performing an appropriate orbifold projection? 
In contrast to the gauging procedure an orbifold will not change the critical central charge, 
but will potentially break the 10D diffeomorphism symmetry to a subgroup. 

\begin{figure}[tbh]
	\begin{center}
		\begin{tikzpicture}[scale=1.7]
			
			\node[draw, rectangle, rounded corners](IIA++) at (0,2.5) {\bf IIA$^{++}$};
			\node[draw, rectangle, rounded corners](IIA+-) at (0,-2.5) {\bf IIA$^{+-}$};
			\node[draw, rectangle, rounded corners](IIB+-) at (2.5,0) {\bf IIB$^{+-}$};
			\node[draw, rectangle, rounded corners](IIB++) at (-2.5,0) {\bf IIB$^{++}$};
			
			\draw (IIA++) -- node[above,pos=0.55]{\footnotesize x} 
				(IIB++) -- node[below,pos=0.45]{\footnotesize t}
				(IIA+-) -- node[below,pos=0.55]{\footnotesize x}
				(IIB+-) -- node[above,pos=0.45]{\footnotesize t}
				(IIA++);
				
			\node[anchor=north] (IIA++91) at (0,2) {\scriptsize $(9,1)$};
			\node[anchor=north] (IIA++55) at (0,1.6) {\scriptsize $(5,5)/I_4$};
			\node[anchor=north] (IIA++19) at (0,1.2) {\scriptsize $(1,9)/I_8$};
			
			\node[anchor=south] (IIA+-91) at (0,-1.1) {\scriptsize $(9,1)/(-1)^{F_L}$};
			\node[anchor=south] (IIA+-55) at (0,-1.5) {\scriptsize $(5,5)/(-1)^{F_L}I_4$};
			\node[anchor=south] (IIA+-19) at (0,-1.9) {\scriptsize $(1,9)/(-1)^{F_L}I_8$};
			
			\node[anchor=west] (IIB++91) at (-1.7,0.4) {\scriptsize $(9,1)$};
			\node[anchor=west] (IIB++55) at (-1.7,0) {\scriptsize $(5,5)/I_4$};
			\node[anchor=west] (IIB++19) at (-1.7,-0.4) {\scriptsize $(1,9)/I_8$};
			
			\node[anchor=west] (IIB+-91) at (0.55,0.4) {\scriptsize $(9,1)/(-1)^{F_L}$};
			\node[anchor=west] (IIB+-55) at (0.55,0) {\scriptsize $(5,5)/(-1)^{F_L}I_4$};
			\node[anchor=west] (IIB+-19) at (0.55,-0.4) {\scriptsize $(1,9)/(-1)^{F_L}I_8$};
			
		\end{tikzpicture}
		\caption{Orbifold projections that remove the
                  massless ghosts for Lorentzian theories.
                       New ghosts in twisted sectors can
                  be avoided by combining these actions with a shift
                  along a spatial direction.}
		\label{fig:projectionsLorentzian}
	\end{center}
\end{figure}
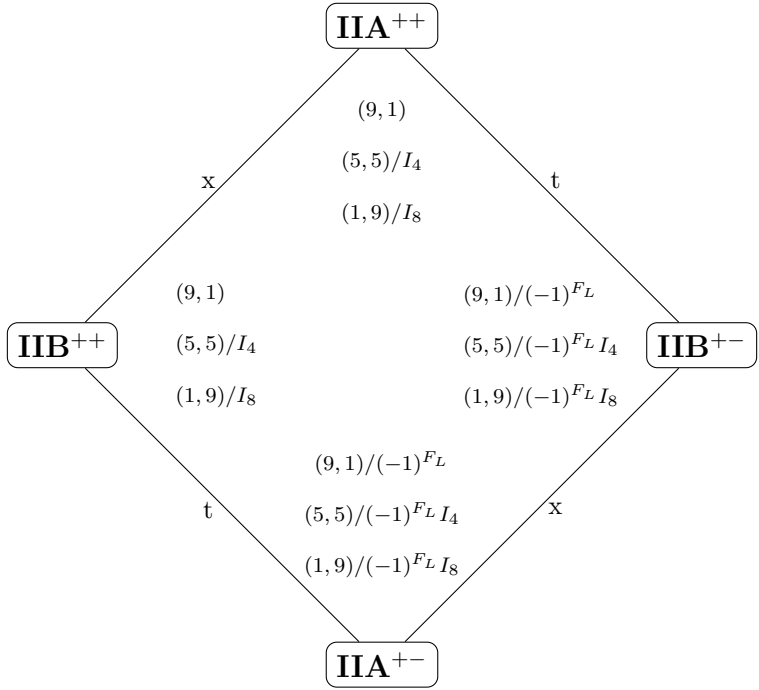

Following the usual recipe for performing an orbifold in string theory, 
the untwisted sector is projected to invariant states and a twisted
sector must be introduced. Let us discuss appropriate orbifold projections to remove
light ghosts from the Lorentzian theories. 
In the IIA/B$^{+-}_{(9,1)}$ theory, the ghost R-R fields 
can be projected out by performing an orbifold by $(-1)^{F_L}$.
To avoid the appearance of new massless ghosts in the $\mathbb Z_2$
twisted sector, one can combine this action with a half-shift 
$S:X\to X+\pi R$ along a compactified spatial direction. 
For the IIA/B$^{++}_{(5,5)}$ theories physical ghosts are related to 
four extra time-like directions. These ghosts can be removed by 
taking the quotient by a $\mathbb Z_2$ reflection $I_4:x^a\to -x^a$ along these
four directions. Similarly, the ghosts in IIA/B$^{++}_{(1,9)}$ are
removed by $I_8$, reflecting the eight extra time-like coordinates.
Finally, the massless ghosts of  IIA/B$^{+-}_{(5,5)}$ and  IIA/B$^{+-}_{(1,9)}$
are projected out by $I_4 (-1)^{F_L}$ and $I_8 (-1)^{F_L}$, 
combining the previous reasoning. 
These results are summarized in figure \ref{fig:projectionsLorentzian}.

\subsubsection*{Compact time-like dimensions}

Eventually, we are interested in compactifications of the exotic
string theories to 4 dimensions with signature $(3,1)$, so in theories with multiple time-like directions
some of them will need to be compactified. 
The standard problem of compact time-like dimensions are closed time-like curves
which violate causality. Since the orbifolds project out massless excitations in these directions, 
one might naively think that the quotient theories are safe.
However, we will see that compact time dimensions in exotic string theories
lead to further complications.

For the  IIA/B$^{+-}_{(9,1)}$ theory the orbifold by  $(-1)^{F_L}$
removes all ghost fields from the untwisted sector. 
In case of the  IIA/B$^{++}_{(5,5)}$ theories however,  
even though the massless mixed graviton modes
$|V^{(ev,odd)}_G(0)\rangle=\epsilon_{m a} \alpha^m_{-1}
\tilde\alpha^a_{-1}|0\rangle$ 
are projected out, for non-vanishing momentum/energy  
the linear combination
\eq{
      |V^{(ev,odd)}_G(p^n,e^b)\rangle  = \epsilon_{m a} \alpha^m_{-1}
         \tilde\alpha^a_{-1}|p^n,e^b\rangle  -
       \epsilon_{m a} \alpha^m_{-1} \tilde\alpha^a_{-1}|p^n,-e^b\rangle
}
remains in the spectrum. 
Here the upper index pair indicates the behavior of the left
and right moving part under the $\mathbb Z_2$ operation.
Here $p^{n}$, $m,n\in\{0,1,\ldots,5\}$ denote the usual energy
$p^0=E$ and space-like momenta, and  $e^{b}$, $a,b\in\{6,\ldots,9\}$
denote the extra time-like energies.
Therefore, there are still massive ghosts in the string
spectrum, a $\mathbb Z_2$ projection does not allow to remove all
of them. 

The on-shell condition for such a state is
\eq{
              E^2+\sum_a (e^a)^2 =\sum_i (p^i)^2 + m^2 \,,
}
where $m$ is the mass of the state. We  interpret this
condition such that for a state of mass $m$ with momenta $p^i$ the
total energy can be distributed among all the time-like energies such
that this quadratic relation is satisfied \cite{Quiros:2007ym}. Only $E$ is the energy that we
have access to. Note that while for negative $E$ we have an
interpretation in terms of anti-particles with positive $E$, the 
additional energies $e^a$ can be both positive and negative. 

Let us now consider a Lorentzian string on a time-like torus  of radius
$R$ .
As for a space-like compactification, 
the time-like momentum (i.e. energy) gets quantized along the compact direction and leads to a mass contribution, 
resulting in a KK tower of massive states. 
Similarly, the winding modes contribute to the mass so that in total
we find the on-shell condition
\eq{
\label{onshelltwo}
   E^2+ \sum_b \left[ \left({m_b\over R}\right)^2
             + 
             \left ( {n_b R\over \alpha'}\right)^2\right]=\sum_{i}
           (p^i)^2+{2\over \alpha'} (N+\ov N-2a) 
}
with $a=1/2$ for the superstring and the level-matching condition
$\sum_a m_a n_a=-(N-\ov N)$.
For $R>\sqrt{\alpha'}$  it is tempting to identify a  UV cutoff  with
the Kaluza-Klein scale  $\Lambda_{\rm UV}=1/R$ that we assume 
to be only a few  orders of magnitude below the string scale.
Let us analyze this on-shell condition in the IR regime  
$|p|<\Lambda_{\rm UV}$.

In the massless sector
$N=\ov N=1/2$, a non-vanishing time-like KK/winding mode
$(m_a,n_a)\ne(0,0)$ already lies outside the IR regime.
Thus all the light on-shell states that we have access to are 
frozen in the extra time directions and feature
$e^a=0$. Then together with the $\mathbb Z_2$ projections there are no 
light ghosts left, so it seems that we are safe.
However, for the tower of massive string excitations $N=\ov N>1/2$
their contribution to the right hand side of \eqref{onshelltwo}
can be balanced against KK/winding contributions.
Therefore, these massive excitations combine with time-like KK/winding modes
to appear as extremely light particles from a 4D perspective.
As already observed in \cite{Dijkgraaf:2016lym}, even for irrational values of the radius
there will always be integers $N,\ov N,m_a,n_a$ such that their 4D mass
lies below any cut-off.
Relatedly, there exist kinematically allowed scattering processes like
\eq{
         |V_{m_1=0}(p_1^m,e_1^a=0)\rangle &+
         |V_{m_2=0}(p_2^m,e_2^a=0)\rangle \\[0.1cm]
         &\longrightarrow
         |V_{m_3>0}(p_3^m,e_f^a)\rangle +
           |V_{m_4>0}(p_4^m,-e_f^a)\rangle 
}
with the extra energies in the final state $e_f^a\ne 0$. Thus, 
the ultralight states with $N=\ov N>1/2$ do not decouple 
in the scattering amplitudes of massless states with $N=\ov N=1/2$.
We can summarize these findings by saying that  the dimensionally reduced 10D Lorentzian
supergravity actions cannot be considered as Wilsonian effective
actions of a 4D theory.

\subsection{Ghosts for the Euclidean string}
\label{sec_orient}

The quantization of the Euclidean fundamental string has been
investigated in~\cite{Dijkgraaf:2016lym} and features a couple of new aspects and pathologies.
Note that this theory is different from the Wick rotated Lorentzian
string. In section \ref{sec_cft} we will review and continue this analysis,
where our special focus will be on the construction of boundary states,
providing the CFT description of D-branes for these exotic string
theories.

One new aspect of the quantization is that factors of $i=\sqrt{-1}$
appear at various places. For instance, the mode algebra  for the
bosonic fields $X^\mu$  now reads
\eq{
 \label{EuclideanModeAlgebraa}
[\hat{\alpha}_m^\mu,\hat{\alpha}_n^\nu]=- i\,m\, \eta^{\mu\nu}_{(10-p,p)} \,\delta_{m,-n}\;.
}
As a consequence,  the diagonal graviton/B-field  states
$|V^{ss}_G(0)\rangle$ and $|V^{tt}_G(0)\rangle$ have negative norm and
the 
off-diagonal ones $|V^{st}_G(0)\rangle$ positive norm (for $\langle 0|0\rangle=1$). However, this
is  not consistent with the normalization of the Einstein-Hilbert term
for the Euclidean string SUGRA actions \eqref{eqn:sugraaction}. 
This can be remedied by choosing the correct normalization of the vertex operators. 
These have been worked out in \cite{Dijkgraaf:2016lym}. 
The graviton gets an extra factor of $-i$, rendering its norm positive, while the B-field remains a ghost. 
Of course the time-like ghosts from the previous section also remain in the spectrum.

\subsubsection*{Orientifolding ghosts}

Now we investigate whether there also exist $\mathbb Z_2$ operations
that can mod out all the massless ghost fields for the Euclidean
exotic string theories IIA$^E$/IIB$^E$. Let us start with the
IIB$^{-+}_{(9,1)}$ theory, which is the S-dual of the IIB$^{+-}_{(9,1)}$ theory.
By looking at its SUGRA action \eqref{eqn:sugraaction} we see that 
$H_3$, $F_1$, $F_5$ have the wrong sign of the kinetic terms and $F_3$
the usual sign. These are precisely the p-form fields that are odd and
even under the world-sheet parity transformation $\Omega$, and 
indeed the S-dual of $(-1)^{F_L}$ is known to be $\Omega$.
Therefore, the orientifold IIB$^{-+}_{(9,1)}/\Omega$ has no ghost
fields in the closed string sector. Depending on whether the
orientifold projection has fixed loci or acts freely (after combining it again
with a shift symmetry), there will be a twisted sector in the form of 
appropriate D-branes that need to be introduced to cancel the R-R
tadpole of the O-plane. This open string sector can host additional
ghosts. We will come to this point in section \ref{sec_four}.

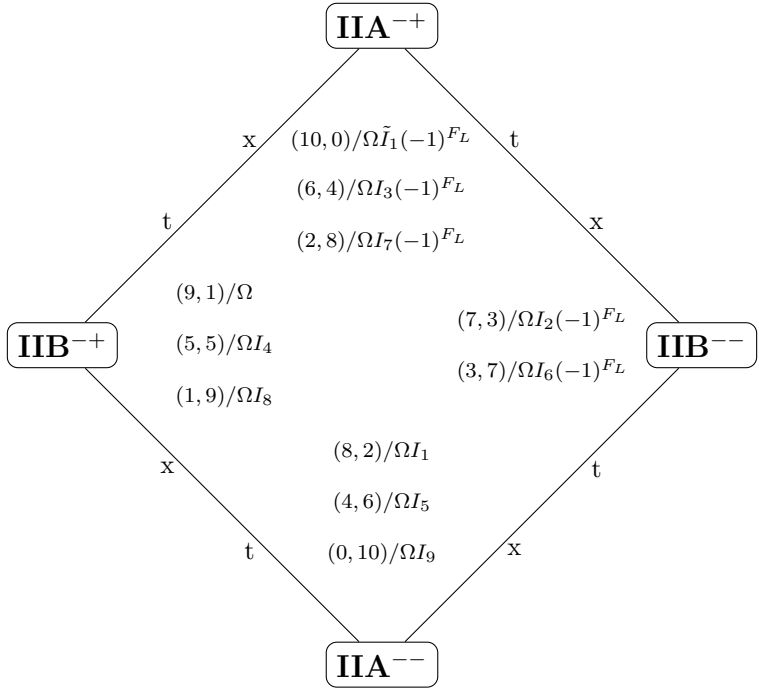
\begin{figure}[tbh]
	\begin{center}
		\begin{tikzpicture}[scale=1.7]
			
			\node[draw, rectangle, rounded corners](IIA-+) at (0,2.5) {\bf IIA$^{-+}$};
			\node[draw, rectangle, rounded corners](IIA--) at (0,-2.5) {\bf IIA$^{--}$};
			\node[draw, rectangle, rounded corners](IIB--) at (2.5,0) {\bf IIB$^{--}$};
			\node[draw, rectangle, rounded corners](IIB-+) at (-2.5,0) {\bf IIB$^{-+}$};
			
			\draw (IIA-+) -- node[above,pos=0.4]{\footnotesize x} node[above,pos=0.7]{\footnotesize t} 
				(IIB-+) -- node[below,pos=0.3]{\footnotesize x} node[below,pos=0.6]{\footnotesize t}
				(IIA--) -- node[below,pos=0.4]{\footnotesize x} node[below,pos=0.7]{\footnotesize t}
				(IIB--) -- node[above,pos=0.3]{\footnotesize x} node[above,pos=0.6]{\footnotesize t}
				(IIA-+);
				
			\node[anchor=north] (IIA-+100) at (0,1.8) {\scriptsize $(10,0)/\Omega \tilde{I}_1 (-1)^{F_L} $};
			\node[anchor=north] (IIA-+64) at (0,1.4)
                        {\scriptsize $(6,4)/\Omega I_3 (-1)^{F_L}$};
			\node[anchor=north] (IIA-+28) at (0,1.0)
                        {\scriptsize $(2,8)/\Omega I_7 (-1)^{F_L}$};
			
			\node[anchor=south] (IIA--82) at (0,-1) {\scriptsize $(8,2)/\Omega I_1 $};
			\node[anchor=south] (IIA--26) at (0,-1.4) {\scriptsize $(4,6)/\Omega I_5 $};
			\node[anchor=south] (IIA--010) at (0,-1.8) {\scriptsize $(0,10)/\Omega I_9 $};
			
			\node[anchor=west] (IIB-+91) at (-1.7,0.4) {\scriptsize $(9,1)/\Omega $};
			\node[anchor=west] (IIB-+55) at (-1.7,0) {\scriptsize $(5,5)/\Omega I_4$};
			\node[anchor=west] (IIB-+19) at (-1.7,-0.4) {\scriptsize $(1,9)/\Omega I_8$};
			
			\node[anchor=west] (IIB--73) at (0.5,0.2)
                        {\scriptsize $(7,3)/\Omega I_2 (-1)^{F_L}$};
			\node[anchor=west] (IIB--37) at (0.5,-0.2) {\scriptsize $(3,7)/\Omega I_6(-1)^{F_L}$};
									
		\end{tikzpicture}
		\caption{Orientifold projections  removing the
                  massless ghosts for  Euclidean theories.}
		\label{fig:projectionsEuclidean}
	\end{center}
\end{figure}

Now by successively applying spatial T-dualities we can find the 
orientifold projections for all the IIA/B$^{-,\beta}_{(10-p,p)}$ theories.
After one T-duality one gets IIA$^{--}_{(8,2)}$ with the projection
$\Omega I_1$, where $I_1$ reflects the new additional time-like 
coordinate.  The corresponding branes are D8-branes localized 
at a point in the new time-like  direction.
Another T-duality leads to the  IIB$^{--}_{(7,3)}/\Omega
I_2 (-1)^{F_L}$  orientifold, etc. All the resulting quotients are shown in the right
hand part of figure \ref{fig:projectionsEuclidean}. 
T-dualizing instead along the time-like direction, we find the appropriate 
orientifold quotient to be IIA$^{-+}_{(10,0)}/\Omega \tilde{I}_1 (-1)^{F_L}$, 
where $\tilde{I}_1$ is a reflection along the space-like direction that was created 
by T-dualizing. 

\subsubsection*{Compact dimensions}

Another new aspect of the Euclidean theories is that the tower of 
string excitations has imaginary squared mass $m^2$. Moreover, since under
T-duality a space-like circle maps to a time-like one and vice versa,
the winding modes contribute with the opposite sign as the KK modes.
Thus, the on-shell relation for a compactification on a torus $T^D$ of
radii $R_k$  with metric  $\eta_k=\pm 1$ now reads
\eq{
\label{onshellthree}
   E^2- \sum_{k=1}^D \eta_k  \left[ \left({m_k\over R_k}\right)^2
             - \left( {n_k R_k\over \alpha'}\right)^2\right]=\sum_{i}
           (p^i)^2-i\, {2\over \alpha'} (N+\ov N-2a) \,.
}
with the level-matching condition $ \sum_k \eta_k m_k n_k = N-\ov N$.
In contrast to the Lorentzian string, here the  KK/winding modes can never 
cancel against the string oscillations. However, 
for both space-like and time-like compactifications 
the KK mode contribution can cancel up to arbitrary precision against the
winding mode contribution, leading again to the conceptual 
problem of interpreting the dimensional reduction of the
10D Euclidean supergravity actions as Wilson effective theories.
As for the Lorentzian string, these ultra-light modes do not decouple
in string scattering amplitudes.

\subsection{Speculations and objectives}

In this section, we will make a couple of general and arguably  speculative
remarks on ghosts.
The appearance of ghosts are generally thought to lead to problems for the
quantized version of a theory as it spoils the probabilistic
interpretation of the wave-function. In the so far discussed
closed string sector the  massless modes are the graviton and other 
form fields. The quantum nature of these states has not really been
tested experimentally so that here we take the pragmatic point of view of being
agnostic whether perturbative ghosts are indeed a disaster for the
theory or only indicate whether we should go beyond our usual
understanding of a quantum (gravity) theory. 

\subsubsection*{Exotic string theories and Krein spaces?}

As we have seen, the exotic string theories give rise
to a space of states that has indefinite norm. Such generalizations
of Hilbert spaces have been considered in the mathematical physics 
literature and are called Krein spaces. They go back to the early days
of quantum mechanics \cite{Dirac:1945cm,Pauli:1949zm} and \cite{Gupta:1949rh,Bleuler:1950cy}.
Here we are not intending to  provide a full review of this field but just
want to touch upon a few issues.
 
A Krein space ${\cal K}$  is a complex vector space with an indefinite metric
$\langle \cdot|\cdot\rangle$ so that ${\cal K}$ splits as 
 ${\cal K}={\cal K}_+\oplus {\cal K}_-$, where  ${\cal K}_+$ and ${\cal K}_-$
are Hilbert spaces.
On such a  space ${\cal K}$ one can introduce an  ``orthonormal'' basis satisfying
$\langle \psi_i|\psi_j\rangle =\eta_{ij}$.
In our case, the Hilbert
space ${\cal K}_+$ is generated by
the non-ghosts and ${\cal K}_-$ by the additional ghosts.  

The adjoint of an operator is defined as usual by the condition
$\langle\psi| A \phi\rangle=\langle A^{\sharp}\psi|  \phi\rangle$ which
gives $A^{\sharp}=\eta^{-1} A^\dagger \eta$. A (pseudo) self-adjoint operator
is defined via $A^{\sharp}=A$ and  gives rise to real expectation
values $\langle\psi| A \psi\rangle\in \mathbb R$. However, the 
eigenvalues of a self-adjoint operator are generically not real. 
An operator preserving the product, i.e. 
$\langle\psi| \phi\rangle=\langle U\psi|  U\phi\rangle$, is called
(pseudo) unitary. It satisfies $U^{\sharp} U=1$ which means 
$U^\dagger=\eta U^{-1} \eta^{-1}$.
On a Krein space the unit operator can be
expressed as $1=|\psi_i\rangle \eta^{ij} \langle \psi_j|$.  
We expect that the (exotic) string S-matrix is such a 
(pseudo) unitary operator.

Whether eventually such a quantum theory on a Krein space   is a viable possibility for a
theory of quantum gravity remains to be seen and goes beyond the scope of this paper. At least
we can state that the quantum theory that originates by allowing extra
time-like directions in string theory  has an already studied mathematical structure.

\subsubsection*{Objectives}

Although the perturbative approach to the exotic string theories is
full of pathologies that are not yet completely understood, it has
been argued that these are just artifacts of the perturbative approach
and will get resolved in the full theories (as they are all dual to
the original type IIA/B superstrings). On the other hand it could also well
be that these pathologies are a result of having compact time
directions (at least at intermediate stages) and that these are simply
not allowed in any reasonable physical theory. In the latter case, 
there would be no point in following these ideas further.

However, we also do not want to miss a potentially interesting new aspect of
string theory, as often applying dualities led to new insights into
the theory. Moreover, it is at least  appealing that despite conceptual pathologies,
the formalism per se seems to go through. Thus, in the following we
still take a positive attitude and further expand the formalism of 
exotic string theories to also include the open string sector. 
The question we are posing is whether a pathological closed string
sector with ghosts that admits dS solutions, can nevertheless host a viable effective D-brane theory
that by itself obeys the usual requirements for a consistent quantum
field theory, i.e. is ghost free  and unitary. This subsector could 
then be considered to be the Standard Model, whose quantum aspects we have
direct experimental access to.

\section{CFT description of Euclidean exotic strings}
\label{sec_cft}

In this rather technical section we will take a closer look on the
Euclidean world-sheet CFTs. After a short review of the closed string
construction of \cite{Dijkgraaf:2016lym} we construct the purely
Euclidean open string theories including the fermionic sector. We use
this to identify the allowed D-branes of the different theories as
well as the tension of the branes. In section \ref{sec_four}  these
results will be 
confirmed  using the  map \eqref{map-offschell} inspired by horizon crossing
in the presence of negative branes.

\subsection{Basics of CFTs with Euclidean world-sheets}

Let us first present a couple of basic results for the structure of
CFTs on purely Euclidean world-sheets. We will see that in this case
some extra factors of $i$ appear.

\subsubsection*{Lorentzian vs. Euclidean world-sheets}

When considering purely Euclidean bosonic closed strings we have to
thoroughly disentangle the differences between a Lorentzian, a
Wick-rotated Euclidean and a purely Euclidean field theory. We could
take several approaches to quantize the theory, but the differences in the
mode algebra of the field become most apparent when using the path
integral formalism. Consider the path integral
\eq{
Z=\int \left[D\gamma\right] \left[D X\right] e^{KS_\varepsilon}
}
where we introduced $K=\left\lbrace i, -1\right\rbrace$ with $K=-1$ 
only in the Wick-rotated case, and ${\varepsilon=\{1,-1\}}$ labels 
the Euclidean or Lorentzian world-sheet action. 
Thus $K=i$, $\varepsilon=-1$  is the Lorentzian string, $K=-1$,
$\varepsilon=1$ is the Wick-rotated Euclidean string and finally $K=i$, 
$\varepsilon=1$ is the purely Euclidean string. Before applying
path integral methods we have to bring the action into a quadratic form
\eq{
	S=-\frac{\varepsilon}{2 \pi \alpha^\prime}\int \mathrm{d}^2 \sigma
	\sqrt{\varepsilon \,\mathrm{det}\gamma}\,  \eta_{\mu \nu} \, X^\mu \partial^a\partial_a X^\nu \,.
}
The kinetic operator depends on the world-sheet metric and
thus differs for Lorentzian and Euclidean field theories. The two
point function has to satisfy the identity
\eq{
	\frac{2\pi \alpha^\prime}{K\varepsilon}\,
        \delta_{\mu_1}^{\mu_2}\, \delta^2(\sigma-\sigma^\prime)
	=\sqrt{\varepsilon \, \mathrm{det}\gamma}\, \eta_{\mu_1\nu} \,\partial^a\partial_a \left\langle X^{\mu_2}(\sigma) X^\nu(\sigma^\prime)\right\rangle
}
where $\sigma,\sigma^\prime$ are coordinates on the
world-sheet. The kinetic operator on the rhs is in the
$\sigma=(\sigma_1,\sigma_2)$ coordinates. 
 Let us concentrate for a moment on the two
Euclidean field theories. Introducing the usual cylinder world-sheet coordinate
\eq{
	z=e^{\sigma_1+i\sigma_2}
}
the equation for the two point function becomes
\eq{\label{twopoint}
	\frac{2\pi \alpha^\prime}{K}\, \delta_{\mu_1}^{\mu_2}\,\delta^2(z-w)=\eta_{\mu_1\nu} \, 
	\partial_z\partial_{\ov{z}}\left\langle X^{\mu_2}(z,\ov{z})X^\nu(w,\ov{w})\right\rangle \, .
}
The solution to this is given by
\eq{
	\left\langle X^\mu(z,\ov{z})X^\nu(w,\ov{w})\right\rangle= \eta^{\mu\nu}\frac{\alpha^\prime}{K} \mathrm{log} |z-w|^2\, .
}
We want to derive the mode algebra for the CFT. The action in the above coordinates reads
\eq{
	S=\frac{1}{2\pi \alpha^\prime} \int \mathrm{d}^2 z \, \partial X(z,\ov{z})\cdot\ov{\partial} X(z,\ov{z})\,, 
}
where the target-space metric is hidden in the multiplication
dot. Going through the usual steps for the above action we find
holomorphic and anti-holomorphic currents $\partial X^\mu(z),\,
\ov{\partial}X^\mu(\ov{z})$. Using \eqref{twopoint} we find for their
two-point functions
\eq{
\left\langle \partial X^\mu(z)\partial X^\nu(w)\right\rangle = \eta^{\mu\nu}\frac{\alpha^\prime}{K}\frac{1}{(z-w)^2}\,.
}
Expanding the current fields $\partial
X^\mu(z)=-\sqrt{\alpha'\over 2\varepsilon}\sum \hat\alpha_m^\mu \, z^{-m-1}$, the
mode algebra reads
\eq{
\left[ \hat\alpha_m^\mu, \hat\alpha_n^\nu \right] = \varepsilon\,\eta^{\mu\nu} \frac{\alpha^\prime}{K}\,  m\,\delta_{m+n,0}\, .
}
Thus we see that for the Wick-rotated Euclidean string $(K=-1)$ the
procedure gives the usual commutation relations,
whereas in the purely Euclidean field theory $(K=i)$ the
commutation relations get an extra factor of $-i$.

\subsubsection*{Closed and open Euclidean strings}

From now on we will only be concerned with the purely Euclidean
theories, i.e. the world-sheet theories of the IIA$^E$ and IIB$^E$
exotic string theories. To construct the world-sheet theory we follow
\cite{Dijkgraaf:2016lym} closely. The action for a free boson is given 
by

\begin{equation}
S_b=\frac{ 1}{2\pi\alpha' }\int\text{d}^2\sigma \sqrt{\text{det}g}\,g^{ab}\,\eta_{\mu\nu}\,\partial_aX^\mu\partial_bX^\nu\,.
\end{equation}
The world-sheet metric $g$ is gauge fixed to the flat metric $g_{\sigma_1\sigma_1}=g_{\sigma_2\sigma_2}=1$ and light cone coordinates are chosen as
\begin{equation}
\label{lightcone}
\sigma_{\pm}=\sigma_1\pm i\sigma_2\,
\end{equation}
such that the derivatives become
\begin{equation}
\partial_{\pm}={1\over 2}(\partial_{\sigma_1}\mp i\partial_{\sigma_2})\,.
\end{equation}
We now choose a convenient mode expansion, simplifying the mode algebra
as much as possible. In this framework the oscillators will behave as
in the usual string theories. The zero modes will be solely
responsible for the changes in the physics. The mode expansion of the
closed string sector is given by
\eq{
\label{modeExpansion}
X^{\mu}(\sigma_1,\sigma_2) &=x^{\mu}+\alpha'p^{\mu}\sigma_1+\sqrt{\alpha'\over 2i}\sum_{n \not = 0}
	\left({\alpha^\mu_n\over n}e^{-n\sigma^+}+{\overline{\alpha}^\mu_n\over n}e^{-n\sigma^-}\right)\\
&=x^{\mu}+\frac{\alpha'}{2}p^{\mu}\,\mathrm{log}(|z|^2)+\sqrt{\alpha'\over 2i}\sum_{n \not = 0}
	\left({\alpha^\mu_n\over n}\,z^{-n}+{\overline{\alpha}^\mu_n\over n}\,\bar{z}^{-n}\right)\,,
}
so that the mode algebra becomes
\begin{equation} 
\label{EuclideanModeAlgebra}
[x^\mu,p^\nu]=i\eta^{\mu\nu}\,,\qquad\quad  
[\alpha_m^\mu,\alpha_n^\nu]=[\ov\alpha_m^\mu,\ov\alpha_n^\nu]=m\,\delta_{m,-n}\,\eta^{\mu\nu}\,
\end{equation}
for $m,n\ne 0$. Moreover, one has as usual 
$[\alpha_m^\mu,\ov\alpha_n^\nu]=0$, and the oscillators $\alpha^\mu_m,\,\ov\alpha_m^\mu$
commute with the zero modes $x^\mu$ and $p^\mu$.

Let us make a couple of remarks. To arrive at  this standard mode algebra, we
have  effectively rescaled 
the standard oscillator modes $\hat{\alpha}$ 
by a factor of $\sqrt{i}$.
As a consequence, one needs to be very careful when computing overlaps 
of states $\langle \phi^1|\phi^2\rangle$.
Indeed, taking the general definition  of the conjugate $(\phi_n)^\dagger=(\phi^\dagger)_{-n}$ 
for a field $\phi$ in Euclidean CFT into account, 
the rescaling leads to phase factors, as some of the  fields won't be purely real anymore. 
On the one hand, in this paper we are mostly concerned with partition functions
where these phases do not matter as one simply counts 
the number of states at each level. On the other hand, in the boundary state
overlaps (to be introduced later in \eqref{overlap}), due to loop-channel tree-channel equivalence
 the (suitably generalized) CPT operator $\Theta$ has to remove these
factors.
 These two facts make this basis very useful for our computations. 

If one wants to calculate the low energy effective action and determine 
for instance the sign of the kinetic terms, one also needs to know the normalization
of the corres\-ponding vertex operators.
In fact in    \cite{Dijkgraaf:2016lym} 
  the normalizations for the metric and the B-field
  vertex operators have been determined. The  graviton
state turned out to be 
\eq{    
  |V_G(p)\rangle=i \epsilon_{\mu\nu} \, \hat\alpha^{\mu}_{-1}
  \hat{\ov\alpha}^{\nu}_{-1} |p\rangle=\epsilon_{\mu\nu}\,  \alpha^{\mu}_{-1}
  {\ov\alpha}^{\nu}_{-1} |p\rangle\,
}
 whereas the Kalb-Ramond state has a different normalization  
\eq{    
   |V_B(p)\rangle= -b_{\mu\nu} \, \hat\alpha^{\mu}_{-1}
  \hat{\ov\alpha}^{\nu}_{-1} |p\rangle=-i b_{\mu\nu} \, \alpha^{\mu}_{-1}
  {\ov\alpha}^{\nu}_{-1} |p\rangle\,.
}
Thus, working with the modes $\alpha^\mu,\ov\alpha^\mu$ and treating them in the
same way as the usual oscillators in string theory makes it evident that
the $B$-field is a ghost.

In a similar fashion one can expand the open string into modes.
For Neumann-Neumann (NN)  and Dirichlet-Dirichlet (DD) boundary conditions the mode expansion reads
\eq{
X^{\mu}_{\rm NN}&=x^\mu+2\alpha'p^\mu\sigma_1+\sqrt{-2i\alpha^\prime}\,\sum_{n\neq 0}{\alpha^\mu_n\over 2n}\Big(e^{-n\sigma_-}+e^{-n\sigma_+}\Big),\\
X^{\mu}_{\rm DD}&=x^\mu+{\Delta x^\mu\over \pi}\sigma_2+\sqrt{-2i\alpha^\prime}\,\sum_{n\neq 0}{\alpha^\mu_n\over 2n}\Big(e^{-n\sigma_-}-e^{-n\sigma_+}\Big),
}
with the distance between the branes $\Delta x^\mu=x^\mu_a-x^\mu_b$.
Taking the  derivatives one gets 
\eq{
\partial_{\pm}X^{\mu}_{\rm NN}&={\alpha'p^\mu}-\sqrt{\alpha'\over 2i}\sum_{n\neq 0}{\alpha^\mu_n}\,e^{-n\sigma_\pm}=-\sqrt{\alpha'\over 2i}\sum_{n\in \mathbb{Z}}{\alpha^\mu_n}\,e^{-n\sigma_\pm}\,,\\
\partial_{\pm}X^{\mu}_{\rm DD}&=\mp i{\Delta x^\mu\over 2\pi}\pm\sqrt{\alpha'\over 2i}\sum_{n\neq 0}{\alpha^\mu_n}\,e^{-n\sigma_\pm}=\pm\sqrt{\alpha'\over 2i}\sum_{n\in \mathbb{Z}}{\alpha^\mu_n}\,e^{-n\sigma_\pm}\,,
}
where we have defined the zero modes as
\eq{
\alpha_0^{\mu,{\rm NN}}=-p^{\mu}\sqrt{2i\alpha'}\,,\qquad
\alpha_0^{\mu,{\rm DD}}=-{\sqrt{-i}\over \sqrt{2\alpha ' }\pi}{\Delta x^\mu}\,.
}
For completeness we also present the mode expansion for mixed boundary
conditions
\eq{
X^{\mu}_{\rm ND}&=x^\mu+\sqrt{-2i\alpha'}\sum_{n\in \mathbb{Z}+1/2}{\alpha^\mu_n\over 2n}\Big(e^{-n\sigma_-}+e^{-n\sigma_+}\Big)\,,\\
X^{\mu}_{\rm DN}&=x^\mu+\sqrt{-2i\alpha'}\sum_{n\in \mathbb{Z}+1/2}{\alpha^\mu_n\over 2n}\Big(e^{-n\sigma_-}-e^{-n\sigma_+}\Big)\,.
}
Next we want to define the closed and open string partition functions.
For that purpose we first focus just on a single direction $X(\sigma_1,\sigma_2)$ and recall that  in the Sugawara construction the energy momentum tensor reads
\eq{
\label{Virasorodef}
T(z)&= {i\over \alpha'} \eta_{\mu\nu}  : \partial X^\mu(z)\partial X^\nu(z):\,.
}
With this, the normal ordered Hamiltonian becomes
\eq{
\label{Hamiltonian}
H&=\int_0^{2\pi} {\text{d}\sigma\over 2\pi \alpha'}
\Big((\partial_+X)^2+(\partial_-X)^2\Big)=-i\left({L_0+\overline{L}_0}-{\textstyle
  {c\over 12}}\right),
}
where the factor of $i$ originates in the mode expansion.
The explicit form of the energy momentum tensor's zero mode $L_0$ is
\eq{
              L_0=i{\alpha' p^2\over 4} + \sum_{n>0} \eta_{\mu\nu}\, \alpha_{-n}^\mu \alpha_n^\nu
}
and similarly for $\ov L_0$. The second term is just the number
operator which has non-negative integer eigenvalues. In
contrast to the usual case, the zero mode contribution is purely
imaginary.

The momentum $P$ which generates $\sigma_2$ translations is now given by
\eq{
P&=-i\int_0^{2\pi} \frac{\mathrm{d}\sigma}{2\pi \alpha'}   \Big(
(\partial_+X)^2-(\partial_-X)^2\Big)=-(L_0-\overline{L}_0)\;.
}
In this case the normal ordering constant cancels out.
As a consequence  the torus and cylinder amplitudes receive  additional factors of $i$. 
The torus amplitude with complex structure $\tau=\tau_1+i \tau_2$ 
is constructed by taking a field theory on a circle,
translating in $\sigma_1$ direction by $\tau_2$, in $\sigma_2$ direction by
$\tau_1$ and identifying the ends, producing the trace. With 
\begin{equation}
q=e^{2 \pi i (\tau_1+i \tau_2)}
\end{equation}
the torus partition function can be written as
\eq{
Z^\text{torus}&={\rm Tr}(e^{-2\pi i \tau_2 H-2\pi i \tau_1 P})={\rm
  Tr}\left(q^{L_0-{c\over 24}}\,\overline{q}^{\overline{L}_0-{c\over
      24}}\right)\,.
}
Note that due to the missing Wick rotation for the Euclidean CFT, the coefficient in front of the Hamiltonian is  $-2\pi i$
instead of the usual $-2 \pi$. But
this factor gets multiplied by the additional factor of $-i$ in the Hamiltonian
\eqref{Hamiltonian}, such that the expression for the partition
function is still the usual one. Evaluating the amplitude for a single
direction  one obtains
\begin{equation}
Z^\text{torus}={e^{i\pi/4} \over  \sqrt{ 4\pi\alpha' \tau_2} \, |\eta(\tau)|^{2}}\;,
\end{equation}
reproducing the result of \cite{Dijkgraaf:2016lym}. 

Now we turn to the open string cylinder amplitude which is defined as
\begin{equation}
\label{cylinderamplitude}
Z^\mathcal{C}(t)={\rm Tr} \left( e^{-2 \pi i t H_{\rm open}}\right)={\rm Tr} \left(e^{-2\pi t (L_0-{c\over 24})}\right),
\end{equation}
with $t$ the circumference of the cylinder.
The explicit form of $L_0$ for NN and DD boundary conditions is
\eq{
              L^{\rm NN}_0=i{\alpha' p^2} + \sum_{n>0} 
              \, \eta_{\mu\nu}\,  \alpha_{-n}^\mu \alpha_n^\nu\,,\qquad
             L^{\rm DD}_0={i\over 4\pi^2 \alpha'} Y^2 + \sum_{n>0}
              \, \eta_{\mu \nu}\, \alpha_{-n}^\mu \alpha_n^\nu\,.
}
The total distance between the Dirichlet loci is defined as $Y^2=\eta_{\mu \nu} \Delta x^\mu \Delta x^\nu$.
Now, considering only a single direction of either NN or DD type, the
open string partition functions can be evaluated straightforwardly  with the result
\eq{
Z^{\mathcal{C}({\rm NN})}(t)&={e^{-i\pi/4}\over \sqrt{2\alpha't}\;\eta(it)}\,,\qquad
Z^{\mathcal{C}({\rm DD})}(t)=e^{-{it\over 2 \pi\alpha'} Y^2}{1\over \eta(it)}\,.
}
The additional factor of $e^{-i\pi/4}$ in the Neumann-Neumann case
arises from the analytic continuation of the Gaussian integral for the
zero mode\footnote{We often employed the Gaussian integral
$\int_{-\infty}^{\infty} \text{d}x\, e^{-ax^2+bx}={\sqrt{\pi\over a}\cdot e^{b^2\over 4a}}$
and its analytic continuation. This is where most of the phases arise.
}.
For mixed boundary conditions one finds
\eq{
Z^{\mathcal{C}({\rm ND})}(t)&=\sqrt{ 2\eta(it)\over \theta_4(it)}\,.
}
As usual, the open string (loop-channel) cylinder amplitude is closely related to the (tree-channel) overlap
of boundary states
\begin{equation}
\label{overlap}
\tilde{Z}(l)=\bra{\Theta B}e^{2 \pi i l H_{\rm closed} }\ket{B}=\bra{\Theta B}e^{-2\pi l (L_0+\overline{L}_0-{c\over 12})}\ket{B}\,,
\end{equation}
with $l$ the length of the cylinder formed by the closed strings exchanged between the boundaries.
We will construct the appropriate boundary states in the next section.

\subsection{Boundary states}

Next we analyze the  construction of  boundary states in a Euclidean
world-sheet  CFT\footnote{Our analysis follows that of \cite{DiVecchia:1999mal,DiVecchia:1999fje,Blumenhagen:2009zz,Recknagel:2013uja} for
  Lorentzian strings.}. For the moment we assume also a purely Euclidean
space-time and postpone the treatment of  the effects of the
target-space  metric signature to the next section.
The boundary conditions are unaffected by the signature of the 
world-sheet. Despite the now Euclidean signature we will think of the
coordinate $\sigma_1 \in (0,l)$ as the time coordinate and $\sigma_2 \in
(0,\pi)$ as the space component. The conformal map exchanging the open
and closed channels acting on the complexified coordinate $\xi=\sigma_2
+i \sigma_1$ is then given 
by
\begin{equation}
f(\xi)=-i{\pi\over l}\xi\,
\end{equation}
which is the same as in the Lorentzian case  exchanging world-sheet
time  $\tau$ and space $\sigma$. 
The Neumann and Dirichlet gluing conditions are given by
\eq{
\partial_{\sigma_1}X^\mu|_{\sigma_1=0}\ket{B_N}=0\,,
\qquad
\partial_{\sigma_2}X^\mu|_{\sigma_1=0}\ket{B_D}=0\,.
}
Inserting the mode expansion \eqref{modeExpansion} results in 
\begin{equation}
p^{\mu}\ket{B_N}=0\,,\qquad (\alpha_n^\mu+\overline{\alpha}^\mu_{-n})\ket{B_N}=0
\end{equation}
as well as
\begin{equation}
x^{\mu}\ket{B_D}=y^{\mu}\,,\qquad (\alpha_n^\mu-\overline{\alpha}_{-n}^\mu)\ket{B_D}=0\,,
\end{equation}
where $y^\mu$ is the position of the brane. These are exactly the same
conditions as in the Lorentzian case. Defining a matrix
$S_{\mu\nu}=\pm \eta_{\mu\nu}$, with the $+$ sign for Neumann
directions and the $-$ sign for Dirichlet directions, the non-zero mode
conditions are given  by
\begin{equation}
\left(\alpha^{\mu}_n+S^{\mu}{}_{\nu}\,\overline{\alpha}^{\nu}_{-n}\right)\ket{B}=0\,.
\end{equation} 
As usual the solution to these gluing conditions is
\begin{equation}
\label{boundarystatesol}
\ket{B}={1\over
  \mathcal{N}}\,\text{exp}\left(-\sum_{n=1}^{\infty}{1\over
    n}\alpha^\mu_{-n}\,
   S_{\mu\nu}\,\overline{\alpha}^\nu_{-n}\right) |0\rangle\,.
\end{equation}
Using the explicit form of the boundary states \eqref{boundarystatesol}, the overlap \eqref{overlap} becomes
\eq{
\tilde Z^{\mathcal{C}({\rm NN})}(l)
={1\over \mathcal{N}_N^2}{1\over \eta(2il)}\,.
}
Mapping the open string  loop-channel result with $t={1/2l}$
and a modular S transformation  to the closed string tree-channel, 
the normalization constant can be determined via 
\begin{equation}
Z^{\mathcal{C}({\rm NN})}(t)={e^{-i\pi/4}\over \sqrt{2 \alpha't}\,\eta(it)}={e^{-i\pi/4}\over \sqrt{\alpha'} \,\eta (2il)}\overset{!}{=}{1\over \mathcal{N}_N^{2}\eta(2il)}\;\Rightarrow\; \mathcal{N}_N=(\alpha')^{1/4}e^{i\pi /8}\,.
\end{equation}
Turning to the DD case, the only thing that  changes is the $\alpha_0$ 
zero-mode contribution which is now given by
\eq{
&\int_{-\infty}^{\infty}{dk_adk_b\over 2\pi} e^{i k_a x_a}\,e^{i k_b x_b}\,\bra{k_a}e^{-2\pi l (\alpha_0)^2}\ket{k_b}\\
=&\int_{-\infty}^{\infty}{dk_adk_b}\, e^{i k_a x_a}\,e^{i k_b x_b}\,e^{-\pi i l\alpha' k_b^2}\,\delta(k_a+k_b)\\
=&\int_{-\infty}^{\infty}{dk_a}\, e^{i k_a (x_a-x_b)}\,e^{-\pi i l\alpha' k_a^2}
={e^{i\pi/4}\over\sqrt{ \alpha'l}}\,e^{-i{Y^2\over 4 \pi\alpha' l}}\,,
}
where we used $\braket{0|0}=2\pi \delta(0)$ and that the CPT operator
$\Theta$ in \eqref{overlap} involves a complex conjugation. Therefore the total DD overlap is
\begin{equation}
\tilde Z^{\mathcal{C}({\rm DD})}(l)={e^{-i\pi/4}\over
  \mathcal{N}_D^2}{1\over\sqrt{ 4\pi^2\alpha'l} \,  \eta(2il)}\,
e^{-i{Y^2\over 4\pi\alpha'  l}} \,.
\end{equation}
 Comparing this to the open string amplitude we obtain
 \begin{equation}
 \mathcal{N}_D=(\alpha')^{-1/4}\, e^{-i\pi/8}\,.
 \end{equation}
Finally, as a cross-check for the normalization factors, we evaluate
the mixed case as
\eq{
\tilde Z^{\mathcal{C}({\rm ND})}(l)
={1\over \mathcal{N}_N\mathcal{N}_D}\sqrt{2\eta(2il)\over \theta_2(2il)}
=\sqrt{2\eta(it)\over \theta_4(it)}=Z^{\mathcal{C}({\rm ND})}(t)\,,
}
featuring that the normalizations of the boundary states are indeed
consistent. 

\subsubsection*{The total cylinder amplitudes}

After having studied the open string amplitude for just a single
direction, we now combine the separate contributions into a total
cylinder amplitude of two parallel $d$-dimensional   branes in $D$ space-time
dimensions\footnote{Note that in this convention  a Dp-brane has $d=p+1$.}.
For the open string loop-channel amplitude   one obtains
 \begin{equation}
\label{annulusopen}
 \mathcal{A}={V_{d}}\int\limits_0^\infty {dt\over 2t}\left({e^{-i\pi/4}\over \sqrt{8\pi^2\alpha't}}\right)^{d}{1\over \eta^{D-2} (it)}e^{-{it\over 2 \pi \alpha'}Y^2}\,,
 \end{equation}
where the additional $\eta^{2}(it)$ factor originates from the ghost
contribution.
The total closed string tree-channel amplitude is
\eq{
 \tilde{\mathcal{A}}&={V_d\over \mathcal{N}^2}\int\limits_0^{\infty}
 dl\, e^{i\pi(D-d)/4}\,\left({1\over 4\pi^2\alpha' l}\right)^{(D-d)/2}
{1\over \eta^{D-2}(2il)}\,e^{-{i\over 4\pi \alpha'  l}Y^2 }\,.
}
Applying a modular S-transformation, this amplitude is mapped to the
loop-channel and comparing it to \eqref{annulusopen} one can read off the normalization
\begin{equation}
\label{normalization}
\mathcal{N}^{-1}=2^{{D-2\over4}}e^{{i\pi\over 8}(D-2d)}
(4\pi^2\alpha')^{{1\over 4} (D-2d-2)}\,.
\end{equation} 
Finally, the tension of the branes is determined by the coupling of the
boundary state to a graviton with polarization $\epsilon_{\mu\nu}$
 \eq{
 \braket{V_g|B}&=
-{1\over \mathcal{N}} \bra{0} \epsilon_{\mu\nu} S^{\mu\nu}\ket{0}
 =-{1\over \mathcal{N}}\,\epsilon_{\mu\nu}S^{\mu\nu}\,V_{d+1}
 \overset{!}{=} -T_d\,\epsilon_{\mu\nu}S^{\mu\nu}\,V_{d+1}\,,
}
so that the tension is given by the normalization of the boundary state as $T_d=\mathcal{N}^{-1}$.
For connecting a D-brane theory to phenomenology, 
we require the tension to be real, so that the 
normalization of the boundary state also has to be real. 
Inserting $D=10$ into \eqref{normalization}, 
we see that there are exactly three cases fulfilling this condition, 
$d\in\{1,5,9\}$ with tension
\eq{
\label{Tension}
T_d=\pm 2^{2}(4\pi^2\alpha')^{(4-d)/2}\;,
}
 with the minus sign for $d\in\{1,9\}$ and the plus sign for $d=5$.
 
\subsection{Fermionic boundary states}

So far we have only discussed the contribution of the world-sheet
bosons to the boundary states. Let us now also discuss the inclusion
of the world-sheet fermions.
The action for a free fermion is 
\begin{equation}
\label{fermact}
S_f={i\epsilon\over 4 \pi}\int\text{d}^2\sigma\sqrt{\epsilon\,
  \text{det} g}\;\,\eta_{\mu\nu}\, \overline{\Psi}^\mu\,\gamma^\alpha\partial_\alpha\Psi^\nu \,,
\end{equation}
where the $2\times2$ matrices $\gamma^\alpha$ satisfy the Clifford algebra
with respect to the world-sheet metric $g_{\alpha\beta}$
\begin{equation}
\label{Clifford}
\{\gamma^\alpha,\gamma^\beta\}=2g^{\alpha\beta}\,\mathbb{1}_2\;.
\end{equation}
Moreover, one defines 
\eq{
\overline{\Psi}^\mu &=\Psi^\mu\gamma^0
\qquad \text{ in the Lorentzian case and}\\
\overline{\Psi}^\mu &=\Psi^\mu C
\qquad \;\;\text{in the Euclidean case,}
}
with $C$ the charge conjugation matrix
\eq{\label{Ccond}
	(\gamma^\alpha)^\mathsf{T}=C\gamma^\alpha
	C^{-1}\,, \quad C^\mathsf{T}=C^{\dagger}=C^{-1}=C\,.
}
For  Lorentzian signature we choose the representation
\begin{gather}
\gamma^0= \begin{pmatrix} 0 & 1 \\ 1 & 0 \end{pmatrix}=\hat\sigma_1\,,\qquad \gamma^1= \begin{pmatrix} 0 & 1 \\ -1 & 0 \end{pmatrix}.
\end{gather}
Under Wick rotation $\tau= i\sigma_1$ one has
$\partial_\tau\rightarrow -i\partial_{\sigma_1}$, so  effectively 
$\gamma^1$ is  replaced with
\begin{equation}
\gamma^1=\begin{pmatrix} 0 & -i \\ i & 0 \end{pmatrix}=\hat\sigma_2\,. 
\end{equation}
Therefore, the Wick rotation has the effect of replacing the Lorentzian
gamma matrices with the Euclidean gamma matrices.
Then, the kinetic term of the Wick rotated theory  is the
same as in the purely Euclidean theory up to an overall sign. Choosing
the same Pauli matrices $\hat\sigma_1,\hat\sigma_2$ also as a representation
for the Euclidean Clifford algebra, the conditions \eqref{Ccond} 
uniquely determine $C$ to be
\begin{equation}
C=\begin{pmatrix} 0 & -1 \\ -1 & 0 \end{pmatrix}=-\gamma^0\,.
\end{equation}
Denoting the components of a 2D spinor as  
$\Psi^\mu= (  \Psi^\mu_+,\, \Psi^\mu_- )^\mathsf{T}$, 
the action reduces to
\begin{equation}
S_f=-{K\over 2\pi}\int\text{d}^2\sigma\; \eta_{\mu\nu} \Big(\Psi^\mu_+\partial_-\Psi^\nu_++\Psi^\mu_-\partial_+\Psi^\nu_-\Big)\,,
\end{equation}
where $K=-1$ for the (Wick-rotated) Lorentzian world-sheet and $K=i$ for the
Euclidean case. 
Here we have used again the coordinates $\sigma^{\pm}=\sigma_1\pm i\sigma_2$.
The equations of motion are
\begin{equation}
\partial_-\Psi^\mu_+=\partial_+\Psi^\mu_-=0\;
\end{equation}
with the usual (anti-)holomorphic solutions 
$\Psi^\mu_+=\Psi^\mu_+(\sigma^+)$ and $\Psi^\mu_-=\Psi^\mu_-(\sigma^-)$, 
which can be expanded as
\eq{
\Psi_+^\mu=\sqrt{-K }\sum_r b_r^\mu\, e^{-2\pi i r \sigma^+}\,,\qquad
\Psi_-^\mu=\sqrt{-K }\sum_r \ov b_r^\mu\, e^{-2\pi i r \sigma^-}\,.
}
As in the bosonic case, the factor $\sqrt{-K}$ ensures that 
the mode algebra takes  the usual form 
\begin{equation}
\{b_r^\mu,b_s^{\nu}\}=\delta_{r,-s}  \,\eta^{\mu\nu}\,,\qquad
\{\ov{b}_r^\mu,\ov{b}_s^{\nu}\}=\delta_{r,-s} \,\eta^{\mu\nu}\qquad
\{{b}_r^\mu,\ov b_s^{\nu}\}=0\,.
\end{equation}
The energy momentum tensor is obtained by the Sugawara construction, 
resulting in the explicit expression for the zero mode 
\begin{equation}
L_0=\sum_{r\ge 1/2}   \left(r+{\textstyle {1\over 2}}\right) \eta_{\mu \nu}\,b^\mu_{-r}\, b^\nu_{r}\,.
\end{equation}
Now that we have the algebra of the fermions we turn to the construction of the boundary state.
We will work again in the Euclidean formalism. The exchange of $\sigma_1$
and $\sigma_2$ acts on the Euclidean light cone variables as 
\begin{equation}
\sigma^{\pm}\rightarrow\sigma'^{\pm}= \mp i \sigma^{\pm}\;.
\end{equation}
The fermions transform under this conformal transformation as
\begin{equation}
\label{trafo}
\Psi'_\pm(\sigma'^\pm)=\left({\partial \sigma'^\pm\over \partial\sigma^\pm}\right)^{-1/2}\Psi_\pm(\sigma^\pm)\;.
\end{equation}
Imposing the open string boundary conditions on the boundary state,
and taking the transformation behavior into account one obtains 
conditions on the boundary states
\begin{equation}
\Big( \Psi^{\mu}_+(\sigma^+)+i\eta S^{\mu}{}_{\nu}\,\Psi^{\nu}_-(\sigma^-)\Big)\ket{B,\eta}=0\,,
\end{equation}
where $\eta=\pm 1$ labels periodic/antiperiodic boundary conditions. 
Expanding into  modes  results in the fermionic gluing conditions
\begin{equation}
\Big( b^{\mu}_r+i\eta S^{\mu}{}_\nu\,\overline{b}^{\nu}_{-r}\Big)\ket{B,\eta}=0\,.
\end{equation}
As usual, these gluing conditions are solved by the state
\begin{equation}
\label{boundarystatesolfermNS}
\ket{B,\eta}_{\rm NS}={1\over
  \mathcal{N}}\,\text{exp}\left(-i\eta\sum_{r=1/2}^{\infty} b^\mu_{-r}\,
   S_{\mu\nu}\,\overline{b}^\nu_{-r}\right) \ket{0}\,
\end{equation}
in the NS sector and by
\begin{equation}
\label{boundarystatesolfermR}
\ket{B,\eta}_{\rm R}={1\over
  \mathcal{N}}\,\text{exp}\left(-i\eta\sum_{n=1}^{\infty} b^\mu_{-n}\,
   S_{\mu\nu}\,\overline{b}^\nu_{-n}\right)\ket{0}_{\rm R}
\end{equation}
in the R sector, where $\ket{0}_R$ is the Ramond ground state which 
satisfies the gluing conditions for the zero modes.
The resulting tree-channel annulus amplitudes for a single fermion
read
\eq{
\bra{B,\eta}_{} e^{-2\pi l (L_0+\overline{L}_0-{c\over
    12})}\ket{B,\eta}_{\rm NS}&=\sqrt{\theta_3(2il)\over\eta(2il)}\,,\\[0.1cm]
\bra{B,\eta}_{} e^{-2\pi l (L_0+\overline{L}_0-{c\over
    12})}\ket{B,-\eta}_{\rm NS}&=\sqrt{\theta_4(2il)\over\eta(2il)}\,,\\[0.1cm]
\bra{B,\eta}_{} e^{-2\pi l (L_0+\overline{L}_0-{c\over
    12})}\ket{B,\eta}_{\rm
  R}&=\sqrt{\theta_2(2il)\over\eta(2il)}\,,\\[0.1cm]
\bra{B,\eta}_{} e^{-2\pi l (L_0+\overline{L}_0-{c\over
    12})}\ket{B,-\eta}_{\rm R}&=0\,.
}
Applying a modular S-transformation  leads to the open channel amplitudes
\eq{
{\rm Tr}_{\rm NS} \Big(e^{-2\pi t ( L_0-{c\over 24} )} \Big)&=\sqrt{\theta_{3}(it)\over\eta({it})}\,,\\
{\rm Tr}_{\rm NS}\Big( (-1)^{F}  e^{-2\pi t ( L_0-{c\over 24} )} \Big)&=\sqrt{\theta_{4}(it)\over\eta({it})}\,,\\
{\rm Tr}_{\rm R}\Big(e^{-2\pi t ( L_0-{c\over 24} )} \Big)&=\sqrt{\theta_{2}(it)\over\eta({it})}\,,\\
{\rm Tr}_{\rm R}\Big((-1)^{F} e^{-2\pi t ( L_0-{c\over 24} )} \Big)&=0\,,
}
which are the same as for Lorentzian superstrings.

Now let us construct the boundary state for a full d-dimensional D-brane in 10D. 
As usual, invariance of the boundary states under the left and right GSO
projections and stability requires the presence of all sectors. 
Then the total loop-channel annulus amplitude for two parallel $d$-dimensional branes 
becomes 
\begin{equation}
{A}={V_{d}}\int\limits_0^\infty {dt\over 2t}\left({1\over \sqrt{8\pi^2\alpha't}}\right)^{d}e^{i\pi d/ 4}\,{{\theta_3^{4}(it)-\theta_4^{4}(it)-\theta_2^{4}(it)}\over \eta^{12}(it)}\,e^{-{it\over 2 \pi \alpha'}Y^2}\,.
\end{equation}
Transforming this amplitude to the closed tree-channel amplitude
\begin{equation}
\tilde{A}={V_{d}}\int\limits_0^\infty {dl}\left({\sqrt{l}\over
    \sqrt{4\pi^2\alpha'}}\right)^{d}e^{i\pi d/ 4}\,
{{\theta_3^{4}(2il)-\theta_2^{4}(2il)-\theta_4^{4}(2il)}\over
  \eta^{12}(2il)\cdot (\sqrt{2l})^{8}}\,e^{-{i\over 4 \pi
    \alpha' l}Y^2}
\end{equation}
allows us to fix the relative contribution from the boundary states as
\begin{equation}
\label{BPSboundary}
\ket{D}={1\over 2\,\mathcal{N}}\Big(\ket{B,-}_{\rm NS}-\ket{B,+}_{\rm
    NS}+i\ket{B,+}_{\rm R}+i\ket{B,-}_{\rm R}\Big)\;,
\end{equation}
where the normalization $\mathcal{N}$ is the same as in the bosonic
case.

\subsection{The influence of space-time signature}

In the CFT approach the signature of the space-time merely appears as
a sign change in the commutation relations. This is hidden in most
formulas we have written down so far. In this section we will take a
closer look how the signature influences the amplitudes and boundary
states. 

We have seen that after absorbing 
the factor $K=i$ in a redefinition of the modes, we essentially get back the 
results for the Lorentzian string. The only difference resides in the
zero mode contribution.
As we will be concerned with branes wrapping various amounts of time dimension, in the following a
D$^{(s,t)}_{(10-p,p)}$-brane will fill $t$ time and $s$ space dimensions in a
$\mathbb{R}^{10-p,p}$ target-space with $p$ time and $10-p$ space dimensions.
Thus the system we are concerned with consists of
\begin{itemize}
\item $N_t=t$ time dimensions with Neumann boundary conditions,
\item $D_t=p-t$ time dimensions with Dirichlet boundary conditions,
\item $N_s=s$ space dimensions with Neumann boundary conditions,
\item $D_s=10-p-s$ space dimensions with Dirichlet boundary conditions.
\end{itemize}

In  the analysis so far all directions were assumed to be
space-like. Let us now analyze what changes in case some of the
directions become time-like.
First, recall that  the oscillator part of the boundary state
\eqref{boundarystatesol} involves the matrix $S_{\mu\nu}$. For a
D$^{(s,t)}_{(10-p,p)}$-brane
this takes the form
\eq{
                 S=\left(\begin{matrix}  \mathbb{1}_{N_s} & & & \\
                                                    & -\mathbb{1}_{N_t} & & \\
                                                      & & -\mathbb{1}_{D_s} & \\
                                           & & & \mathbb{1}_{D_t}  \end{matrix}\right)\,.
}
Thus, we see that the oscillators of a space-like N/D direction contribute to the
boundary state like a D/N time-like direction. However, these signs in
$S_{\mu\nu}$  cancel anyway when computing the overlap.

Now, let us consider the zero mode contribution, where some phase
factors appeared from the zero mode integrals. 
For a Neumann boundary condition in a space-like direction this phase is
\eq{
\mathcal{N}^{-2}_{N,\text{space}}\propto\int_{0}^\infty\text{d}p\; e^{-\pi i p^2}=e^{-i\pi /4}\;.
}
Changing the signature replaces $p^2$ by $-p^2$, so that
\eq{
\mathcal{N}^{-2}_{N,\text{time}}\propto\int_{0}^\infty\text{d}p\; e^{\pi i p^2}=e^{i\pi /4}\,.
}
Similarly, for a Dirichlet direction the exact same integrals appear
in the overlap of the zero modes of the boundary states, 
i.e. for a space-like  direction the phase
\eq{
\mathcal{N}^{\;2}_{D,\text{space}}&\propto\int_{0}^\infty\text{d}p\; e^{-\pi i p^2}=e^{-i\pi /4}\\
\Rightarrow \mathcal{N}^{\;2}_{D,\text{time}}&\propto\int_{0}^\infty\text{d}p\; e^{\pi i p^2} \;\;=e^{i\pi /4}
}
appears.
This implies that changing the signature, the only effect on the
normalization of the boundary state is  a change of the phase factor such that 
\eq{
\arg\left(\mathcal{N}_{N,\text{space}}\right)&=\arg\left(\mathcal{N}_{D,\text{time}}\right),\\
\arg\left(\mathcal{N}_{N,\text{time}}\right)&=\arg\left(\mathcal{N}_{D,\text{space}}\right).
}
Effectively this means that the formula for the normalization \eqref{normalization}
holds in all signatures, one just has to adjust the phase factor as
\begin{equation}
\label{Tension2}
T^{(s,t)}_{(10-p,p)}=2^{2}  e^{{i\pi\over 4}(5+t-p-s)}
(4\pi^2\alpha')^{{1\over 2} (4-s-t)}\,.
\end{equation} 
Note that we have simply replaced $d\to \tilde d=d+D_t-N_t=p+s-t$
in the phase factor to account for the additional phases. This formula is now valid for all branes in Euclidean world-sheet theories.

As a check let us consider a (positive) Dp-brane on top of a negative $\widehat{\rm Dp}$-brane
from section \ref{sec_map}. Crossing the interface of
the negative brane, the type II Dp-brane becomes a D$^{(1,p)}_{(10-p,p)}$-brane in the 
respective exotic superstring theory.
According to \eqref{Tension2},
the tension $ T^{(1,p)}_{(10-p,p)}=-T_{\rm Dp}$
changes sign, which is consistent with the map
\eqref{map-offschell}. The map also implies that  the R-R charge of the
brane has changed sign\footnote{The precise sign of the coupling
  $\mu^{(s,t)}_{(10-p,p)}$ of  the corresponding boundary state to
the correctly normalized R-R form is not so easy to determine. Since
we do not need it in the following, we refrain from going through the
exercise.\label{foota}}.

Taking now into account that the tension is real only for
$\tilde d\in\{1,5,9\}$,  it is straightforward to iterate all possible (real)
branes for a given  space-time signature. In the appendix we present
an exhaustive list of all D-branes in  all possible Euclidean
string theories. 
Here, let us  just discuss   two examples of space-time
signature $(7,3)$ and $(5,5)$.

In the first case there are $3$ time directions, thus
$p=3$. Then, $\tilde d\in\{1,5,9\}$  requires that $s-t$ is either $-2$, $2$ or
$6$. Moreover $s$ and $t$ count the number of longitudinal dimensions
of the brane, which cannot exceed the available dimensions, i.e. in
this case $0\le t\le 3$, $0\le s\le 7$. Iterating over all
possibilities one finds the allowed branes and tensions as shown on
the left in table \ref{table73}. Note that the tensions are given by \eqref{Tension2}, here we just
list the signs.
Now we turn to the second example with signature $(5,5)$. As $p=5$, from
$\tilde d\in\{1,5,9\}$ follows that $s-t$ is either equal to $-4$, $0$ or
$4$. Moreover, $s$ and $t$ are integers in the interval
$[0,5]$. Iterating over all possibilities we find the brane spectrum
listed on the right in table \ref{table73}. 

\begin{table}[ht]
\centering
\begin{tabular}{cccccc}
  \toprule
	s & t   & Tension& Dp&E/L	\\
	\midrule
       0 & 2 &$-$&D1&E  \\ \addlinespace
       1 & 3 &$-$&D3&L  \\ \addlinespace
	   6 & 0 &$-$&D5&E \\ \addlinespace
       7 & 1 &$-$&D7&L\\ \addlinespace
       2 & 0 &$+$&D1&E\\ \addlinespace
       3 & 1&$+$&D3&L\\ \addlinespace
       4 & 2 &$+$&D5&E  \\ \addlinespace
       5 & 3 &$+$&D7&L  \\ \addlinespace
	\bottomrule
 \end{tabular}	
\hspace{2cm}
\begin{tabular}{cccccc}
  \toprule
	s & t  & Tension& Dp&E/L	\\
	\midrule
       0 & 4 &$-$&D3&E\\ \addlinespace
	   1 & 5 &$-$&D5&L \\ \addlinespace
       5 & 1&$-$&D5&L\\ \addlinespace
       4 & 0 &$-$&D3&E\\ \addlinespace
       0 & 0 &$+$&D(-1)&E  \\ \addlinespace
       1 & 1 &$+$&D1&L  \\ \addlinespace
       2 & 2 &$+$&D3&E  \\ \addlinespace
       3 & 3 &$+$&D5&L  \\ \addlinespace
       4 & 4 &$+$&D7&E  \\ \addlinespace
       5 & 5 &$+$&D9&L  \\ \addlinespace
	\bottomrule
 \end{tabular}	
 \caption{Brane spectrum for  signature ${(7,3)}$ (left)
   and signature ${(5,5)}$ (right).}
  \label{table73}
\end{table}

As one can see, only even dimensional
branes exist, implying that we are in a type IIB setup. Note that this
information was not put in by hand, but is enforced by the signature
of space-time.
Let us already comment that these tables agree precisely with the
results from the next section where  a different target-space argument
is given for the existence of D-branes in Euclidean exotic superstring theories 
(see tables \ref{table_D3} and \ref{table_IIB55}). Moreover, the 
tables are consistent with the classification of
D-branes  reviewed in tables \ref{table_HullIIA} and \ref{table_HullIIB}.

As a final remark we note that  in our derivation the constraints for
the allowed D-branes
followed directly from the bosonic normalization factor. We have not explicitly discussed the GSO
projections in the fermionic sector, but as usual the constraint on even/odd
dimension of the branes follows  directly from  the Clifford algebra of the fermionic zero modes.
This computation  does not change in the Euclidean case so that the D-branes
obtained from the bosonic normalization are also GSO invariant.

\subsection{Orientifolds of Euclidean strings}
\label{sec_CFToplanes}

In this section,  we will take a look at orientifold projections of
the Euclidean exotic superstring theories. 
As the calculation strongly resembles the usual one, we will be very
brief and refer to standard textbooks \cite{Angelantonj:2002ct,Blumenhagen:2009zz} for more details.
Here we only show that in the computation of the loop-channel Klein-bottle and
M\"obius strip amplitudes, the same phase factors appear as for the
corresponding annulus amplitude. 

Thus, let us consider a single  bosonic direction $X(\sigma_1,\sigma_2)$. The orientifold projection
$\Omega:(\sigma_1,\sigma_2)\to (\sigma_1,-\sigma_2)$
acts on the modes as 
\begin{equation}
\Omega \;\alpha_n\; \Omega^{-1} =\ov{\alpha}_n\,.
\end{equation}
One can also combine $\Omega$ with the reflection $I_1: X\to -X$  so
that
 \begin{equation}
(\Omega I_1) \;\alpha_n\; (\Omega I_1)^{-1} =-\ov{\alpha}_n\,.
\end{equation}
Moreover, we choose the action of $\Omega$ on the vacuum as
$\Omega\ket{0}=\ket{0}$. 
Recall that the Klein bottle amplitude is
defined as
\eq{
Z_\Omega^{\mathcal{K}}&={\rm Tr}\left(\Omega\, q^{L_0-c/24}\,\ov{q}^{\ov{L}_0-c/24}\right)={\rm Tr}_{\rm sym}\left(e^{-4\pi t(L_0-c/24)}\right)\,.
}
The non-zero mode contribution again agrees with the usual result,
while the zero modes contribute a phase due to the additional factor
of $i$ in
the Gaussian integral. Thus for a single boson we get 
\begin{equation}
\label{Klein}
Z_\Omega^{\mathcal{K}}={e^{-i \pi /4}\over \sqrt{\alpha' t}}{1\over \eta(2it)}\,.
\end{equation}
The Klein Bottle amplitude for the orientifold projection $\Omega I_1$
does not receive any  zero mode contribution so that one obtains
\begin{equation}
\label{Kleinb}
Z_{\Omega I_1}^{\mathcal{K}}={\rm Tr}\left(\Omega I_1\,
  q^{L_0-c/24}\,\ov{q}^{\ov{L}_0-c/24}\right)= e^{i\pi \over
  24}\sqrt{2}\sqrt{\eta(2it)\over \theta_2(2it)}\;.
\end{equation}
Turning to the open string sector, the action of the orientifold on the modes is
\begin{equation}
\Omega \;\alpha^\mu_n\; \Omega^{-1} =\pm(-1)^n\alpha^\mu_n\,,
\end{equation}
with the plus sign for NN boundary conditions and the
minus sign for DD conditions. Again the non-zero
modes agree with the usual expressions. As in the DD sector
there is no zero mode contribution in the open string channel,
the M\"obius strip amplitude is as  usual
\begin{equation}
Z^{\mathcal{M}(DD)}=e^{i\pi \over
  24}\sqrt{2}\sqrt{\eta(it+\textstyle{1\over 2})\over \theta_2(it+\textstyle{1\over 2})}\;.
\end{equation}
The NN amplitude receives  an additional phase from the Gaussian integral
so that  
\begin{equation}
Z^{\mathcal{M}(NN)}=e^{i\pi \over 24}{e^{-i\pi /4}\over \sqrt{2\alpha'
  t}}{1\over\eta(it+\textstyle{1\over 2})}\,.
\end{equation}
Therefore, both the former  annulus amplitudes and  these additional
non-oriented one-loop amplitudes 
differ from the usual ones for Lorentzian signature by the same relative phases.
The next step is to introduce the corresponding crosscap states 
satisfying the usual crosscap gluing conditions
and allowing  the description of the amplitudes in tree-channel. 
Moreover, one can add the contributions from the world-sheet fermions.
However, also here the only difference to the standard case is the appearance of
the same phases as already experienced for the D-brane boundary
states. Thus, we refrain from presenting  the explicit form.

Performing now a full orientifold  projection\footnote{As already  shown in
  figure \ref{fig:projectionsEuclidean} there will be extra factors of $(-1)^{F_L}$ in certain
cases.} $\Omega I_{9-p}$
of the Euclidean type IIA/B superstring theories, the tadpole
cancellation conditions go through as usual, the Op-planes will have tension
\begin{equation}
 T_{Op}=-2^{p-4}T_{Dp}\,.
 \end{equation}
Introducing time-like  directions has the same effect on the
phase of the tension as for the corresponding boundary states.
To cancel the  tadpole induced by the orientifold projection one can introduce  stacks of Dp-branes
on top of the orientifold planes.


\section{D-branes  for  exotic string theories}
\label{sec_four}

In this section we further investigate D-branes in the exotic string theories.
The main question is which of these branes carry a ghost-free
low-energy effective action. Here we will  not analyze the full action 
for an in general intersecting brane system, but as a first
step we will restrict to the kinetic term of the gauge field itself.

As mentioned previously, the motivation behind this analysis 
is that, being agnostic about the fate and meaning of ghosts in 
gravity (closed string) theories, the experimentally  accessible 
gauge theory (open string) sector should satisfy the usual 
requirements that we impose on quantum field theories like the 
Standard Model, namely freedom from physical ghosts and unitarity. 
Indeed, should we find a ghost-free gauge sector in a theory
with closed string ghosts, our pragmatic approach
would open a window for dS in string theory.

\subsection{D-branes for Euclidean exotic strings}

First, we consider the Euclidean exotic string theories studied in section \ref{sec_cft}
and their D-branes. Having already constructed the corresponding boundary
states in a CFT approach, we will now determine their effective action
by applying the map \eqref{map-offschell} derived from the negative brane scenario
to the DBI action of the D-branes in  type IIA/B superstrings. 
Note that the negative tension $\widehat{\rm Dp}$-brane should be considered just as a nice tool to identify the
correct map from regular to exotic theories as in \eqref{TheoryMappings}. 
In the following, we will call this brane the {\it defining $\widehat{Dp}$-brane}.

\subsubsection*{Exotic DBI actions}

The usual DBI+CS action for a Dq-brane (in IIA/IB$^{++}$ string theories) can be expanded as:  
\eq{
S_{\rm DBI+CS}= &-T_q \int_{\Sigma_{q+1}} \de^{q+1} x \,\sqrt{|g|}\, e^{-\phi}
\left[1+\frac{1}{4}(2\pi\alpha')^2F_{\mu\nu}F^{\mu\nu} + ...
\right]\\[0.2cm]
&+\mu_q \, \int_{\Sigma_{q+1}}  \left[C_{q+1}+ F\wedge C_{q-1}+... \right]
\label{DBI}
}
where $F$ denotes the gauge field strength on the brane and  $C_p$
bulk R-R p-forms. Of course, for a BPS Dq-brane the
tension is the same as the R-R charge, i.e. $T_q=\mu_q>0$.

The above action is  ghost-free, since the gauge kinetic term
has the expected overall minus sign. When one performs the mapping
to the exotic string theories, there are two places in the above
action where factors of $i$ (or signs) will arise. The first  is the
relative sign between the two terms in the DBI part: since ${|F|^2}$
contains two inverse metric factors, it is clear that under
\eqref{map-offschell}  it will pick up a  minus factor.
This relative sign change happens always,
regardless of the number of dimensions that change signature or the
dimension of the Dq-brane. The second place 
is the overall sign of the DBI part due to the rescaling of the dilaton  
as well as the rescaling of the measure.
The factor coming from the dilaton depends on the number $p$ of
space-time dimensions that change sign, while the rescaling of the measure
now depends on the position (number of parallel and transverse 
dimensions $n_\parallel$, $n_\perp$) 
of the Dq-brane in the signature-changing space-time directions.
Since the topological CS term does not contain dilaton or metric
factors, the only change there can come from the transformation of the
R-R form $C_{q+1}$ which we haven't fully determined here (see also footnote \ref{foota}).

It is worth noting that even
though we are dealing with factors of $i$, all of them nicely cancel out
for  BPS configurations, giving at most an overall sign
change. Here BPS means that the Dq-brane is
supersymmetric relative to the defining $\widehat{\rm Dp}$-brane.
The  requirement for a Dq-brane to be BPS  can be translated to
the condition  
\eq{
n_\perp+(p+1)-n_\parallel=0 \mod 4\,.
} 
As a consequence, the DBI action for a  BPS Dq-brane in the exotic
theory can only take one of the  two  forms 
  \begin{equation}
    S_{\rm DBI}=\!
    \begin{cases}
      -T_q \int \de^{q+1} x\, \sqrt{|g|}\,e^{-\phi}\!
      \left[1-\frac{1}{4} (2\pi\alpha')^2 F_{\mu\nu}F^{\mu\nu} + ...
      \right]
    \pm\mu_q  \int  \left[C_{q+1}+...\right] \\[0.4cm]
      +T_q \int \de^{q+1} x \,\sqrt{|g|}\,e^{-\phi}\!
      \left[1-\frac{1}{4} (2\pi\alpha')^2 F_{\mu\nu}F^{\mu\nu} + ...
      \right]
    \pm\mu_q  \int \left[ C_{q+1}+...\right].
    \end{cases}
  \end{equation}

Note that the relative sign in front of the kinetic term of the gauge
field changed in both cases. We believe that this reflects the generic
sign change reported in \cite{Hull:1999mt} for all D-branes in
Euclidean exotic string theories.
In addition, the methods employed in this paper also
allow us   to determine the sign of the overall normalization (tension).

In the upper case the overall sign is the usual minus. 
The action is of the same form as the usual DBI+CS action, with
the significant difference that the sign in front of the gauge kinetic
term is now altered. Hence, the gauge field comes with a kinetic term
of the wrong overall sign so that  this brane sector is \emph{not ghost-free}. 
The physical interpretation of the second action is
also clear. The gauge kinetic term comes with the usual negative
sign, so the gauge sector is \emph{ghost-free}. However, the first
term in the bracket now carries  a relative negative sign with respect
to the usual case.  Therefore, such an exotic  Dq-brane has \emph{negative tension}.

\subsubsection*{Classification of BPS branes}

We will now move forward and  present a comprehensive
classification of the BPS  branes that appear in the 
Euclidean  exotic string theories. This will enable us to verify the
involved CFT construction from the previous section.

We start  with the regular  type IIB theory and consider  a
defining $\widehat{Dp}$-brane (with p odd) that is Lorentzian. Then the 
map \eqref{map-offschell}  gives the exotic Euclidean IIB theory
\eq{
\rm{IIB}^{++}_{(9,1)}\longrightarrow \rm{IIB}^{-(-)^{\frac{p-1}{2}}}_{(10-p,p)}\,
}
and the corresponding mirror theories \eqref{eqn:mirrors}. 
Next, we introduce all possible relatively BPS, Lorentzian Dq-branes in the original type IIB 
theory and map them via \eqref{map-offschell} to the corresponding
Dq-brane in the exotic $\rm{IIB}^{-(-)^{(p-1)/2}}_{(10-p,p)}$ theory\footnote{While we will use a notation inspired
by the negative brane horizon crossing of section \ref{sec_three},
let us stress  once again that we only utilize a negative brane in an
intermediate step but are finally interested in the result of mapping 
a usual BPS type IIB Dq-brane to  the exotic string theories.}.
Hence, $(p+1)$ is  the number of space-time
directions $x_i$ which will change signature, while the signature of the 
other $(9-p)$ directions $y_j$ stays the same. 

Let us mention again that  we  denote by $n_{\parallel}$ the number of
dimensions along the signature changing $x_i$'s, and $n_{\perp}$ the number of dimensions
along the $y_i$'s, with $n_\parallel + n_\perp = q+1$. 
A Dq-brane in the exotic string theory is denoted as
Dq$^{(s,t)}_{(10-p,p)}$, where the pair $(s,t)$ adds up to $q+1$
and indicates the signature of the brane.

Then applying the map to the metric $g$ on Dq, the measure picks up a 
factor of $\omega^{-1/2}$ for each signature changing direction, and a factor
of $\omega^{1/2}$ for the others. 
The metric and the dilaton transform exactly as in \eqref{map-offschell}.
It is then straightforward to determine how the DBI
action for the Dq-brane transforms 
\eq{
S_{\rm DBI}\rightarrow &-T_q \int \de^{q+1} x\, \sqrt{|g|}\,
\omega^{\frac{n_\perp-n_\parallel}{2}} \omega^{\frac{p-3}{2}}\,e^{-\phi}
\left[1-\frac{1}{4}(2\pi\alpha')^2F_{\mu\nu}F^{\mu\nu} + ...
\right]\\[0.2cm]
 &=-\omega^{\frac{p+n_\perp-n_\parallel-3}{2}} T_q  \int \de^{q+1} x\, \sqrt{|g|}\,e^{-\phi} \left[1-\frac{1}{4}(2\pi\alpha')^2F_{\mu\nu}F^{\mu\nu} + ...  \right]\,,
\label{DBImapping}
}
which allows to read-off the tension of the brane in the exotic string
theory.
We note  that depending on the position of the Dq-brane, we might 
get branes of the same dimension which nevertheless have different
tensions. As long as
all branes of the same dimension are either Lorentzian or
Euclidean, it is still consistent with our general framework. 
Conveniently, this will be  satisfied automatically.

\subsubsection*{Example with defining $\widehat{\rm D3}$-brane}

Let us present  an illustrative example. We pick a defining
$\widehat{\rm D3}$-brane and present 
all the consistent branes in the corresponding exotic theories, 
namely IIB$^{--}_{(7,3)}$ and its ``mirror" IIB$^{--}_{(3,7)}$. 
For  $p=3$, either the first four directions
($x^i,i=0,..,3$) will get their signature reversed, leading to a $(7,3)$
space-time,  or the six last ($x^i,i=4,\ldots,9$), leading to a $(3,7)$
space-time. 
Then we successively consider D1-, D3-,$\ldots$, D9-branes and put
them in relatively supersymmetric positions to the defining
$\widehat{\rm D3}$-brane. 

For example, one can see from table \ref{table_D3} that for the D1-brane
there exists essentially only one possibility.
In the ``brane positioning'' column of the table,
we denote by the superscript whether a direction is
space-like (s) or time-like (t) in the $(7,3)$ theory. 
The subscript denotes the same for the ``mirror" $(3,7)$ theory.
One can read off that the type IIB D1-brane maps to a Euclidean 
D1$_{(7,3)}^{(2,0)}$-brane with positive tension in the 
exotic $(7,3)$ theory. Being Euclidean, 
this is consistent with a IIB$^{--}$ theory.
Moreover, being of positive tension implies that 
the gauge field on the brane is a ghost field. 

Similarly, one can analyze the higher dimensional Dq-branes and
fill out the entire table \ref{table_D3}.
As expected, the signatures (Lorentzian/Euclidean) of the branes alternate. 
There are 7 different BPS configurations allowed. Out of these, 3 have negative
tension and are therefore ghost-free. We should note here that the table only includes the overall sign
of the brane tension, as the precise value is irrelevant for the present discussion.
Let us also stress that while there exist negative tension (ghost-free) D3-, D5- and D7-
branes, not all are of this type since the arrangement of the branes in 
space-time plays a crucial role. We observe that the brane spectrum of table
\ref{table_D3} could be  partially incomplete, since additional branes
can occur with the mirror dual 
$p=7$ mapping. 

\begin{table}[bht]
\centering
\begin{adjustbox}{width=\textwidth}
  \begin{tabular}{ccccccccc|ccccccccc}
  \toprule
	\multirow{2}{*}{{Dq} } &\multirow{2}{*}{n$_{\perp}$}  & \multirow{2}{*}{n$_{\parallel}$}  & \multirow{2}{*}{k} &\multirow{2}{*}{Tension} &  	
	 \multicolumn{10}{c}{Brane positioning}& IIB$^{--}_{(7,3)}$ &IIB$^{--}_{(3,7)}$ &{Brane} \\
		&&&&&0$^s_t$&1$^t_s$&2$^t_s$&3$^t_s$&4$^s_t$&5$^s_t$&6$^s_t$&7$^s_t$&8$^s_t$&9$^s_t$& Brane & Brane&Type(E/L) \\
	\midrule
	D1 &  1 & 1  &1 &$+$ &\checkmark &- &- &- &\checkmark  &- &- &- &- &- &D1$_{(7,3)}^{(2,0)}$  &D1$_{(3,7)}^{(0,2)}$&E \\ \addlinespace
	D3 &  0 & 4  &0 &$-$ &\checkmark  &\checkmark  &\checkmark  &\checkmark  &- &- &- &- &- &- &D3$_{(7,3)}^{(1,3)}$  &D3$_{(3,7)}^{(3,1)}$& L\\ \addlinespace
	D3 &  2 & 2  &1 &$+$ &\checkmark  &\checkmark  &- &- &\checkmark  &\checkmark  &- &- &- &- &D3$_{(7,3)}^{(3,1)}$&D3$_{(3,7)}^{(1,3)}$&L   \\ \addlinespace
	D5 &  3 & 3  &1 &$+$ &\checkmark  &\checkmark  &\checkmark  &- &\checkmark  &\checkmark  &\checkmark  &- &- &- &D5$_{(7,3)}^{(4,2)}$ &D5$_{(3,7)}^{(2,4)}$&E  \\ \addlinespace
	D5 &  5 & 1  &2 &$-$ &\checkmark  &- &- &- &\checkmark  &\checkmark  &\checkmark  &\checkmark  &\checkmark  &- &D5$_{(7,3)}^{(6,0)}$  &D5$_{(3,7)}^{(0,6)}$ &E \\  \addlinespace
	D7 &  4 & 4  &1 &$+$ &\checkmark  &\checkmark  &\checkmark  &\checkmark  &\checkmark  &\checkmark  &\checkmark  &\checkmark  &- &- &D7$_{(7,3)}^{(5,3)}$ &D7$_{(3,7)}^{(3,5)}$& L \\ \addlinespace
 	D7 &  2 & 6  &2 &$-$ &\checkmark  &\checkmark  &- &- &\checkmark  &\checkmark  &\checkmark  &\checkmark  &\checkmark  &\checkmark  &D7$_{(7,3)}^{(7,1)}$ &D7$_{(3,7)}^{(1,7)}$& L \\    \addlinespace
 	D9 &\multicolumn{14}{c}{\emph{No consistent D9-brane configuration}}&-&-&-\\
	\bottomrule
  \end{tabular}
 \end{adjustbox}  
 \caption{Brane spectrum of IIB$^{--}_{(7,3)/(3,7)}$ theories.}
 \label{table_D3}
\end{table}


We can now perform  the classification also for $p=1,5,7,9$.
In table \ref{table_IIB91} of appendix \ref{app_IIBtable}  
we present the brane spectrum of IIB$^{-+}_{(9,1)}$ and its mirror dual
 IIB$^{-+}_{(1,9)}$.  In table \ref{table_IIB73} we give the result for
IIB$^{--}_{(7,3)/(3,7)}$ and in table \ref{table_IIB55} that for
IIB$^{-+}_{(5,5)/(5,5)}$.
In a similar fashion, we computed the
D-brane spectrum of the exotic Euclidean type IIA theories in the various consistent
signatures. We present the results in
tables~\ref{table_IIA100},~\ref{table_IIA82},~\ref{table_IIA64} of
appendix~\ref{app_IIAtable}. 
Note that there is one major difference
to the type IIB case: While for type IIB the space-time mirror
theories are of the same type, in type IIA the space-time mirror also
affects the type of the theory, in particular whether the branes are
Euclidean or Lorentzian. The Euclidean type IIA space-time mirrors are 
IIA$^{--}_{(10-p,p)}\!\leftrightarrow\,$IIA$^{-+}_{(p,10-p)}$.
Finally, we want to stress that the results of the present section \ref{sec_four}
are in complete accord with the brane spectrum of section \ref{sec_cft}, 
which was acquired using Euclidean CFT techniques. 
Our results are also compatible with the spectra reviewed in 
tables \ref{table_HullIIA} and \ref{table_HullIIB}.
In addition we found the missing D$8$ and D$9$ 
branes, and more importantly derived the signs of the tensions
for all the branes.

\subsection{Ghost-free D-brane theories}

Scanning through the tables we extract all D-branes that are
ghost-free (i.e. have negative tension) and contain a $(3,1)$
subspace. These are the D-branes that have a chance to lead to a 
viable phenomenology. These ghost-free branes are
\eq{
\label{branesghostfree}
&{\rm type\, IIB} :\; {\rm D9}^{(9,1)}_{(9,1)}\,,\quad {\rm D7}^{(7,1)}_{(7,3)}\,,\quad {\rm D5}^{(5,1)}_{(5,5)}\,,\quad {\rm D3}^{(3,1)}_{(3,7)}\,,\\[0.3cm]
&{\rm  type\, IIA} :\; {\rm{D8}}^{(8,1)}_{(8,2)}\,,\quad
{\rm D6}^{(6,1)}_{(6,4)}\,,\quad {\rm D4}^{(4,1)}_{(4,6)}\,.
}
In the following, we discuss this class of branes in more detail, as
they share  a couple of common features.

First, all these branes have in common that they are space-filling, but localized in the
extra time-like directions. For instance, as can be seen from 
table \ref{table_D3}, the D7$^{(7,1)}_{(7,3)}$ brane is localized in
the $t_2$ and $t_3$ directions and longitudinal along $s_0,t_1,s_4,\ldots,s_9$.
Compactifying the extra time-like directions and all space-like
directions except the three large ones that are to make our world, 
an open string ending on the brane will have KK modes along 
the compact space-like directions and winding modes in the compact
time-like directions. As a consequence, employing \eqref{onshellthree}
the mass spectrum of such an
open string reads
\eq{
\label{onshellopen}
   E^2= \sum_{i}
           (p^i)^2+\sum_s \left({m_s\over R_s}\right)^2
             + \sum_t \left( {n_t R_t\over \alpha'}\right)^2-{i\over \alpha'}\,  (N-a) \,,
}
where the indices $s (t)$ indicate space(time)-like directions. 
Therefore, for these particular branes both KK and winding modes
contribute positively to the right hand side of \eqref{onshellopen}.
This is the same behavior as for D-branes in the usual type IIA/IIB
theories. This implies that in contrast to closed strings, for such D-branes there is no issue
with an infinite number of open string modes becoming arbitrarily light.

Being localized in the extra time-like directions, the transversal
deformations of the D-branes in \eqref{branesghostfree} will be ghosts. On a torus such
deformations will exist but on a more general background they can be
absent, if the brane wraps a rigid cycle. There will certainly exist
massive open string ghosts, but they are expected to kinematically decouple from the
massless open string states below a cut-off  $\Lambda_{\rm UV}$.
Whether also the 
ultra-light closed string states decouple is a more intricate
question. Since they  couple gravitationally, they are expected to
decouple in the large Planck-mass limit. However, there are in principle
infinitely many such states, so it is not a trivial question whether they will
have a negligible overall effect on the low-energy scattering of massless open string
modes.

The second common feature of the ghost-free D-branes in \eqref{branesghostfree}
is that they are all directly related to the orientifold projections
discussed in section \ref{sec_orient}  and summarized in figure \ref{fig:projectionsEuclidean}.
If  transversal directions of a D-brane are compactified, its R-R
charge has to be cancelled, a feature known as  tadpole
cancellation. Therefore,
one is forced  to introduce also oppositely charged objects in the
backgrounds.
These are  the orientifold planes constructed in section \ref{sec_CFToplanes}. They arise by performing an orientifold
projection $\Omega I_{\perp}$, where $I_{\perp}$ reflects the coordinates transversal
to the brane. As an example consider D7$^{(7,1)}_{(7,3)}$, whose
related orientifold quotient  is IIB$^{--}_{(7,3)}/\Omega I_2
(-1)^{F_L}$, 
where $I_2$ reflects the two extra time-like coordinates.
However, this is precisely the orientifold projection that removes
all the 10D massless ghosts in the closed string sector. The same
behavior
arises for all the ghost-free branes in \eqref{branesghostfree}. 
Summarizing,  the
required orientifold projection that allows us to introduce these branes
in the given background in the first place, is also the orientifold 
that we encountered in figure \ref{fig:projectionsEuclidean} 
which projects out all the massless
ghosts appearing in the exotic 10D supergravity actions.

Recall now that the de Sitter solutions of the type presented in section \ref{sec_two}
were only possible because of the existence of massless ghosts in 10D
closed string action. Without this loop-hole in the classical dS no-go
theorem, dS vacua will very likely not be possible. Therefore, for the
Euclidean exotic string theories with in general multiple times, there
is a strong correlation between the presence of a phenomenologically
viable D-brane (gauge theory) sector and the existence  of dS solutions:
  
\begin{quotation}
\noindent
{\bf Conjecture:}
{\it A  compactified Euclidean  exotic string theory contains a 4D 
ghost-free gauge theory with signature $(3,1)$, iff the closed string
sector does not admit classical dS vacua.}
\end{quotation}

\noindent
This can be interpreted as an extension of the (classical) dS
swampland conjecture to Euclidean exotic superstring theories.
We still consider it a conjecture, as in principle there could be
other orientifold projections still admitting dS solutions, whose 
tadpole could be cancelled by introducing more involved configurations
of intersecting D-branes. The latter will likely contain besides the 
ghost-free branes \eqref{branesghostfree} also some other ones with
ghosts. Moreover, there will be additional massless states on the intersection of
branes that could also be ghosts. Whether such configurations could
lead to a viable  ghost-free standard-model subsector remains to be
seen, though we doubt it.

\subsection{Orientifolds of Lorentzian exotic strings}

Let us now consider the Lorentzian exotic string theories and  analyze
whether ghost-free D-branes can be introduced there. Here the  $\mathbb Z_2$
projections from figure \ref{fig:projectionsLorentzian} that project out the massless 10D
ghosts are not orientifold projections but just $\mathbb Z_2$ orbifolds.

Thus, first we investigate what kind of $\mathbb Z_2$ projections the various kinds
of such theories do admit. In the second part we will analyze
whether there exist orientifolds that support D-branes with a
ghost-free kinetic term for the gauge field while still potentially admitting
de Sitter solutions in the closed string sector.

\subsubsection*{Generalities on  $\mathbb Z_2$ projections}

In the following we will be interested in  $\Omega I_{m,n}$
orientifolds for the type IIA/B$^{(+,\beta)}_{(10-p,p)}$ theories, where $I_{m,n}$ denotes the
reflection of $m$ space-like and $n$ time-like directions. 
As is known already for the usual type II theories  there appears a subtlety
in the Ramond sector of the theory. 

For all theories of signature $(q,p)\in\{(9,1),(5,5),(1,9)\}$ the Clifford algebra
\eq{
              \{\Gamma^A,\Gamma^B\}=2\, \eta^{AB}\,,\qquad
}
has similar properties (as $q-p=0\ {\rm mod\ 8}$). Here, $\eta^{AB}=\pm 1$ for space/time-like
directions. Let us recall some of the salient properties of the
$\Gamma$-matrices. In the following $\Gamma^a$  denote space-like
directions and $\Gamma^\alpha$ time-like ones. 
All $\Gamma$-matrices are unitary, if they satisfy the following
hermiticity conditions
\eq{
                 (\Gamma^a)^\dagger = \Gamma^a\,,\qquad (\Gamma^\alpha)^\dagger = -\Gamma^\alpha\,.
}
Moreover, one can define the chirality operator
\eq{
                \Gamma^{10}=\prod_A \Gamma^A
}
which anti-commutes with all $\Gamma^A$, is Hermitian
$(\Gamma^{10})^\dagger=\Gamma^{10}$ and satisfies $(\Gamma^{10})^2=1$. One can choose
the $\Gamma$-matrices to be purely imaginary in which case
$\Gamma^{10}$ is real.
In this representation, a Majorana spinor is real.

The Ramond ground state in both the left and the right-moving sector is a
Majorana-Weyl spinor in 10D, thus it is chiral and real. Type IIB has
two spinors of the same chirality and type IIA two spinors of opposite chirality.
These spinors of positive and negative chirality are denoted as usual by $S^+$ and $S^-$.

The reflection  along a single space-like direction $x^a$  acts on the
spinors as
\eq{
              I^a: S \to  i \Gamma^{10} \Gamma^a S \,,         
}
guaranteeing $\{I^a,\Gamma^a\}=0$ and
$[I^a,\Gamma^B]=0$ (for $a\ne B$) . 
This operation is Hermitian, real and changes the chirality of the
spinor. Moreover, it satisfies $(I^a)^2=1$.
The reflection along a time-like
direction can also be chosen to be Hermitian but then it becomes purely
imaginary
\eq{
             I^\alpha: S \to   \Gamma^{10} \Gamma^\alpha S \,.         
}
Thus, $I^\alpha$ is Hermitian, imaginary, changes the chirality
and satisfies $(I^\alpha)^2=1$. Let us now consider a general  $\mathbb Z_2$ reflection 
\eq{ 
I_{m+n}=\prod_{i=1}^m  I^{a_i}  \, \prod_{j=1}^n  I^{\alpha_j}
}
along $m$ space-like and $n$
time-like directions. Its action on a spinor is summarized 
in table \ref{table_actspinor}.

\begin{table}[ht]
\centering
  \begin{tabular}{c c}
  \toprule
  $I_{m,n}$    & {\rm \ \ action on spinor}    \\
    \midrule
 &  \\[-0.4cm]
$m$ even, $n$ even & $S^\pm \to S^\pm$\\[0.1cm]
$m$ odd, $n$ odd & $S^\pm \to i S^\pm$\\[0.1cm]
 \midrule
$m$ odd, $n$ even & $S^\pm \to S^\mp$\\[0.1cm]
$m$ even, $n$ odd & $S^\pm \to i S^\mp$\\[0.1cm]
\bottomrule
  \end{tabular}
  \caption{Action of $I_{m+n}$ reflection on spinors.}
  \label{table_actspinor}
\end{table}

\noindent
Using that  $\{I^A,I^B\}=2\delta^{AB}$,
one can show that for $m+n=2k$ or $m+n=2k+1$
the square of $I_{m+n}$ on
the Ramond ground state is $ I_{m+n}^2=(-1)^{k}$.
We are interested in the consistent orientifold projections
of type $\Omega I_{m+n}$ for the type IIB/IIA$^{+,\beta}$ string theories.
Requiring that  the full orientifold projection squares to $+1$ one
obtains the admissible possibilities listed in table
\ref{table_orientIIAB} for the type IIB/IIA string.
Here, as usual the factor
$(-1)^{F_L}$ is introduced to compensate for $(\Omega I_{m+n})^2=-1$.

\begin{table}[ht]
\centering
  \begin{tabular}{c c}
  \toprule
  Type IIA/B$^{\alpha\beta}$    & {\rm orientifold }    \\
    \midrule
 &  \\[-0.4cm]
Type IIB$^{++}$ & $\Omega I_{2m,2n} \left[ (-1)^{F_L}\right]^{m+n}$
\\[0.1cm]
Type IIB$^{+-}$ & $\Omega I_{2m-1,2n-1} \left[ (-1)^{F_L}\right]^{m+n-1}$
\\[0.1cm]
Type IIA$^{++}$ & $\Omega I_{2m-1,2n} \left[ (-1)^{F_L}\right]^{m+n-1}$
\\[0.1cm]
Type IIA$^{+-}$ & $\Omega I_{2m,2n-1} \left[ (-1)^{F_L}\right]^{m+n-1}$
\\[0.1cm]
\bottomrule
  \end{tabular}
  \caption{Admissible orientifold projections.}
  \label{table_orientIIAB}
\end{table}

\subsubsection*{Ghost-free D-branes and dS}

Let us now  analyze whether orientifolds of  type IIA/B$^{+,\beta}$ 
can support D-branes of at least signature $(3,1)$ and without gauge field ghosts while
still admitting  dS-type solutions in the closed string sector.
The type IIA/B$^{+,\beta}_{(1,9)}$ theories can be dismissed right
away, as they do not have at least three space-like directions.
Moreover, type IIA/B$^{++}_{(9,1)}$ are just the usual type IIA/B
theories  for which the dS swampland conjecture is
supposed to hold. The type IIA/B$^{+-}_{(9,1)}$ theories only contain
Euclidean D-branes that cannot support a gauge theory in $(3,1)$ 
dimensions. Thus we are left with the type IIA/B$^{+,\beta}_{(5,5)}$
theories. 

Let us have a closer look at the type
IIA$^{++}_{(5,5)}$ theory. This theory still contains Lorentzian
fundamental strings so that the CFT is like the usual type IIA
theory, only the signature changes from $(9,1)$ to $(5,5)$. All
D-branes have positive tension and the usual sign of the kinetic term
for the gauge field. Of course, this implies that the time-like
components of the gauge field $A^\mu$ are ghosts.

As we have seen, it is the $\mathbb Z_2$ projection
$I_4$ or $I_4 (-1)^{F_L}$ reflecting the four extra time-like directions that removes
all closed string ghosts from the action. Clearly, this is not
an orientifold so that it could well be that e.g. an orientifold
$\Omega I_3 (-1)^{F_L}$ with O6-planes and corresponding D6-branes
gives a ghost-free gauge theory, while still allowing closed string ghosts.

For concreteness, let us consider a compactification on a six torus $T^6$.
Let us denote the two compact space- and the four compact
time-directions as $\{x_1,x_2;t_1,t_2,t_3,t_4\}$.
Therefore, we can group the six-coordinates
in three pairs $\{(x_1,x_2),(t_1,t_2),(t_3,t_4)\}$ and
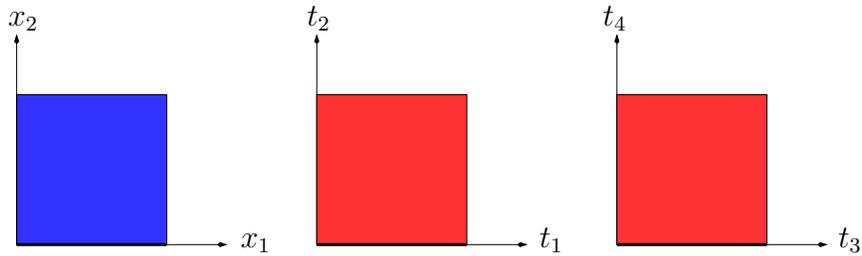
\begin{figure}[ht]
  \centering
   \includegraphics[width=0.7\textwidth]{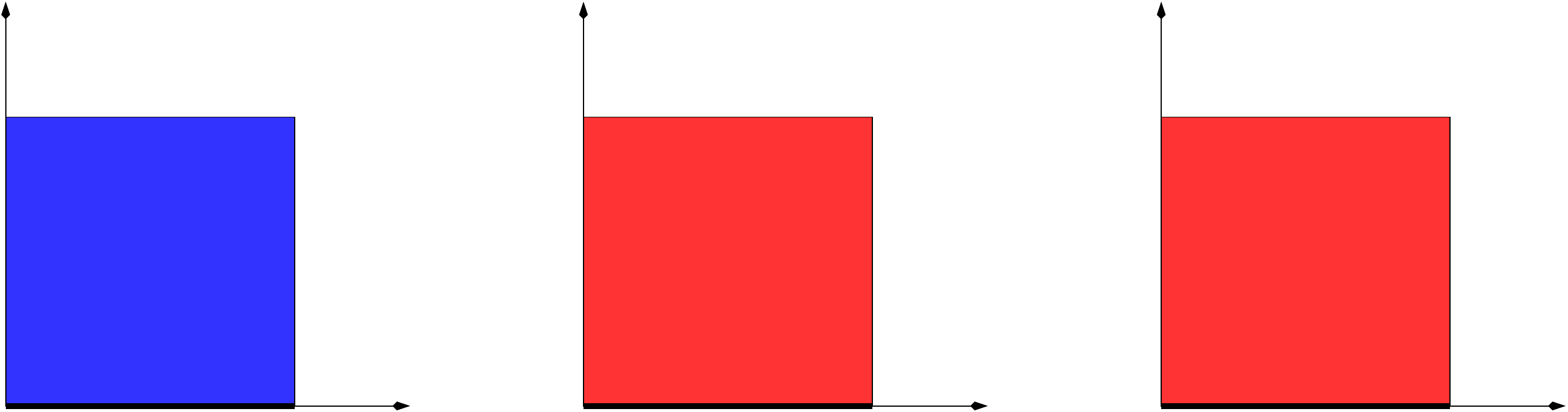}
\begin{picture}(0,0)
    \put(-314,84){$x_2$}
    \put(-201,84){$t_2$}
    \put(-89,84){$t_4$}
    \put(-226,0){$x_1$}
    \put(-113,0){$t_1$}
    \put(0,0){$t_3$} 
    \end{picture}
 \caption{The six-torus of signature $(2,4)$ with O6-plane along $(x_1,t_1,t_3)$.}
  \label{fig:six-torus}
\end{figure}
choose the orientifold projection to be of type $\Omega I_{1,2} (-1)^{F_L}$  reflecting the three
coordinates $\{x_2,t_2,t_4\}$. This  leads to an  O6-planes
parallel to the plane $\{x_1,t_1,t_3\}$. This is shown in figure \ref{fig:six-torus}.
The induced tadpole can be
cancelled by D6-branes on top of the O6-planes. 
Note that these  D6-branes are Lorentzian in the sense that there
is an odd number (namely three) of longitudinal time-like directions.

Moreover, the 4D gauge field
on these D6-branes has the usual kinetic term and is ghost-free. However,
in this toroidal example the Wilson-lines along the  $\{t_1,t_3\}$
directions and the deformations of the brane in the $\{t_2,t_4\}$
directions will be ghosts in the effective 4D theory. However, for
more general internal spaces (something like a CY of signature $(2,4)$)
these open string moduli could be avoided if the D6-branes wrap
a rigid 3-cycle.

In order to see whether dS vacua are in principle possible,
let us investigate which flux components survive the orientifold
projection. For this purpose we recall the general result about the
cohomological classification of the orientifold even fluxes shown in
table~\ref{table_fluxcohom}.

\begin{table}[ht]
\centering
  \begin{tabular}{c c}
  \toprule
  Flux    & {\rm Cohomology}    \\
    \midrule
 &  \\[-0.4cm]
$H$ & $H_-^3(X)$  \\[0.1cm]
$\{ F_0, F_2,F_4 ,F_6\}$ & $\{H^0_+, H^2_-, H^4_+, H^6_-    \}$ \\
\bottomrule
  \end{tabular}
  \caption{Equivariant cohomology groups of orientifold even fluxes.}
  \label{table_fluxcohom}
\end{table}

\noindent
Let us look at $F_2$, for which the flux
\eq{ 
            F_2=f \, dx_1\wedge dt_2 
}
is in $ H^2_-$ and therefore survives the orientifold projection.
Now, since $F_2$ is supported along one space-like and one time-like leg, the kinetic term of this two-form flux
\eq{ 
                |F_2|^2 \sim g^{i_1 i_2}\, g^{j_1 j_2}\,
                  (F_{2})_{ i_1 j_1} \,(F_2)_{i_2 j_2}
}
has the opposite sign to the usual one. For the other fluxes one finds
similar ghost-like components, as well. 
Therefore, this type IIA  model features fluxes with the wrong sign of their
kinetic terms and thus  the usual dS no-go theorem does not apply and 
dS vacua might be possible.
Since  we will see below that these exotic orientifolds have other
problems, it is beyond the scope of this paper to work out in detail a dS model 
on a fully fledged ``CY'' space. At least we can state that up
to this point there is no
immediate obstacle for dS solutions with a ghost-free massless gauge
field  on the brane.

\subsubsection*{Ultralight open string modes}

Recall that for the ghost-free D-branes in Euclidean exotic theories,
the KK and winding modes were such that they contributed like a
positive mass squared $m^2$ to the right hand side of the on-shell relation \eqref{onshellopen}.
This is different for the D-branes in Lorentzian exotic theories.
For such a D-brane the on-shell condition now reads
\eq{
\label{onshellopenLorentz}
   E^2 +\sum_{t_\parallel} \left({m_t\over R_t}\right)^2
             &+ \sum_{t_\perp} \left( {n_t R_t\over \alpha'}\right)^2\\[0.0cm]
     &= \sum_{i}
           (p^i)^2+\sum_{s_\parallel} \left({m_s\over R_s}\right)^2
             + \sum_{s_\perp} \left( {n_s R_s\over \alpha'}\right)^2+{1\over \alpha'}\,  (N-a) \,,
}
where the indices $(s/t)_\parallel$ and $(s/t)_\perp$  indicate space-/time-like
directions parallel and perpendicular to the D-brane world-volume.  
Therefore, here both time-like KK and time-like winding modes contribute always
to the left hand side of this relation. As for the closed string,
these time-like modes can cancel against oscillatory modes yielding
infinitely many arbitrarily light open string modes. This questions the 
role of the Wilsonian effective gauge theory action on the D-brane
and seems to be a  general problem with a potential phenomenological
application of orientifolds of Lorentzian exotic superstring theories
(with multiple times).

\section{Conclusions}

In this paper we performed a detailed study of the brane sector in
exotic string theories. Driven by the motivating example of dS
solutions in supergravity theories with more than one time direction,
we investigated the behavior of exotic string theories
upon torus compactification. Despite being able to get rid of massless
ghosts in the 10D theory by performing suitable orientifold
projections, we found that torus compactifications with time-like
circles lead to an unsatisfactory infinite tower of arbitrarily light
states in the lower dimensional effective theories. One can ask if
this is a generic behavior of time-like compactifications. In
particular there exist pseudo-Riemannian analogs of Calabi-Yau manifolds
with reduced holonomy (see the article of H. Baum in
\cite{pseudoriemann})  which would partially preserve supersymmetry
upon compactifying exotic string theories on them.

Although the closed string sector seems to show unavoidable
pathologies, there might still be a phenomenologically consistent
brane sector coupling to those ``bizarre'' quantum gravity
theories. After all, experiments today don't test quantum gravity,
hence we adopted an agnostic point of view and merely asked for a well
behaved massless open string sector. Taking a perturbative point of
view (after all, the dS solutions arise for the leading order
supergravity actions) we  continued along the line of \cite{Dijkgraaf:2016lym} and constructed a 
CFT description of closed and in particular open Euclidean exotic
string theories. Guided by
consistency of the mathematical formalism we derived all allowed 
D-branes (of real tension) in exotic string theories with all possible metric
signatures. Notably we identified  in every exotic theory  a
phenomenologically consistent D-brane lacking massless ghosts and having
a $(3,1)$-signature subspace. 

The same classification of allowed
branes was obtained by considering negative tension branes in type IIA/B
supergravity theories and performing an analytic continuation of
D-brane actions beyond the horizon in the singular space-time. 
As a special feature of those branes we found that  from a space-time
point of view, their existence is related to precisely the orientifold projections used to discard
massless closed string ghosts. We formulated this result  as an ``exotic"
de Sitter no-go conjecture.

Of course we cannot claim to have given  a viable 
interpretation/description of quantum gravity with multiple time-like
directions. While the formalism of conformal field theory and
supergravity appears to go through for such theories, their conceptual
interpretation still remains elusive and we have not much to add to
that. 
Admittedly, we also  left open a couple of important and interesting
technical questions, the most pressing of which is the role of
supersymmetry. Related to this is the question of general
compactifications on manifolds with pseudo-Riemannian metrics that go 
beyond the toroidal case.


\vspace{1.5cm}
\emph{Acknowledgements:}
We thank Chris Hull for very useful comments about a former version of this paper. 
Furthermore we thank Brage Gording for discussions and Ben Heidenreich for useful comments following a talk by  RB
given at the String Phenomenology 2019 conference at CERN.


\clearpage
\appendix

\section{(A)dS spaces of signature $(p,q)$}
\label{appendix_AdS}

First we introduce the notion of anti-de Sitter and de
Sitter spaces with signature $(p,q)$ where $p$ denotes the number
of time-like directions.\\

\paragraph{$AdS_{p,q}$ spaces:} This space is defined as the real hypersurface 
\eq{
      -\sum_{i=0}^{p}  t_i^2 +\sum_{j=1}^q x_j^2 =-\alpha^2
}
in $\mathbb R_{p+1,q}$, where $\alpha$ denotes a real number.
A solution to this equation can be written as 
\eq{
          t_0&=\alpha \cosh( \rho/\alpha)+{e^{\rho/\alpha}\over
            2\alpha} ( -\sum_{i=1}^p \hat t_i^2 + \sum_{j=2}^q \hat
          x_j^2)\\
          x_1&=\alpha \sinh( \rho/\alpha)-{e^{\rho/\alpha}\over
            2\alpha} ( -\sum_{i=1}^p \hat t_i^2 + \sum_{j=2}^q \hat
          x_j^2)\\
          t_i&= e^{\rho/\alpha} \hat t_i\,,\quad i=1,\ldots p\\
          x_j&= e^{\rho/\alpha} \hat x_j\,,\quad j=2,\ldots q\,.
}
The metric on $AdS_{p,q}$ is then given as
\eq{
\label{metricAdS}
              ds^2=d\rho^2 +e^{2\rho/\alpha}(  -\sum_{i=1}^{p}  d\hat
              t_i^2 +\sum_{j=2}^q d\hat x_j^2 )
}
which is the so-called flat slicing of $AdS_{p,q}$.
One can introduce a corresponding $(p+q)$-bein $E^A$ on $AdS_{p,q}$ 
so that  the metric \eqref{metricAdS} takes the simple form
$ds^2=\eta^{(p,q)}_{AB} E^A\, E^B$.
The resulting Ricci-tensor in this frame reads
\eq{ 
                       R_{AB}=-\eta^{(p,q)}_{AB}\, {(p+q-1)\over
                         \alpha^2}
}
so that the Ricci-scalar becomes $R=-(p+q)(p+q-1)/\alpha^2$.

\paragraph{$dS_{p,q}$ spaces:} Similarly, one can define and describe de Sitter spaces with signature $(p,q)$.
It  is defined as the real hypersurface 
\eq{
      -\sum_{i=0}^{p-1}  t_i^2 +\sum_{j=1}^{q+1} x_j^2 =\beta^2
}
in $\mathbb R_{p,q+1}$ with  a solution given as 
\eq{
          t_0&=\beta \sinh( \tau/\beta)+{e^{\tau/2\beta}\over
            2\beta} ( -\sum_{i=1}^{p-1} \hat t_i^2 + \sum_{j=2}^{q+1} \hat
          x_j^2)\\
          x_1&=\beta \cosh( \tau/\beta)-{e^{\tau/2\beta}\over
            2\beta} ( -\sum_{i=1}^{p-1} \hat t_i^2 + \sum_{j=2}^{q+1} \hat
          x_j^2)\\
          t_i&= e^{\tau/\beta} \hat t_i\,,\quad i=1,\ldots p-1\\
          x_j&= e^{\tau/\beta} \hat x_j\,,\quad j=2,\ldots q+1\,
}
yielding the metric 
\eq{
\label{metricdS}
 ds^2=-d\tau^2 +e^{2\tau/\beta}(  -\sum_{i=1}^{p-1}  d\hat
              t_i^2 +\sum_{j=2}^{q+1} d\hat x_j^2 )\,,
}
from which one can read of a $(p+q)$-bein ${\cal E}_A$.
The resulting Ricci-tensor  reads
\eq{ 
                       R_{AB}=\eta^{(p,q)}_{AB}\, {(p+q-1)\over
                         \beta^2}
}
with the Ricci-scalar $R=(p+q)(p+q-1)/\beta^2$.

\section{Table of branes in exotic IIB theories}
\label{app_IIBtable}

\begin{table}[ht]
\centering
  \begin{adjustbox}{width=\textwidth}
  \begin{tabular}{cccccc|cccccc}
  \toprule
	Theory & Dp  & Branes  & Type(E/L) & Tension  & Ghost-free   &Theory & Dp  & Branes  & Type(E/L) & Tension  & Ghost-free 	\\
	\midrule
	\multirow{8}{*}{IIB$^{-+}_{(9,1)}$} 
	&  D(-1) & D(-1)$^{(0,0)}_{(9,1)}$ &E &$-$  &\checkmark & \multirow{8}{*}{IIB$^{-+}_{(1,9)}$} & D(-1) & D(-1)$^{(0,0)}_{(1,9)}$ &E &$-$  &\checkmark \\ \addlinespace
    &  D1 & D1$^{(1,1)}_{(9,1)}$ &L &$-$  &\checkmark &  & D1 & D1$^{(1,1)}_{(1,9)}$ &L &$-$  &\checkmark \\ \addlinespace
	&  D3 & D3$^{(4,0)}_{(9,1)}$ &E &$+$  & - &&  D3 & D3$^{(0,4)}_{(1,9)}$ &E &$+$  & $-$   \\ \addlinespace
    &  D5 & D5$^{(5,1)}_{(9,1)}$  &L &$+$ &-  &&  D5 & D5$^{(1,5)}_{(1,9)}$  &L &$+$   &$-$   \\ \addlinespace
    &  D7 & D7$^{(8,0)}_{(9,1)}$  &E &$-$ &\checkmark & &  D7 & D7$^{(0,8)}_{(1,9)}$  &E &$-$ &\checkmark  \\ \addlinespace
    &  D9 & D9$^{(9,1)}_{(9,1)}$  &L &$-$ &\checkmark &&  D9 & D9$^{(1,9)}_{(1,9)}$  &L &$-$ &\checkmark  \\  
	\bottomrule
  \end{tabular}
   \end{adjustbox}  
 \caption{Brane spectrum of mirror IIB$^{-+}_{(9,1)/(1,9)}$ theories.}
 \label{table_IIB91}
\end{table}

\begin{table}[ht]
\centering
  \begin{adjustbox}{width=\textwidth}
  \begin{tabular}{cccccc|cccccc}
  \toprule
	Theory & Dp  & Branes  & Type(E/L) & Tension  & Ghost-free   &Theory & Dp  & Branes  & Type(E/L) & Tension  & Ghost-free 	\\
	\midrule
	\multirow{12}{*}{IIB$^{--}_{(7,3)}$} &  D1 & D1$^{(2,0)}_{(7,3)}$ &E &$+$  &- & \multirow{12}{*}{IIB$^{--}_{(3,7)}$} & D1 & D1$^{(0,2)}_{(3,7)}$ &E &$+$  &- \\ \addlinespace
	&    &  D1$^{(0,2)}_{(7,3)}$ &E &$-$  &\checkmark & &    & D1$^{(2,0)}_{(3,7)}$ &E & $-$ &\checkmark \\ \addlinespace
	& D3 &  D3$^{(3,1)}_{(7,3)}$ &L &$+$  & -         & & D3 & D3$^{(1,3)}_{(3,7)}$ &L & $+$ &- \\ \addlinespace
	&    &  D3$^{(1,3)}_{(7,3)}$ &L &$-$  &\checkmark & &    & D3$^{(3,1)}_{(3,7)}$ &L & $-$ &\checkmark \\ \addlinespace
	& D5 &  D5$^{(4,2)}_{(7,3)}$ &E &$+$  & -         & & D5 & D5$^{(2,4)}_{(3,7)}$ &E & $+$ &- \\ \addlinespace
	&    &  D5$^{(6,0)}_{(7,3)}$ &E &$-$  &\checkmark & &    & D5$^{(0,6)}_{(3,7)}$ &E & $-$ &\checkmark \\ \addlinespace
	& D7 &  D7$^{(5,3)}_{(7,3)}$ &L &$+$  & -         & & D7 & D7$^{(3,5)}_{(3,7)}$ &L & $+$ &- \\ \addlinespace
	&    &  D7$^{(7,1)}_{(7,3)}$ &L &$-$  &\checkmark & &    & D7$^{(1,7)}_{(3,7)}$ &L & $-$ &\checkmark \\
	\bottomrule
  \end{tabular}
   \end{adjustbox}  
 \caption{Brane spectrum of mirror IIB$^{--}_{(7,3)/(3,7)}$ theories.}
 \label{table_IIB73}
\end{table}


\begin{table}[bht]
\centering
  \begin{adjustbox}{height=80pt}
\begin{tabular}{cccccc}
  \toprule
	Theory & Dp  & Branes  & Type(E/L) & Tension  & Ghost-free  	\\
	\midrule
	\multirow{15}{*}{IIB$^{-+}_{(5,5)}$} 
	 &  D(-1) & D(-1)$^{(0,0)}_{(5,5)}$  &E &$+$ & - \\ \addlinespace
     &  D1 & D1$^{(1,1)}_{(5,5)}$  &L &$+$ & - \\ \addlinespace
	 &  D3 & D3$^{(2,2)}_{(5,5)}$  &E &$+$ & -   		\\ \addlinespace
	    &      & D3$^{(0,4)}_{(5,5)}$  &E &$-$ &\checkmark \\ \addlinespace
   		&      & D3$^{(4,0)}_{(5,5)}$  &E &$-$ &\checkmark \\ \addlinespace
     &  D5 & D5$^{(3,3)}_{(5,5)}$  &L &$+$ & -   		\\ \addlinespace
        &      & D5$^{(1,5)}_{(5,5)}$  &L &$-$ &\checkmark \\ \addlinespace
   &      & D5$^{(5,1)}_{(5,5)}$  &L &$-$ &\checkmark \\ \addlinespace
     &  D7 & D7$^{(4,4)}_{(5,5)}$  &E &$+$ & -  \\ \addlinespace
     &  D9 & D9$^{(5,5)}_{(5,5)}$  &L &$+$ & - \\ 
     \bottomrule
 \end{tabular}	
    \end{adjustbox}
 \caption{Brane spectrum of IIB$^{-+}_{(5,5)}$.}
  \label{table_IIB55}
\end{table}

\newpage

\section{Table of branes in exotic IIA theories}
\label{app_IIAtable}

\begin{table}[ht!]
\centering
  \begin{adjustbox}{width=\textwidth}
  \begin{tabular}{cccccc|cccccc}
  \toprule
	Theory & Dp  & Branes  & Type(E/L) & Tension  & Ghost-free   &Theory & Dp  & Branes  & Type(E/L) & Tension  & Ghost-free 	\\
	\midrule
	\multirow{8}{*}{IIA$^{-+}_{(10,0)}$} &  D0 &  D0$^{(1,0)}_{(10,0)}$ &E &$-$  &\checkmark & \multirow{8}{*}{IIA$^{--}_{(0,10)}$} & D0 &  D0$^{(0,1)}_{(0,10)}$ &L &$-$  &\checkmark  \\ \addlinespace
	& D2 &\multicolumn{4}{c}{\emph{No consistent D2-brane configuration}} & & D2 &\multicolumn{4}{c}{\emph{No consistent D2-brane configuration}} \\ \addlinespace
	& D4 &  D4$^{(5,0)}_{(10,0)}$ &E &$+$  & -         & & D4 & D4$^{(0,5)}_{(0,10)}$ &L & $+$ &- \\ \addlinespace
	& D6 &\multicolumn{4}{c}{\emph{No consistent D6-brane configuration}}& & D6 &\multicolumn{4}{c}{\emph{No consistent D6-brane configuration}} \\ \addlinespace
	& D8 &  D8$^{(9,0)}_{(10,0)}$ &E &$-$  &\checkmark & & D8 & D8$^{(0,9)}_{(0,10)}$ &L & $-$ &\checkmark \\ 
	\bottomrule
  \end{tabular}
   \end{adjustbox}  
 \caption{Brane spectrum of mirror IIA$^{-+}_{(10,0)}$ and IIA$^{--}_{(0,10)}$ theories.}
 \label{table_IIA100}
\end{table}

\begin{table}[ht!]
\centering
  \begin{adjustbox}{width=\textwidth}
  \begin{tabular}{cccccc|cccccc}
  \toprule
	Theory & Dp  & Branes  & Type(E/L) & Tension  & Ghost-free   &Theory & Dp  & Branes  & Type(E/L) & Tension  & Ghost-free 	\\
	\midrule
	\multirow{12}{*}{IIA$^{--}_{(8,2)}$} &  D0 &  D0$^{(0,1)}_{(8,2)}$ &L &$-$  &\checkmark &\multirow{12}{*}{IIA$^{-+}_{(2,8)}$} & D0 &  D0$^{(1,0)}_{(2,8)}$ &E &$-$  &\checkmark  \\ \addlinespace
	& D2 &  D2$^{(1,2)}_{(8,2)}$ &E &$-$  &\checkmark & & D2 & D2$^{(2,1)}_{(2,8)}$ &L & $-$ &\checkmark \\ \addlinespace
	&    &  D2$^{(3,0)}_{(8,2)}$ &E &$+$  & -         & &    & D2$^{(0,3)}_{(2,8)}$ &L & $+$ &- \\ \addlinespace
	& D4 &  D4$^{(4,1)}_{(8,2)}$ &L &$+$  & -         & & D4 & D4$^{(1,4)}_{(2,8)}$ &E & $+$ &- \\ \addlinespace
	& D6 &  D6$^{(5,2)}_{(8,2)}$ &E &$+$  & -         & & D6 & D6$^{(2,5)}_{(2,8)}$ &L & $+$ &- \\ \addlinespace
	&    &  D6$^{(7,0)}_{(8,2)}$ &E &$-$  &\checkmark & &    & D6$^{(0,7)}_{(2,8)}$ &L & $-$ &\checkmark \\ \addlinespace
	& D8 &  D8$^{(8,1)}_{(8,2)}$ &L &$-$  &\checkmark & & D8 & D8$^{(1,8)}_{(2,8)}$ &E & $-$ &\checkmark \\ 
	\bottomrule
  \end{tabular}
   \end{adjustbox}  
 \caption{Brane spectrum of mirror IIA$^{--}_{(8,2)}$ and IIA$^{-+}_{(2,8)}$ theories.}
 \label{table_IIA82}
\end{table}

\begin{table}[ht!]
\centering
  \begin{adjustbox}{width=\textwidth}
  \begin{tabular}{cccccc|cccccc}
  \toprule
	Theory & Dp  & Branes  & Type(E/L) & Tension  & Ghost-free   &Theory & Dp  & Branes  & Type(E/L) & Tension  & Ghost-free 	\\
	\midrule
	\multirow{15}{*}{IIA$^{-+}_{(6,4)}$} &  D0 &  D0$^{(1,0)}_{(6,4)}$ &E &$+$  &- &\multirow{15}{*}{IIA$^{--}_{(4,6)}$}& D0 &  D0$^{(0,1)}_{(4,6)}$ &L &$+$  &- \\ \addlinespace
	& D2 &  D2$^{(0,3)}_{(6,4)}$ &L &$-$  &\checkmark & & D2 & D2$^{(1,2)}_{(4,6)}$ &E & $-$ &\checkmark \\ \addlinespace
	&    &  D2$^{(2,1)}_{(6,4)}$ &L &$+$  & -         & &    & D2$^{(3,0)}_{(4,6)}$ &E & $+$ &- \\ \addlinespace
	& D4 &  D4$^{(1,4)}_{(6,4)}$ &E &$-$  &\checkmark & & D4 & D4$^{(4,1)}_{(4,6)}$ &L & $-$ &\checkmark \\ \addlinespace
	&    &  D4$^{(3,2)}_{(6,4)}$ &E &$+$  & -         & &    & D4$^{(2,3)}_{(4,6)}$ &L & $+$ &- \\ \addlinespace
	&    &  D4$^{(5,0)}_{(6,4)}$ &E &$-$  &\checkmark & &    & D4$^{(0,5)}_{(4,6)}$ &L & $-$ &\checkmark \\ \addlinespace
	& D6 &  D6$^{(4,3)}_{(6,4)}$ &L &$+$  & -         & & D6 & D6$^{(3,4)}_{(4,6)}$ &E & $+$ &- \\ \addlinespace
	&    &  D6$^{(6,1)}_{(6,4)}$ &L &$-$  &\checkmark & &    & D6$^{(1,6)}_{(4,6)}$ &E & $-$ &\checkmark \\ \addlinespace
	& D8 &  D8$^{(5,4)}_{(6,4)}$ &E &$+$  &-          & & D8 & D8$^{(4,5)}_{(4,6)}$ &L & $+$ & - \\ 
	\bottomrule
  \end{tabular}
   \end{adjustbox}  
 \caption{Brane spectrum of mirror IIA$^{-+}_{(6,4)}$ and IIA$^{--}_{(4,6)}$ theories.}
 \label{table_IIA64}
\end{table}

\clearpage

\bibliography{references}

\providecommand{\href}[2]{#2}\begingroup\raggedright\begin{thebibliography}{10}

\bibitem{Vafa:2005ui}
C.~Vafa, ``{The String landscape and the swampland},''
\href{http://arxiv.org/abs/hep-th/0509212}{{\ttfamily arXiv:hep-th/0509212
  [hep-th]}}.

\bibitem{Palti:2019pca}
E.~Palti, ``{The Swampland: Introduction and Review},''
  \href{http://dx.doi.org/10.1002/prop.201900037}{{\em Fortsch. Phys.}
  {\bfseries 67} no.~6, (2019) 1900037},
\href{http://arxiv.org/abs/1903.06239}{{\ttfamily arXiv:1903.06239 [hep-th]}}.

\bibitem{ArkaniHamed:2006dz}
N.~Arkani-Hamed, L.~Motl, A.~Nicolis, and C.~Vafa, ``{The String landscape,
  black holes and gravity as the weakest force},''
  \href{http://dx.doi.org/10.1088/1126-6708/2007/06/060}{{\em JHEP} {\bfseries
  06} (2007) 060},
\href{http://arxiv.org/abs/hep-th/0601001}{{\ttfamily arXiv:hep-th/0601001
  [hep-th]}}.

\bibitem{Ooguri:2006in}
H.~Ooguri and C.~Vafa, ``{On the Geometry of the String Landscape and the
  Swampland},'' \href{http://dx.doi.org/10.1016/j.nuclphysb.2006.10.033}{{\em
  Nucl. Phys.} {\bfseries B766} (2007) 21--33},
\href{http://arxiv.org/abs/hep-th/0605264}{{\ttfamily arXiv:hep-th/0605264
  [hep-th]}}.

\bibitem{Klaewer:2016kiy}
D.~Kl{\"a}wer and E.~Palti, ``{Super-Planckian Spatial Field Variations and
  Quantum Gravity},'' \href{http://dx.doi.org/10.1007/JHEP01(2017)088}{{\em
  JHEP} {\bfseries 01} (2017) 088},
\href{http://arxiv.org/abs/1610.00010}{{\ttfamily arXiv:1610.00010 [hep-th]}}.

\bibitem{Ooguri:2016pdq}
H.~Ooguri and C.~Vafa, ``{Non-supersymmetric AdS and the Swampland},''
  \href{http://dx.doi.org/10.4310/ATMP.2017.v21.n7.a8}{{\em Adv. Theor. Math.
  Phys.} {\bfseries 21} (2017) 1787--1801},
\href{http://arxiv.org/abs/1610.01533}{{\ttfamily arXiv:1610.01533 [hep-th]}}.

\bibitem{Palti:2017elp}
E.~Palti, ``{The Weak Gravity Conjecture and Scalar Fields},''
  \href{http://dx.doi.org/10.1007/JHEP08(2017)034}{{\em JHEP} {\bfseries 08}
  (2017) 034},
\href{http://arxiv.org/abs/1705.04328}{{\ttfamily arXiv:1705.04328 [hep-th]}}.

\bibitem{Obied:2018sgi}
G.~Obied, H.~Ooguri, L.~Spodyneiko, and C.~Vafa, ``{De Sitter Space and the
  Swampland},''
\href{http://arxiv.org/abs/1806.08362}{{\ttfamily arXiv:1806.08362 [hep-th]}}.

\bibitem{Andriot:2018wzk}
D.~Andriot, ``{On the de Sitter swampland criterion},''
  \href{http://dx.doi.org/10.1016/j.physletb.2018.09.022}{{\em Phys. Lett.}
  {\bfseries B785} (2018) 570--573},
\href{http://arxiv.org/abs/1806.10999}{{\ttfamily arXiv:1806.10999 [hep-th]}}.

\bibitem{Cecotti:2018ufg}
S.~Cecotti and C.~Vafa, ``{Theta-problem and the String Swampland},''
\href{http://arxiv.org/abs/1808.03483}{{\ttfamily arXiv:1808.03483 [hep-th]}}.

\bibitem{Garg:2018reu}
S.~K. Garg and C.~Krishnan, ``{Bounds on Slow Roll and the de Sitter
  Swampland},'' \href{http://dx.doi.org/10.1007/JHEP11(2019)075}{{\em JHEP}
  {\bfseries 11} (2019) 075},
\href{http://arxiv.org/abs/1807.05193}{{\ttfamily arXiv:1807.05193 [hep-th]}}.

\bibitem{Ooguri:2018wrx}
H.~Ooguri, E.~Palti, G.~Shiu, and C.~Vafa, ``{Distance and de Sitter
  Conjectures on the Swampland},''
  \href{http://dx.doi.org/10.1016/j.physletb.2018.11.018}{{\em Phys. Lett.}
  {\bfseries B788} (2019) 180--184},
\href{http://arxiv.org/abs/1810.05506}{{\ttfamily arXiv:1810.05506 [hep-th]}}.

\bibitem{Gautason:2018gln}
F.~F. Gautason, V.~Van~Hemelryck, and T.~Van~Riet, ``{The Tension between 10D
  Supergravity and dS Uplifts},''
  \href{http://dx.doi.org/10.1002/prop.201800091}{{\em Fortsch. Phys.}
  {\bfseries 67} no.~1-2, (2019) 1800091},
\href{http://arxiv.org/abs/1810.08518}{{\ttfamily arXiv:1810.08518 [hep-th]}}.

\bibitem{Klaewer:2018yxi}
D.~Kl{\"a}wer, D.~L{\"u}st, and E.~Palti, ``{A Spin 2 Conjecture on the
  Swampland},'' \href{http://dx.doi.org/10.1002/prop.201800102}{{\em Fortsch.
  Phys.} {\bfseries 67} no.~1-2, (2019) 1800102},
\href{http://arxiv.org/abs/1811.07908}{{\ttfamily arXiv:1811.07908 [hep-th]}}.

\bibitem{Heckman:2019bzm}
J.~J. Heckman and C.~Vafa, ``{Fine Tuning, Sequestering, and the Swampland},''
  \href{http://dx.doi.org/10.1016/j.physletb.2019.135004}{{\em Phys. Lett.}
  {\bfseries B798} (2019) 135004},
\href{http://arxiv.org/abs/1905.06342}{{\ttfamily arXiv:1905.06342 [hep-th]}}.

\bibitem{Lust:2019zwm}
D.~L{\"u}st, E.~Palti, and C.~Vafa, ``{AdS and the Swampland},''
  \href{http://dx.doi.org/10.1016/j.physletb.2019.134867}{{\em Phys. Lett.}
  {\bfseries B797} (2019) 134867},
\href{http://arxiv.org/abs/1906.05225}{{\ttfamily arXiv:1906.05225 [hep-th]}}.

\bibitem{Bedroya:2019snp}
A.~Bedroya and C.~Vafa, ``{Trans-Planckian Censorship and the Swampland},''
\href{http://arxiv.org/abs/1909.11063}{{\ttfamily arXiv:1909.11063 [hep-th]}}.

\bibitem{Kehagias:2019akr}
A.~Kehagias, D.~L{\"u}st, and S.~L{\"u}st, ``{Swampland, Gradient Flow and
  Infinite Distance},''
\href{http://arxiv.org/abs/1910.00453}{{\ttfamily arXiv:1910.00453 [hep-th]}}.

\bibitem{Blumenhagen:2019vgj}
R.~Blumenhagen, M.~Brinkmann, and A.~Makridou, ``{Quantum Log-Corrections to
  Swampland Conjectures},''
\href{http://arxiv.org/abs/1910.10185}{{\ttfamily arXiv:1910.10185 [hep-th]}}.

\bibitem{Heidenreich:2017sim}
B.~Heidenreich, M.~Reece, and T.~Rudelius, ``{The Weak Gravity Conjecture and
  Emergence from an Ultraviolet Cutoff},''
  \href{http://dx.doi.org/10.1140/epjc/s10052-018-5811-3}{{\em Eur. Phys. J.}
  {\bfseries C78} no.~4, (2018) 337},
\href{http://arxiv.org/abs/1712.01868}{{\ttfamily arXiv:1712.01868 [hep-th]}}.

\bibitem{Grimm:2018ohb}
T.~W. Grimm, E.~Palti, and I.~Valenzuela, ``{Infinite Distances in Field Space
  and Massless Towers of States},''
  \href{http://dx.doi.org/10.1007/JHEP08(2018)143}{{\em JHEP} {\bfseries 08}
  (2018) 143},
\href{http://arxiv.org/abs/1802.08264}{{\ttfamily arXiv:1802.08264 [hep-th]}}.

\bibitem{Heidenreich:2018kpg}
B.~Heidenreich, M.~Reece, and T.~Rudelius, ``{Emergence of Weak Coupling at
  Large Distance in Quantum Gravity},''
  \href{http://dx.doi.org/10.1103/PhysRevLett.121.051601}{{\em Phys. Rev.
  Lett.} {\bfseries 121} no.~5, (2018) 051601},
\href{http://arxiv.org/abs/1802.08698}{{\ttfamily arXiv:1802.08698 [hep-th]}}.

\bibitem{Lee:2018urn}
S.-J. Lee, W.~Lerche, and T.~Weigand, ``{Tensionless Strings and the Weak
  Gravity Conjecture},'' \href{http://dx.doi.org/10.1007/JHEP10(2018)164}{{\em
  JHEP} {\bfseries 10} (2018) 164},
\href{http://arxiv.org/abs/1808.05958}{{\ttfamily arXiv:1808.05958 [hep-th]}}.

\bibitem{Lee:2018spm}
S.-J. Lee, W.~Lerche, and T.~Weigand, ``{A Stringy Test of the Scalar Weak
  Gravity Conjecture},''
  \href{http://dx.doi.org/10.1016/j.nuclphysb.2018.11.001}{{\em Nucl. Phys.}
  {\bfseries B938} (2019) 321--350},
\href{http://arxiv.org/abs/1810.05169}{{\ttfamily arXiv:1810.05169 [hep-th]}}.

\bibitem{Lee:2019wij}
S.-J. Lee, W.~Lerche, and T.~Weigand, ``{Emergent Strings from Infinite
  Distance Limits},''
\href{http://arxiv.org/abs/1910.01135}{{\ttfamily arXiv:1910.01135 [hep-th]}}.

\bibitem{Danielsson:2018ztv}
U.~H. Danielsson and T.~{Van Riet}, ``{What if string theory has no de Sitter
  vacua?},'' \href{http://dx.doi.org/10.1142/S0218271818300070}{{\em Int. J.
  Mod. Phys.} {\bfseries D27} no.~12, (2018) 1830007},
\href{http://arxiv.org/abs/1804.01120}{{\ttfamily arXiv:1804.01120 [hep-th]}}.

\bibitem{Dvali:2014gua}
G.~Dvali and C.~Gomez, ``{Quantum Exclusion of Positive Cosmological
  Constant?},'' \href{http://dx.doi.org/10.1002/andp.201500216}{{\em Annalen
  Phys.} {\bfseries 528} (2016) 68--73},
\href{http://arxiv.org/abs/1412.8077}{{\ttfamily arXiv:1412.8077 [hep-th]}}.

\bibitem{Dvali:2017eba}
G.~Dvali, C.~Gomez, and S.~Zell, ``{Quantum Break-Time of de Sitter},''
  \href{http://dx.doi.org/10.1088/1475-7516/2017/06/028}{{\em JCAP} {\bfseries
  1706} (2017) 028},
\href{http://arxiv.org/abs/1701.08776}{{\ttfamily arXiv:1701.08776 [hep-th]}}.

\bibitem{Dvali:2018jhn}
G.~Dvali, C.~Gomez, and S.~Zell, ``{Quantum Breaking Bound on de Sitter and
  Swampland},'' \href{http://dx.doi.org/10.1002/prop.201800094}{{\em Fortsch.
  Phys.} {\bfseries 67} no.~1-2, (2019) 1800094},
\href{http://arxiv.org/abs/1810.11002}{{\ttfamily arXiv:1810.11002 [hep-th]}}.

\bibitem{Dvali:2018fqu}
G.~Dvali and C.~Gomez, ``{On Exclusion of Positive Cosmological Constant},''
  \href{http://dx.doi.org/10.1002/prop.201800092}{{\em Fortsch. Phys.}
  {\bfseries 67} no.~1-2, (2019) 1800092},
\href{http://arxiv.org/abs/1806.10877}{{\ttfamily arXiv:1806.10877 [hep-th]}}.

\bibitem{Dasgupta:2019gcd}
K.~Dasgupta, M.~Emelin, M.~M. Faruk, and R.~Tatar, ``{de Sitter Vacua in the
  String Landscape},''
\href{http://arxiv.org/abs/1908.05288}{{\ttfamily arXiv:1908.05288 [hep-th]}}.

\bibitem{Dasgupta:2019vjn}
K.~Dasgupta, M.~Emelin, M.~M. Faruk, and R.~Tatar, ``{How a four-dimensional de
  Sitter solution remains outside the swampland},''
\href{http://arxiv.org/abs/1911.02604}{{\ttfamily arXiv:1911.02604 [hep-th]}}.

\bibitem{Hull:1998vg}
C.~M. Hull, ``{Timelike T duality, de Sitter space, large N gauge theories and
  topological field theory},''
  \href{http://dx.doi.org/10.1088/1126-6708/1998/07/021}{{\em JHEP} {\bfseries
  07} (1998) 021},
\href{http://arxiv.org/abs/hep-th/9806146}{{\ttfamily arXiv:hep-th/9806146
  [hep-th]}}.

\bibitem{Hull:1998ym}
C.~M. Hull, ``{Duality and the signature of space-time},''
  \href{http://dx.doi.org/10.1088/1126-6708/1998/11/017}{{\em JHEP} {\bfseries
  11} (1998) 017},
\href{http://arxiv.org/abs/hep-th/9807127}{{\ttfamily arXiv:hep-th/9807127
  [hep-th]}}.

\bibitem{Hull:1998fh}
C.~M. Hull and R.~R. Khuri, ``{Branes, times and dualities},''
  \href{http://dx.doi.org/10.1016/S0550-3213(98)00691-9}{{\em Nucl. Phys.}
  {\bfseries B536} (1998) 219--244},
\href{http://arxiv.org/abs/hep-th/9808069}{{\ttfamily arXiv:hep-th/9808069
  [hep-th]}}.

\bibitem{Hull:1999mt}
C.~M. Hull and R.~R. Khuri, ``{World volume theories, holography, duality and
  time},'' \href{http://dx.doi.org/10.1016/S0550-3213(00)00057-2}{{\em Nucl.
  Phys.} {\bfseries B575} (2000) 231--254},
\href{http://arxiv.org/abs/hep-th/9911082}{{\ttfamily arXiv:hep-th/9911082
  [hep-th]}}.

\bibitem{Bobev:2018ugk}
N.~Bobev, P.~Bomans, and F.~F. Gautason, ``{Spherical Branes},''
  \href{http://dx.doi.org/10.1007/JHEP08(2018)029}{{\em JHEP} {\bfseries 08}
  (2018) 029},
\href{http://arxiv.org/abs/1805.05338}{{\ttfamily arXiv:1805.05338 [hep-th]}}.

\bibitem{Bobev:2019bvq}
N.~Bobev, P.~Bomans, F.~F. Gautason, J.~A. Minahan, and A.~Nedelin,
  ``{Supersymmetric Yang-Mills, Spherical Branes, and Precision Holography},''
\href{http://arxiv.org/abs/1910.08555}{{\ttfamily arXiv:1910.08555 [hep-th]}}.

\bibitem{Dijkgraaf:2016lym}
R.~Dijkgraaf, B.~Heidenreich, P.~Jefferson, and C.~Vafa, ``{Negative Branes,
  Supergroups and the Signature of Spacetime},''
  \href{http://dx.doi.org/10.1007/JHEP02(2018)050}{{\em JHEP} {\bfseries 02}
  (2018) 050},
\href{http://arxiv.org/abs/1603.05665}{{\ttfamily arXiv:1603.05665 [hep-th]}}.

\bibitem{Hertzberg:2007wc}
M.~P. Hertzberg, S.~Kachru, W.~Taylor, and M.~Tegmark, ``{Inflationary
  Constraints on Type IIA String Theory},''
  \href{http://dx.doi.org/10.1088/1126-6708/2007/12/095}{{\em JHEP} {\bfseries
  12} (2007) 095},
\href{http://arxiv.org/abs/0711.2512}{{\ttfamily arXiv:0711.2512 [hep-th]}}.

\bibitem{Ooguri:1991ie}
H.~Ooguri and C.~Vafa, ``{N=2 heterotic strings},''
\href{http://dx.doi.org/10.1016/0550-3213(91)90042-V}{{\em Nucl. Phys.}
  {\bfseries B367} (1991) 83--104}.

\bibitem{Maldacena:2000mw}
J.~M. Maldacena and C.~Nunez, ``{Supergravity description of field theories on
  curved manifolds and a no go theorem},''
  \href{http://dx.doi.org/10.1142/S0217751X01003935,
  10.1142/S0217751X01003937}{{\em Int. J. Mod. Phys.} {\bfseries A16} (2001)
  822--855}, \href{http://arxiv.org/abs/hep-th/0007018}{{\ttfamily
  arXiv:hep-th/0007018 [hep-th]}}.
[,182(2000)].

\bibitem{Junghans:2018gdb}
D.~Junghans, ``{Weakly Coupled de Sitter Vacua with Fluxes and the
  Swampland},'' \href{http://dx.doi.org/10.1007/JHEP03(2019)150}{{\em JHEP}
  {\bfseries 03} (2019) 150},
\href{http://arxiv.org/abs/1811.06990}{{\ttfamily arXiv:1811.06990 [hep-th]}}.

\bibitem{Ooguri:1991fp}
H.~Ooguri and C.~Vafa, ``{Geometry of N=2 strings},''
\href{http://dx.doi.org/10.1016/0550-3213(91)90270-8}{{\em Nucl. Phys.}
  {\bfseries B361} (1991) 469--518}.

\bibitem{Quiros:2007ym}
I.~Quiros, ``{Time-like versus space-like extra dimensions},''
\href{http://arxiv.org/abs/0707.0714}{{\ttfamily arXiv:0707.0714 [gr-qc]}}.

\bibitem{Dirac:1945cm}
P.~A.~M. Dirac, ``{Unitary Representations of the Lorentz Group},''
\href{http://dx.doi.org/10.1098/rspa.1945.0003}{{\em Proc. Roy. Soc. Lond.}
  {\bfseries A183} (1945) 284--295}.

\bibitem{Pauli:1949zm}
W.~Pauli and F.~Villars, ``{On the Invariant regularization in relativistic
  quantum theory},''
\href{http://dx.doi.org/10.1103/RevModPhys.21.434}{{\em Rev. Mod. Phys.}
  {\bfseries 21} (1949) 434--444}.

\bibitem{Gupta:1949rh}
S.~N. Gupta, ``{Theory of longitudinal photons in quantum electrodynamics},''
\href{http://dx.doi.org/10.1088/0370-1298/63/7/301}{{\em Proc. Phys. Soc.}
  {\bfseries A63} (1950) 681--691}.

\bibitem{Bleuler:1950cy}
K.~Bleuler, ``{A New method of treatment of the longitudinal and scalar
  photons},''
{\em Helv. Phys. Acta} {\bfseries 23} (1950) 567--586.

\bibitem{DiVecchia:1999mal}
P.~Di~Vecchia and A.~Liccardo, ``{D Branes in String Theory, I},''
  \href{http://dx.doi.org/10.1007/978-94-011-4303-5_1}{{\em NATO Sci. Ser. C}
  {\bfseries 556} (2000) 1--60},
\href{http://arxiv.org/abs/hep-th/9912161}{{\ttfamily arXiv:hep-th/9912161
  [hep-th]}}.

\bibitem{DiVecchia:1999fje}
P.~Di~Vecchia and A.~Liccardo, ``{D-branes in string theory. 2.},'' in {\em
  {YITP Workshop on Developments in Superstring and M Theory Kyoto, Japan,
  October 27-29, 1999}}, pp.~7--48.
\newblock 1999.
\newblock
\href{http://arxiv.org/abs/hep-th/9912275}{{\ttfamily arXiv:hep-th/9912275
  [hep-th]}}.
\newblock

\bibitem{Blumenhagen:2009zz}
R.~Blumenhagen and E.~Plauschinn, ``{Introduction to conformal field theory},''
\href{http://dx.doi.org/10.1007/978-3-642-00450-6}{{\em Lect. Notes Phys.}
  {\bfseries 779} (2009) 1--256}.

\bibitem{Recknagel:2013uja}
A.~Recknagel and V.~Schomerus,
  \href{http://dx.doi.org/10.1017/CBO9780511806476}{{\em {Boundary Conformal
  Field Theory and the Worldsheet Approach to D-Branes}}}.
\newblock Cambridge Monographs on Mathematical Physics. Cambridge University
  Press, 2013.
\newblock
\url{http://www.cambridge.org/9780521832236}.
\newblock

\bibitem{Angelantonj:2002ct}
C.~Angelantonj and A.~Sagnotti, ``{Open strings},''
  \href{http://dx.doi.org/10.1016/S0370-1573(02)00273-9,
  10.1016/S0370-1573(03)00006-1}{{\em Phys. Rept.} {\bfseries 371} (2002)
  1--150}, \href{http://arxiv.org/abs/hep-th/0204089}{{\ttfamily
  arXiv:hep-th/0204089 [hep-th]}}.
[Erratum: Phys. Rept.376,no.6,407(2003)].

\bibitem{pseudoriemann}
V.~Cort{\'e}s, ed., {\em {Handbook of Pseudo-Riemannian Geometry and
  Supersymmetry}}, vol.~16 of {\em {IRMA Lectures in Mathematics and
  Theoretical Physics}}.
\newblock European Mathematical Society, Z{\"u}rich, Switzerland, 2010.

\end{thebibliography}\endgroup
\bibliographystyle{utphys}


\end{document}